\RequirePackage{amsmath}
\documentclass[runningheads]{llncs}
\usepackage[table,xcdraw]{xcolor}
\usepackage{longtable}
\usepackage{pdfpages}
\usepackage[utf8]{inputenc}

\usepackage{graphicx}
\usepackage{vdmlisting}
\usepackage{hyperref}
\usepackage{times}
\usepackage{url}
\usepackage{tcolorbox}

\usepackage{graphicx}
\usepackage{vdmlisting}
\usepackage{todonotes}
\usepackage{pgfplotstable}
\usepackage{pgfplots}
\usepackage{booktabs}
\usepackage{tikz}
\usetikzlibrary{trees,shapes,arrows}
\usetikzlibrary{shapes.multipart}
\usetikzlibrary{positioning}
\usepackage{amssymb}
\usepackage{amsmath}
\usepackage{multirow}

\usepackage{graphicx}

\usepackage{paralist}
\usepackage{graphicx}
\usepackage{subcaption}

\usepackage{xspace}
\newcommand{\eg}{e.\,g.,\ }
\newcommand{\ie}{i.\,e.,\ }

\usepackage{soul}

\usepackage[acronym]{glossaries} 
\newacronym{ast}{AST}{Abstract Syntax Tree}
\newacronym{ct}{CT}{Combinatorial Testing}
\newacronym{czt}{CZT}{Community Z Tools}
\newacronym{dap}{DAP}{Debug Adapter Protocol}
\newacronym{dbgp}{DBGP}{Common DeBugGer Protocol}
\newacronym{gui}{GUI}{Graphical User Interface}
\newacronym{ide}{IDE}{Integrated Development Environment}
\newacronym{lsp}{LSP}{Language Server Protocol}
\newacronym{poc}{PoC}{Proof of Concept}
\newacronym{pog}{POG}{Proof Obligation Generation}
\newacronym{slsp}{SLSP}{Specification Language Server Protocol}
\newacronym{vdm}{VDM}{Vienna Development Method}
\newacronym{vscode}{VS Code}{Visual Studio Code}

\usepackage{chngcntr}

\usepackage{varioref}

\usepackage[capitalise,nameinlink]{cleveref}
\crefname{section}{Sect.}{Sect.}
\Crefname{section}{Section}{Sections}
\crefname{lstlisting}{Listing}{Listing}
\Crefname{lstlisting}{Listing}{Listing}

\usepackage{listings} 
\usepackage{times}    

\lstdefinelanguage{json}{
    basicstyle=\ttfamily\small, 
    numbers=left,
    stepnumber=1,
    numbersep=8pt,
    breaklines=true,
    frame=single,
    xleftmargin=.11\textwidth, 
    xrightmargin=.11\textwidth
}

\usepackage{subcaption}

\usepackage{graphicx}

\usepackage{algorithm}
\usepackage{algpseudocode}
\usepackage{pseudocode}

\usepackage{tikz}
\usetikzlibrary{positioning}

\usepackage{cprotect}
\usepackage{multirow}
\usepackage{graphicx}
\usepackage{xspace}
\usepackage{paralist}
\usepackage{fancybox}
\usepackage{calc}
\usepackage{verbatim}
\usepackage{marginnote}
\usepackage{todonotes}

\usepackage{booktabs}
\usepackage[scaled]{helvet}

\usepackage{times}

\usepackage{listingsVDM}

\newcommand{\keywords}[1]{\par\addvspace\baselineskip
  \noindent\keywordname\enspace\ignorespaces#1}

\usepackage{graphicx}
\usepackage{svg}
\usepackage{amsmath}
\usepackage{subcaption}
\usepackage{url}
\usepackage{tablefootnote}
\usepackage{dirtytalk} 
\newcommand{\keepvalues}{%
  \edef\restorevalues{%
    \textfloatsep=\textfloatsep
    \intextsep=\the\intextsep
    \floatsep=\the\floatsep
    \abovecaptionskip=\the\abovecaptionskip
    \belowcaptionskip=\the\belowcaptionskip
  }%
}

\begin{document}

\title{Proceedings of the 18$^{th}$ International Overture Workshop}
\institute{}
\author{John Fitzgerald \and Tomohiro Oda  \and Hugo Daniel Macedo (Editors)}
%
%

\maketitle

\setcounter{page}{1}

\chapter*{Preface}
\markboth{Preface}{Preface}

The 18th in the “Overture” series of workshops on the Vienna Development Method (VDM), its
associated tools and applications, was held online, on 7th December 2020. VDM is one of the longest
established formal methods, and yet has a lively community of researchers and practitioners in
academia and industry grown around the modelling languages (VDM-SL, VDM++, VDM-RT) and tools
(VDMTools, VDMJ, ViennaTalk, Overture, Crescendo, Symphony, and the INTO-CPS chain). Together,
these provide a platform for work on modelling and analysis technology that includes static and
dynamic analysis, test generation, execution support, and model checking.

Research in VDM and Overture is driven by the need to develop industry practice as well as
fundamental theory. The 18th Workshop reflected the breadth and depth of work supporting and
applying VDM. In this technical report, we include two technical sessions with papers. The first
session consists of three papers associated with tool building and cloud computing systems, and the
second consists of two papers on co-simulation and Cyber-Physical Systems.

As a consequence of the pandemic that gripped the world in 2020, this was the first Overture
Workshop to be conducted entirely online. Thinking back over the 17 previous times on which we
had shared our meetings, as well as hospitality, in person in locations around the world, this was a
bittersweet event. It was bitter in the sense that we were separated from colleagues and friends,
and that pressures of other tasks had kept the contributions to a minimum. At the same time, it was
a great pleasure to share a few hours again with colleagues and friends with whom we have, over
decades, come to share scientific interests and enthusiasms.

We would like to thank the authors, PC members, reviewers and participants for their help in making
this a valuable and successful workshop, and we look forward together to meeting once more in a
safer world in 2021.

\medskip
\begin{flushright}\noindent
John Fitzgerald, Newcastle\\
Tomohiro Oda, Tokyo\\
Hugo Daniel Macedo, Aarhus
\end{flushright}

\tableofcontents

\chapter*{Organization}

\section*{Programme Committee}
\begin{longtable}{p{0.3\textwidth}p{0.7\textwidth}}
Keijiro Araki & National Institute of Technology, Kumamoto College, Japan\\[12pt]
Nick Battle & Newcastle University, UK\\[12pt]
Luis D. Couto & Forcepoint, Ireland\\[12pt]
Fuyuki Ishikawa & National Institute of Informatics, Japan\\[12pt]

Hugo Daniel Macedo & Aarhus University, Denmark\\[12pt]

Paolo Masci & National Institute of Aerospace (NIA), USA\\[12pt]

Ken Pierce & Newcastle University, UK\\[12pt]

Marcel Verhoef & European Space Agency, The Netherlands
\end{longtable}

\mainmatter              

\setcounter{page}{5}

\begingroup
\lstdefinelanguage{html}{ 
  backgroundcolor={\color[gray]{1}},
  basicstyle=\small\ttfamily,
  morekeywords={code}, 
  sensitive=true, 
  morestring=[s]{"}{"}, 
  style=HtmlStyle 
} 
\lstdefinestyle{HtmlStyle}{ 
}
\title{Specifying Abstract User Interface in VDM-SL}

\titlerunning{18th Overture Workshop, 2020}

\author{
Tomohiro Oda\inst{1} \and 
Keijiro Araki\inst{2} \and
Yasuhiro Yamamoto\inst{3} \and
Kumiyo Nakakoji\inst{3,1} \and
Han-Myung Chang\inst{4} \and
Peter Gorm Larsen\inst{5}
}

\authorrunning{Oda, T.; Araki, K.; Yamamoto, Y.; Nakakoji, H.; Chang, H.M.; Larsen, P.G.}

\institute{Software Research Associates, Inc.~(\email{tomohiro@sra.co.jp})
\and National Institute of Technology, Kumamoto College~(\email{araki@kyudai.jp})
\and Future University Hakodate~(\email{yxy@acm.org}, \email{kumiyo@acm.org})
\and Nanzan University~(\email{chang@nanzan-u.ac.jp})
\and Aarhus University, DIGIT, Department of Engineering,~(\email{pgl@eng.au.dk}) 
}
\maketitle

\graphicspath{{./1-Oda}}

\begin{abstract}
Interactive systems are often equipped with graphical user interfaces using complex states, constraints and computation to bridge between the rest of the system and the user.
Elements on a graphical user interface are dynamically created, laid out, styled and set up by event handlers according to the internal state of the system.
This paper introduces ViennaVisuals, a framework to construct XML-based abstract graphical user interface models in VDM-SL which is illustrated with an example.
\end{abstract}

\section{Introduction}
\label{sec:Introduction}
A Graphical User Interface (GUI) is a powerful mechanism to allow the user to interact with a software system.
Objects rendered on the screen represent a part of the system's internal states, and also provide cues to trigger the next available operations whose preconditions are satisfied.
In addition to navigation based on the system’s functional model, a user-friendly GUI also guides the user to appropriate operations according to prospective use scenarios.
Thus, GUI design should implement both the prospective use scenarios and the preconditions of operations available to the user.

User validation of the formal specification is a strength of the VDM-SL tool support.
The executable subset of the VDM-SL notation allows the developers to evaluate the validity of the specified functionality using specification animation.
A formal specification in VDM-SL defines each operation to the system by specifying inputs, outputs, and the change of the internal state.
Every operation specified in VDM-SL has its precondition based on the internal state and the inputs, and the internal state can have an invariant assertion.
One can therefore analyse the possible sequence of operations and their inputs based on the preconditions and the changes of internal states.
On the other hand, VDM-SL does not have language constructs dedicated to the explicit definition of user scenarios.
Although one can write a sequence of operation calls inside an operation to express one concrete user scenario, such operation calls are rarely enough to evaluate the validity in general cases.

The actions to be taken by a user and the corresponding sequence of operations that makes sense to the user often depend on the information presented to the user.
The user is not simply performing a predefined sequence of operations using a GUI.
The user does not operate on the GUI to only obtain or express the solution, but also gradually revises and confirms the decisions that the user is making \cite{Nakakoji&13}.

The authors have been studying animation tools to support use scenarios with VDM-SL.
The Overture tool has an interpreter console and a debugger to animate executable specification in VDM-family \cite{1-Oda:Larsen&10a}.
A GUI generator to drive a specification was also developed on the Overture tool \cite{Nielsen&12}.
ViennaTalk is an IDE for VDM-SL designed to support the exploratory process of the early stage of the specification phase and provides a UI prototyping tool called Lively WalkThrough and a WebAPI server called Webly WalkThrough \cite{Oda&17c}.
In those animation tools, GUIs are not included in the formal specification, but they are prototyped outside the formal specification.
However, considering the increasing functional complexity of GUIs for interactive systems, the benefits of specifying GUIs as a part of the formal specification is becoming significant.

This paper proposes ViennaVisuals, a framework that enables VDM-SL to specify internal states and constraints on GUI in Section~\ref{sec:ViennaVisuals}, followed by a concrete example in Section~\ref{sec:Example}.
The design of ViennaVisuals is discussed in Section~\ref{sec:Discussions}.
Afterwards, Section~\ref{sec:RelatedWork} explains other researches on visual components of model-based formal specifications. Finally, Section~\ref{sec:ConcludingRemarks} provides a few concluding remarks.

\section{ViennaVisuals}\label{sec:ViennaVisuals}
VDM-SL is a formal specification language that enables mathematical analysis on computational systems.
A conventional VDM-SL specification typically defines a functional model using internal states and operations that read and/or write the states.
The internal states are constrained by types and invariant assertions.
The operations represent functionalities that the specified system provides.
An operation is typed and also its invocation is constrained by a precondition.
One can analyse the mathematical property of the satisfiability of preconditions and invariants, such as deciding whether or not a particular sequence of operations are possible and safe to call at a given state.
The executable subset of VDM-SL enables the developers to walk-through an expected use scenario.

The user interface is another model of the system that defines a possible sequence of operations and the validity of the system in use scenarios.
A GUI is typically stateful in the sense that the graphics on the screen and the system's response to the user's manipulation changes dynamically.
A GUI should satisfy both the validity in the expected scenarios and the constraints by the system's functional model.
An operation that leads to violation of any of the invariants on the internal system should not be enabled on the GUI.

ViennaVisuals is a framework for VDM-SL that provides a VDM-SL module to specify XML-based GUIs and also drives the GUIs on a web browser.
 Specifying a GUI is not just drawing GUI widgets using a graphics library.
Although an implemented system may have a GUI on a bitmap display, the output of the GUI specification is not pixels, but an abstract presentation of visual objects.
The GUI represents the system's functionality as well as an interpretation of the internal state of the system.
ViennaVisuals represents visual objects in an Abstract Syntax Tree (AST) of geometric shapes in a similar way that  Read–Eval–Print Loop (REPL) interpreters represent the language's expressions in the ASTs.

ViennaVisuals is implemented as a component of ViennaTalk \cite{Oda&17c}.
ViennaTalk has a HTTP server called Webly WalkThrough that can animate VDM specifications and a parser library to convert a VDM expression into various formats.
The implementation of ViennaVisuals consists of an extension to Webly WalkThrough and a library in VDM-SL to construct ASTs of XML elements.
The extension to Webly Walk-Through includes a converter from AST in VDM-SL to the XML format, a JavaScript code to send HTTP requests to the Webly WalkThrough server, and animation manager to drive the UI model specified in VDM-SL.

\begin{figure}
\begin{center}
\includegraphics[width=1\textwidth]{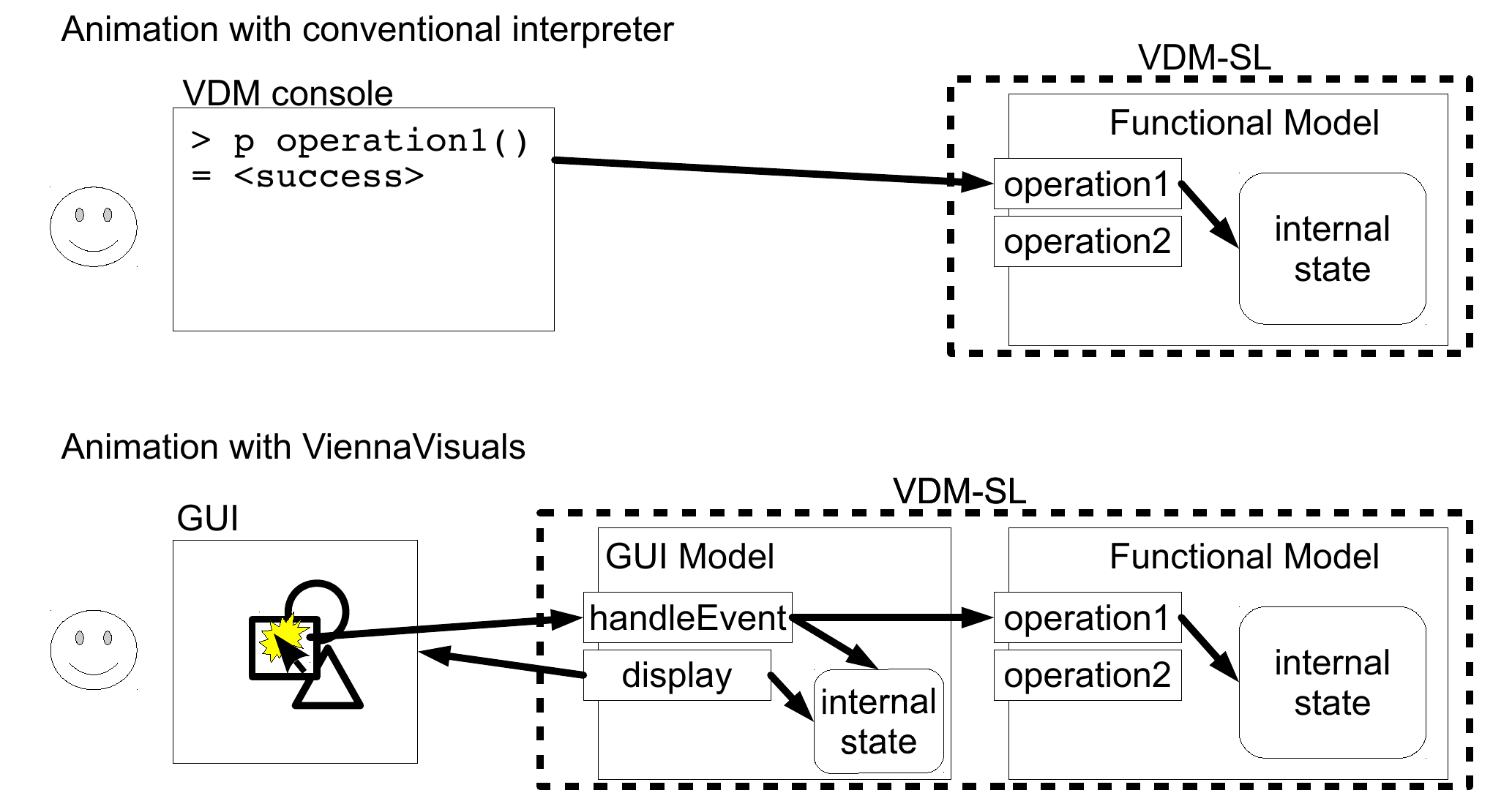}
\end{center}
\caption{Animation with the conventional VDM interpreter and ViennaVIsuals}
\label{fig:ViennaVisuals-interactions}
\end{figure}

Figure \ref{fig:ViennaVisuals-interactions} illustrates the specification animation with ViennaVisuals.
Assuming that the specified system has two public operations {\tt operation1} and {\tt operation2} and the user is validating the {\tt operation1} in a certain scenario.
The system has its internal state in the functional model.
When the user animates the use of the public operation {\tt operation1}, the user types an expression to call {\tt operation1}.
The interpreter evaluates it and shows the resulting value on the screen.
Because the VDM-SL specification is the functional model of the specified system, the specification does not tell how the user gives the command to the system with the user interface.
The user may not know how the result of the operation will be shown to the user.

In the animation with ViennaVisuals, the user sees graphical objects on the screen.
The system has a GUI model along with the functional model.
The GUI model exports one operation for an event handler, typically named {\tt handleEvent}, and another operation to render the GUI, typically named {\tt display} by convention.
When the user clicks on a rectangle, ViennaVisuals executes the associated event handler, in this case {\tt handler1}, which will call the {\tt operation1} operation of the functional model.
ViennaVisuals will then execute the {\tt display} operation to generate the GUI shown as the result to the user.

Both the GUI model and the functional model are written in VDM-SL.
The specifier can use tools other than ViennaVisuals for further analysis, such as proof obligation generators and testing frameworks.
The GUI model has two responsibilities; constructing visual objects to display and handling UI events from the user.
The remainder of this section will explain them one by one.

\subsection{Constructing visual objects}

The construction of visual objects is one of the major features that a GUI framework should provide.
ViennaVisuals provides a mechanism to synthesise and show graphics to the user.
While some GUI frameworks implemented in and for programming languages provide imperative commands to draw graphics on a bitmap screen, ViennaVisuals takes an approach called DOM (Domain Object Model) tree.
Each node in a DOM tree represents a visual object, and one DOM tree represents all visual objects on the screen.
ViennaVisuals defines the DOM in VDM-SL and therefore tools for VDM can process visual objects as values in VDM-SL.

ViennaVisuals expects a GUI module to define an operation, typically named\linebreak {\tt display}, that generates a DOM tree.
{\tt ViennaDOM}\footnote{The source of ViennaDOM is available at \url{https://github.com/overturetool/documentation/blob/editing/examples/VDMSL/xml/ViennaDOM.vdmsl}.} is a module that provides types and operations to generate XML-based DOM elements.
Although {\tt ViennaDOM} can handle arbitrary XML-based formats, ViennaVisuals is designed to use primarily the SVG (Scaled Vector Graphics) format.
SVG is an XML-based format to define graphical shapes, such as rectangles, circles, lines, polygons, and texts.

\begin{figure}
\begin{vdmsl}
types
     Element = TaggedElement| String;
     TaggedElement ::
        name : Name
        attributes : map Name to [String| real]
        eventHandlers : set of EventType
        contents : seq of Element
        tokens : set of token
        identifier_ : nat;
     String = seq of char;
     Name = seq1 of char
     inv [c]^str == c not in set IllegalNameStartChar
            and elems str inter IllegalNameChar = {};
     Point :: x : int y : int;\end{vdmsl}
\caption{The definition of {\tt TaggedElement} and {\tt Element} types for XML elements}
\label{fig:ViennaDOM-TaggedElement}
\end{figure}

Figure \ref{fig:ViennaDOM-TaggedElement} shows the definition of the {\tt TaggedElement} type that represents the AST in XML.
The {\tt attributes} field holds attributes each of whose values are either a string or a real number.
Conversion functions such as integer to string and vice versa are also provided for convenience.
The {\tt contents} field holds its child elements nested in the parent element.
For example, {\tt mk\_TaggedElement("A", \{|->\}, \{\}, [mk\_TaggedElement("B", \{|->\}, \{\}, [], \{\}, 2), "C"], \{\},\linebreak 3)} will be rendered as the following XML extract:\\ {\tt <A identifier="3"><B identifier="2"></B>C</A>}.

The {\tt tokens} fields stores marker tokens so that the element can be retrieved later from the current document tree.
The tokens stored in the {\tt tokens} fields are not rendered in the XML document.

\begin{figure}
\begin{vdmsl}
state DOM of
    current : Element
    nextIdentifier : nat
init s == s = mk_DOM("", 0)
end

operations
    createElement : String ==> TaggedElement
    pure getElementById : String ==> [TaggedElement]
    pure getElementsByToken : token ==> seq of TaggedElement

functions
    setAttribute : TaggedElement * String *
               (String | real | seq of Point) -> TaggedElement
    getAttribute : TaggedElement * String -> [String]
    hasAttribute : TaggedElement * String -> bool
    removeAttribute : TaggedElement * String -> TaggedElement
    appendChild : TaggedElement * Element -> TaggedElement
    addToken : TaggedElement * token -> TaggedElement
    hasToken : TaggedElement * token -> bool
\end{vdmsl}
\caption{The signatures of major operations and functions to manipulate XML elements}
\label{fig:ViennaDOM-operations}
\end{figure}

{\tt ViennaDOM} also provides a set of operations and functions to create, modify, and search for XML elements.
Figure \ref{fig:ViennaDOM-operations} shows the state definition, major operations and functions to construct XML documents.
The operations and functions shown in Figure \ref{fig:ViennaDOM-operations} are designed to conform the DOM API in HTML 5.

The {\tt createElement} operation creates an XML element tagged with the given name.
For example, {\tt createElement("rect")} returns a SVG element that will be rendered as {\tt <rect identifier\_=...></rect>}.
Use of the {\tt createElement} instead of the record constructor {\tt mk\_TaggedElement(...)} is highly recommended to properly manage the {\tt iden\-ti\-fier\_} field.
The {\tt TaggedElement} type has the {\tt identifier\_} field to trace identities of XML elements between VDM-SL and Java\-Script.
The state variable {\tt nextIdentifier} manages to provide values for the {\tt iden\-ti\-fier\_} field.
The invocation of {\tt createElement} operation thus makes a change of state and is defined as an impure operation.

The {\tt getElementById} operation retrieves an XML element from the {\tt current} document tree.
The {\tt getElementsByToken} operation retrieves all the XML elements that hold the marker token given as the argument.

The {\tt setAttribute} function takes the XML element to manipulate, the attribute name, and its value to set, and returns the resulting XML element.
The reason why it is defined as a function is that VDM-SL does not provide mutable objects.
The DOM API of HTML 5 provides XML elements as mutable objects and their methods naturally modify its content.
The values of the {\tt TaggedElement} type are immutable as well as any other values in VDM-SL.
The only state variables are mutable in VDM-SL.
Although it is technically possible to define an operation that modifies the {\tt current} document tree by replacing it with a new XML element, the performance will be quite inefficient.
While programs in HTML 5 often modify the DOM tree incrementally, the use of the document tree in ViennaVisuals would tend to be constructive than mutative.
We therefore designed {\tt setAttribute} as a function.
The functions {\tt getAttribute}, {\tt hasAttribute}, {\tt appendChild} are defined in a similar manner.

The functions {\tt addToken} and {\tt hasToken} are used to set and test marker tokens.
Although marker values could be set into attributes, the type of the attribute values is limited to {\tt String} in XML.
By giving a value of the {\tt\textbf{token}} type, an arbitrary value in VDM-SL could be used as a marker token to trace particular XML elements.

\subsection{Event handling}

This section describes the event handling with ViennaVisuals.
The event handler is a simple mechanism used in various GUI platforms.
When the UI framework detects that the user made an action at a visual object, the event handler is invoked to process the action.
While many UI frameworks associates event handlers with each visual object, ViennaVisuals expects the GUI module to define one operation, typically named {\tt handleEvent}, that processes events triggered in the GUI.

\begin{figure}
\begin{vdmsl}
types
     EventType = <click> | <change>;
     Event = MouseEvent| ChangeEvent;
     MouseEvent :: 
     	type : EventType 
     	target : TaggedElement 
     	x : nat 
     	y : nat;
     ChangeEvent :: 
     	type : EventType 
     	target : TaggedElement 
     	value : String;

functions
    setEventHandler : TaggedElement * EventType -> TaggedElement
    pure getElement : nat ==> [TaggedElement]

\end{vdmsl}
\caption{Type definitions and major functions of the {\tt ViennaDOM} module}
\label{fig:ViennaDOM-events}
\end{figure}

Figure \ref{fig:ViennaDOM-events} shows the type definitions and the signatures of major functions defined in the {\tt ViennaDOM} module.
ViennaVisuals currently supports two types of events; the {\tt <click>} event and the {\tt change} event.
To receive an event triggered at a visual object, the DOM element of the visual object should be registered to handle events.

The registration is carried out in the DOM tree.
The {\tt setEventHandler} function is an auxiliary function to set the registration on the DOM element given as the first argument and the type of the event specified as the second argument.
The event handler fields of DOM elements are rendered as {\tt onclick} or {\tt onchange} attributes in the generated XML document.
An event handler operation should be implemented with the signature {\tt Event ==> ()}.
The argument could be either a mouse event or a change event.
Either event type has the {\tt type} field and the {\tt target} field.
The {\tt target} field stores the XML element where the event occurred.

\begin{figure}
\begin{center}
\includegraphics[width=0.9\textwidth]{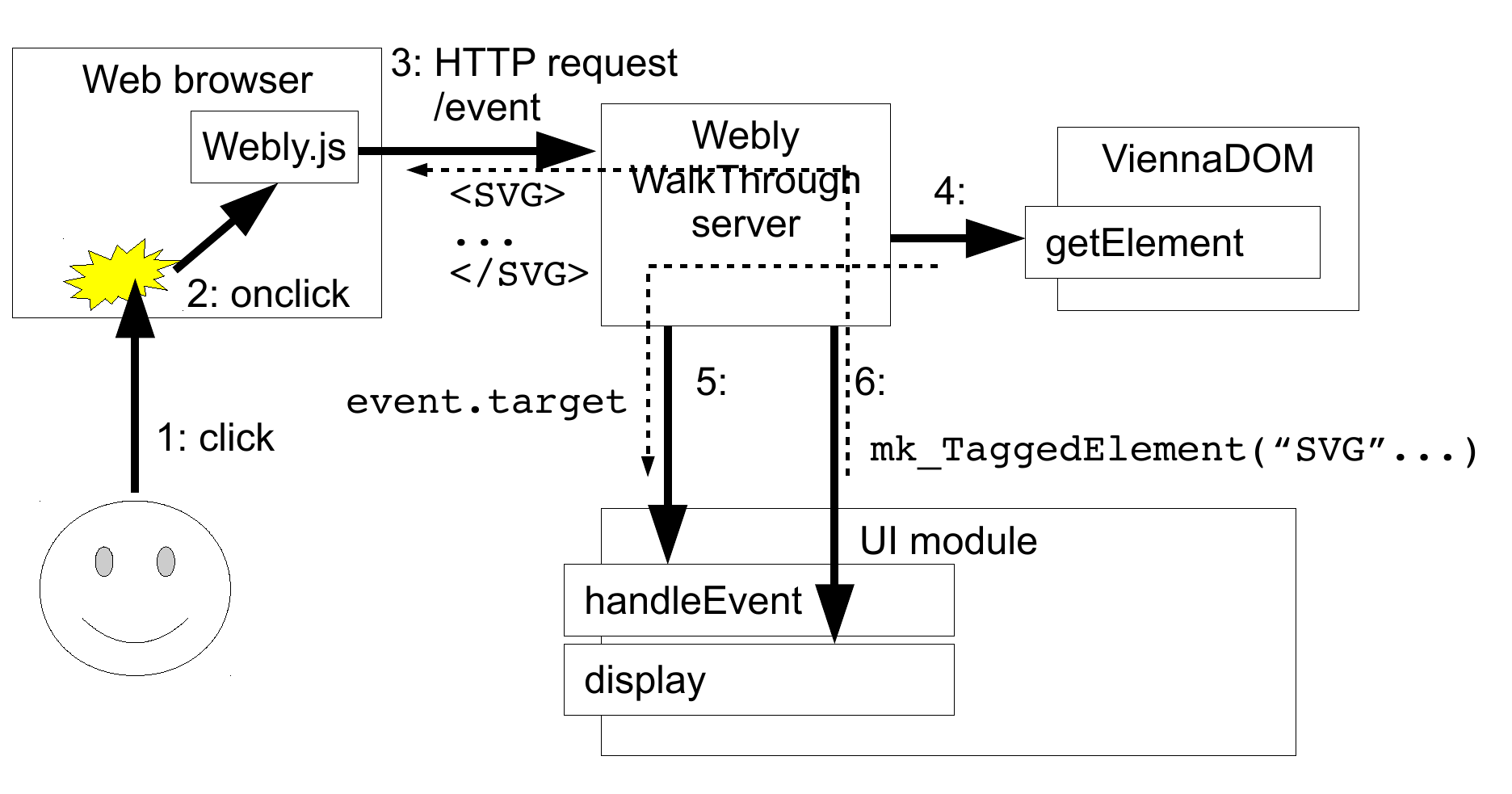}
\end{center}
\caption{Event handling in ViennaVisuals}
\label{fig:event-handling}
\end{figure}

The mechanism to handle a GUI event is illustrated in Figure \ref{fig:event-handling}.
The thick arrows with numbers show the sequence of messages.
The thin dashed arrows indicate the flow of data passed along the messages.

A click at a visual object on the browser (the message 1 in the figure) invokes a function in {\tt Webly.js} (message 2).
The function then sends an HTTP request to the Webly WalkThrough server on ViennaVisuals (message 3).
The {\tt identifier} attribute of the visual object is passed in the request.
The Webly WalkThrough server has to invoke the {\tt handleEvent} operation and gives an event value as an argument to process the event.

Because ViennaVisuals use web browsers as the client platform and VDM-SL as the specification language, a visual object exists both in the web browser and inside the VDM-SL model.
To have the DOM element of the event as the parameter, traceability from the browser's DOM element to the ViennaDOM's element is crucial.
The {\tt getElement} operation retrieves the DOM element with the specified {\tt identifier\_} field from the {\tt current} DOM tree (message 4).
An event value is composed using the information given by the HTTP request and the return value of the {\tt getElement} operation.
The {\tt handleEvent} operation is called with the event value as the argument (message 5).
The event has been processed by this call.

The event handler defined in the GUI module may typically change the internal state of the system.
The visual objects on the GUI need to be updated.
ViennaVisuals calls the {\tt display} operation to obtain the updated DOM tree (message 6).
The Webly WalkThrough server then converts the DOM tree value into the XML format and sends it to the browser as the response to the HTTP request.
The browser will render the SVG document and the user is ready to make a new action.

Both the {\tt display} operation and the {\tt handleEvent} operation are defined in VDM-SL.
They are just conventional operation with no binding with external native libraries so that interpreters and analysis tools for VDM-SL can process the operations.
For example, one can write a unit test to assert the result of operation calls on the {\tt handleEvent} performs an expected side-effect.

\section{Example: Reviewers' bidding}\label{sec:Example}

This section introduces an example application of ViennaVisuals to explain how a dynamic GUI can be specified based on a functional model.
The application is a bidding system for peer-reviewing.

Peer reviewing is widely practiced in academic publications such as proceedings and journals.
Bidding is one method to assign papers to each reviewer.
A reviewer chooses a certain number of papers of which the reviewer is confident of expertise for judging the quality.
The choice is called a {\it bid}.
Reviewers will be then selected for each paper based on the bids from all reviewers.

On the other hand, there are rules that constrain who can review which papers.
In this section, we will respect the following three rules.

\begin{enumerate}
\item  A reviewer can make three bids.
\item A reviewer cannot bid on the papers authored by the reviewer.
\item A reviewer can bid on at most two papers authored by the same person.
\end{enumerate}

Rule 1 defines the number of papers that a reviewer can bid.
Rule 2 prohibits self-reviews while rule 3 is to avoid situations that one reviewer has too much influence on one particular author's work.
It is easy to write the three rules in VDM-SL, and apply the constraint on the ``bid'' operation as the precondition.
Rule 3 defined in VDM-SL is shown in Figure \ref{fig:Bidding-rule3}.
The validation function {\tt isValidBid} can be defined using such functions as shown in Figure \ref{fig:Bidding-isValidBid}, and is used as precondition of the {\tt bid} operation.
The {\tt cancelBid} operation is also defined.
The operations {\tt numPapers}, {\tt getPaper} and {\tt getBids} are also defined similarly.
These functions and operations are core parts of the functional model of the bidding system.
They can be tested using unit testing and combinatorial testing to check their mathematical correctness.

\begin{figure}
\begin{vdmsl}
functions
  exceeds_max_papers_from_same_author : set of Paper -> bool
  exceeds_max_papers_from_same_author(papers) ==
     exists author in set dunion {elems paper.authors
                                 | paper in set papers} &
            card {paper | paper in set papers 
                        & author in set elems paper.authors}
            > MAX_PAPERS_FROM_AUTHOR;
\end{vdmsl}
\caption{An example definition of bidding rules: number of papers from the same author}
\label{fig:Bidding-rule3}
\end{figure}

\begin{figure}
\begin{vdmsl}
functions
  isValidBid : Person * set of Paper -> bool
  isValidBid(reviewer, papers) ==
      not ((has_conflict_of_interest(reviewer, papers)
              or exceeds_max_bids_per_reviewer(papers))
          or exceeds_max_papers_from_same_author(papers));

operations    
   bid : Person * Paper ==> ()
   bid(person, paper) ==
       bids := bids union {mk_(person, paper)}
   pre canBid(person, paper);
\end{vdmsl}
\caption{The definition of validation function for bidding and the {\tt bid} operation}
\label{fig:Bidding-isValidBid}
\end{figure}

However, the mathematical correctness of the definitions in the functional model is just one of the two sides of the coin.
We also need to check whether or not the set of public operations {\tt numPapers}, {\tt getPaper}, {\tt getBids}, {\tt canBid}, {\tt isBidded}, {\tt bid}, {\tt cancelBid} is feasible for the user to make appropriate bids.
The three rules of bidding are not complex, but still require a cognitive burden for the user to operate correctly.

One important responsibility of GUI is to guide the user to operate the system.
The operational feasibility of a functional model partly depends on the GUI.
One advantage of GUI is expressiveness; the GUI can tell the user which paper is legal to bid, which paper is no longer legal to bid, and which paper has already received bids.

\begin{figure}
\begin{vdmsl}
state BiddingUIState of
    reviewer : Person
init s == s = mk_BiddingUIState("Alice")
end
\end{vdmsl}
\caption{State definition of \texttt{BiddingUI} module}
\label{fig:BiddingUI-state}
\end{figure}

A GUI is used by one user at a time.
The GUI module named {\tt BiddingUI} has the {\tt reviewer} variable to capture who is the user as shown in Figure \ref{fig:BiddingUI-state}.
The {\tt display} operation defined in Figure \ref{fig:BiddingUI-display} renders a list of the submitted papers.
Each paper is displayed as a text enclosed by a rounded rectangle.
A token with the paper index is set to the rectangle element so that the event handler can later retrieve the paper from the visual element.
If a paper is legal to bid or has already been bid for, the rounded rectangle is set clickable.
Otherwise, the paper is illegal to bid and thus the text is in {\tt silver} to indicate it is disabled.
The already bidded papers are also marked by a check sign.
These are the mapping rule from each paper to its visual presentation supplemented with the contextual information by the bidding rules.
With the visual objects, the user's task is now reduced to look for a rounded rectangle without a check sign and click on the paper with the most confident of refereeing.

\begin{figure}
\begin{vdmsl}
operations
  display : () ==> TaggedElement
  display() ==
    (dcl list:TaggedElement := createElement("svg");
     for index = 1 to  numPapers()
     do
        let paper : Paper = getPaper(index)
        in
           (dcl itemText:TaggedElement, 
                itemRect:TaggedElement;
           itemRect := frame(index);
           itemText := title(index);
           itemRect := addToken(itemRect, mk_token(index));
           if canBid(reviewer, paper) 
              or isBidded(reviewer, paper)
           then itemRect := 
                   setEventHandler(itemRect, <click>)
           else itemText := 
                    setAttribute(itemText, "fill", "silver");
           list := appendChild(list, itemRect);
           if isBidded(reviewer, paper)
           then list := appendChild(list, check(index));
           list := appendChild(list, itemText));
    return list);
\end{vdmsl}
\caption{The display operation of \texttt{BiddingUI}}
\label{fig:BiddingUI-display}
\end{figure}

\begin{figure}
\begin{vdmsl}
operations
  handleEvent : Event ==> ()
  handleEvent(event) ==
      cases event:
          mk_MouseEvent(<click>, target, x, y) ->
              let index in set {1, ..., numPapers()}
              be st hasToken(target, mk_token(index))
              in toggleBid(getPaper(index)),
          others -> skip
      end;
\end{vdmsl}
\caption{The handleEvent operation of \texttt{BiddingUI}}
\label{fig:BiddingUI-handleEvent}
\end{figure}

The event handler is defined in Figure \ref{fig:BiddingUI-handleEvent}.
The handler accepts any event but responds only to {\tt <click>} events.
The handler then toggles the bidding state of the paper using the token attached to the visual element stored in the {\tt target} field.

\begin{figure}
\begin{center}
\includegraphics[width=1\textwidth]{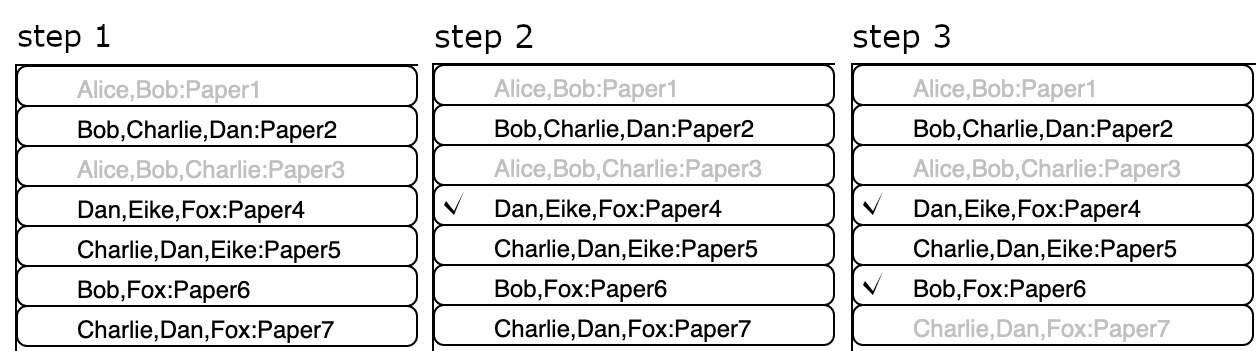}
\end{center}
\caption{Steps in bidding for a reviewer {\it Alice} with the GUI}
\label{fig:BiddingUI-steps}
\end{figure}

Figure \ref{fig:BiddingUI-steps} illustrate the flow of making bids.
A reviewer {\it Alice} opened the UI to make bids.
At the step 1, the system shows the list of submitted papers and two papers (Paper1 and Paper3) are disabled because {\tt Alice} is a co-author.
Alice clicks on Paper4 written by Dan, Eike and Fox.
The system then put a check sign at Paper4 as shown at the step 2.
Paper2, Paper5, Paper6 and Paper7 are available and Alice choose Paper6.
The system again put a check sign at Paper6 and disables Paper7 at the step 3 because bidding on Paper7 would violate the bidding rule 3.
Fox is a co-author of Paper4 and Paper6 and Paper7 is also co-authored by Fox.
Bidding rule 3 prohibits bidding more than two papers authored by the same person.

The specifier can walk through the scenario described above on a web browser.
Also the scenario above can also be written as a unit test using operation calls to the {\tt display} operation and the {\tt handleEvent} operation, and run on a CI server.

\section{Discussions}\label{sec:Discussions}
The major features in the design of ViennaVisuals are:

\begin{itemize}
\item XML as the representation of visual objects,
\item one {\tt display} operation to generate an AST of XML, and
\item one {\tt handleEvent} operation to process UI events.
\end{itemize}

The objective of ViennaVisuals is not to draw fancy graphics, but to {\em specify} the UI of the system: how the system should guide the user by visual objects, and how the system should respond to the user's actions.
This section discusses the design choices of ViennaVisuals in the perspective of engineering tasks in formal modeling.

\subsection{Formal abstraction of GUI}
ViennaVisuals models the GUI as visual objects composed of geometric shapes and discrete events triggered by the user's manipulations.
Comparing to the imperative graphics library with commands such as drawLine, drawCircle, and drawRectangle, ViennaVisuals can handle the result of the rendering as the first-class object.
The {\tt rect} element is a visual object that can be passed to a function, stored into a state variable, and referred to in the proof obligations.

The events are defined as records in VDM-SL.
ViennaVisuals generates an event value to be passed to the {\tt handleEvent} operation.
Although the generation of the event value is done by the runtime of ViennaVisuals, the event value is a first-class object and subject to formal analysis.

\subsection{Centralised specification of event handling}
ViennaVisuals assumes one {\tt handleEvent} operation in a GUI module.
Many GUI frameworks allow each visual object to have their own event handlers while ViennaVisuals expects the only one operation per GUI model.
Because operations are not first-class objects in VDM-SL, an operation cannot be passed as an argument or be assigned to a field of a record value.
Although it is possible to design a GUI framework to register the name of the event handler operation as a string, it would make it difficult to analyse why an event handler is dispatched by a UI event because traceability from the string to the event handler operation would be missed.
Thus passing the event handler name as a string could be problematic in the formal analysis.

With the current design of ViennaVisuals, the event handler operation can take the benefit of the strong pattern matching mechanism of VDM-SL as seen in Figure \ref{fig:BiddingUI-handleEvent}.
Without having each visual object with its own event handler, VDM-SL can concisely abstract the event handling process in its own way.

\subsection{Time granularity of interactions}
The current implementation of ViennaVisuals uses web browsers as the UI platform and HTTP as the protocol between VDM-SL and the selected web browser.
The GUI is not just displaying the result of the computation but provides communication between the user and the system.
Some GUIs use visual effects, such as little vibration of the visual object, to enable subtle interactions between the user and the visual object.
Such visual effects typically happen in the time granularity of milliseconds.
ViennaVisuals is not good for specifying such effects because of the overhead by HTTP and data migrations between JavaScript and VDM-SL.
Although SVG supports some of such visual effects, the effects are not the first-class objects in VDM-SL and thus out of the scope of formal analyses.

ViennaVisuals is designed to specify interactions in the order of seconds.
It should be also noted that VDM-SL does not support temporal features such as the \textbf{\texttt{time} }variable in VDM-RT.
ViennaVisuals is not appropriate for specifying the precise timing of the interactions.

\subsection{Extensibility to other XML-based UIs}
ViennaVisuals provides the DOM construction library named {\tt ViennaDOM}, which can handle general XML elements.
Although the design of ViennaVisuals aims at SVG as the document format, it is possible to specify the {\tt display} operation to generate other XML-based models, such as VoiceXML\footnote{\url{https://www.w3.org/TR/voicexml21/}} for Voice UI and X3D\footnote{\url{https://www.web3d.org/x3d/what-x3d}} for 3D graphics.

\subsection{Limitations}

The {\tt display} operation is impure because the {\tt createElement} operation in the {\tt ViennaDOM} module is impure, which could be obstacle to formal analysis.
The \linebreak {\tt createElement} operation needs to be impure to manage the traceability between the visual elements in XML and in VDM-SL.
The traceability is mandatory to retrieve the target of a UI event when invoking the {\tt handleEvent} operation.
The traceability could be managed within the {\tt display} operation, but a failure of managing the traceability could result in destroying the semantics of the event mechanism.

The design of ViennaVisuals does not fit with multi-threading.
ViennaVisuals assumes that the internal state is kept after the last display() call until the next event callback invokes the {\tt handleEvent} operation.
If the state is changed between the {\tt display} operation and the {\tt handleEvent} operation, the traceability will be broken and thus the semantics of the event mechanism will be unreliable.
VDM-SL does not support multi-threading, but this is a limitation of the design of ViennaVisuals when applying to other formalisms.

\section{Related Work}\label{sec:RelatedWork}

\subsection{PVSio-web}
PVSio-web \cite{Masci&15} is a graphical toolkit to platform to evaluate a formal model defined in PVS and user interface.
It is used in practice to evaluate safety of medical devices with combinations of user interface and a formal model.
PVS is a formal modeling system based on theorem provers and animation.
PVSio-web drives realistic user interfaces so that operations using the interface can be carried out safely.
ViennaVisuals does not aim at operating a realistic GUI but the abstract specification of the user interface based on visual shapes and event handling.

\subsection{Lively WalkThrough}
Lively WalkThrough \cite{Oda&17c} is another GUI prototyping framework in ViennaTalk, which aims at encouraging communication between the formal specification engineers and UI designers.
In Lively WalkThrough, one can compose a UI prototype on a given VDM-SL specification and define event handlers for each visual component in the UI prototype.
By walking through a user scenario on the UI prototype, the formal specification engineer can see how the system's functionality will be used and the UI designer can see how the user scenario can be carried out by the system's functionalities.
Because the focus of Lively WalkThrough is to drive the specification with a prototypical UI, the UI and event handling is not specified in VDM-SL but separately defined on the tool.

\subsection{Crescendo/Symphony/INTO-CPS}
Crescendo, Symphony, and INTO-CPS are co-simulation tools based on the Overture tool \cite{Foldager&17}.
The three co-simulation tools drive two different types of simulations together: functional models with discrete events defined in VDM, and physical models with continuous-time using simulation engines.
Some components of the co-simulation tools provide 3D graphics to visualise the virtual physical model in the simulation model.
Although the co-simulation does not aim at GUI design, the visualisation can be seen as a graphical presentation of the internal state of the co-simulated models.

\section{Concluding Remarks}\label{sec:ConcludingRemarks}
A specification of the GUI is the specification of what the user will see through the system.
A graphical user interface is the meeting point between the user and the system.
The validity of the GUI should take both the views from the user and also from the system into the account.
The GUI from the user's side of the view possibly involves human cognition.
This side of the GUI is hard to be subject to formal modeling.
The other side of the GUI is the computational system.
Formalisms including VDM are designed to capture this side.

ViennaVisuals is a step to combine the two sides.
A GUI model defined with ViennaVisuals can be analysed together with the functional model specified in the same formal specification language.
The GUI model does not represent the user's cognition and emotions, but it could hopefully let the user virtually preview how the realised system would look in situations.

\section*{Acknowledgments}
The authors thank anonymous reviewers for their thoughtful and constructive feedback. 
A part of this research was supported by JSPS KAKENHI Grant Number JP 18K18033 and 19K11911. We would also like to express our thanks to the anonymous reviewers.

\bibliographystyle{splncs03}

\clearpage
\endgroup

\begingroup

\title{Modelling the HUBCAP Sandbox Architecture In VDM: A Study In Security}
%
%
\author{Tomas Kulik\inst{1} \and
Hugo Daniel Macedo\inst{1} \and
Prasad Talasila\inst{1} \and
Peter Gorm Larsen\inst{1}}
\authorrunning{T. Kulik, H. D. Macedo, P. Talasila and P. G. Larsen}
%
\institute{DIGIT, Department of Engineering, Aarhus University}
%
\maketitle              
\setcounter{footnote}{0}

\makeatletter
\def\input@path{{2-Kulik/}}
\makeatother

\graphicspath{{2-Kulik}}
\lstdefinestyle{VDM}
{
  frame=single,
  basicstyle=\small\ttfamily,
  escapechar=!,
  breaklines=true,
  frameround=false,
  linewidth=\columnwidth,
  morekeywords={atomic,is,inv,values,dcl,forall,in,set,nil,and,let,be,st,set1,pure,nat,pre,post,map,to,of,true,false},
  moredelim={[is][keywordstyle]{@}{@}},
}

\lstdefinestyle{TraceOutput}
{
  basicstyle=\small\ttfamily,
  frame=single,
  captionpos=b
}

\newcolumntype{R}[1]{>{\raggedleft\let\newline\\\arraybackslash\hspace{0pt}}m{#1}}

\begin{abstract}

In this paper, we report on the work in progress towards the security analysis
of a cloud based collaboration platform proposing a novel sandboxing concept.
To overcome the intrinsic complexities of a full security analysis, we follow
the formal methods approach and used separate teams: one developing a model and
one implementing the platform. The goal is to create an abstract formal model
of the system, focusing on the critical dataflows and security pain points,
which create results to be shared among teams. The work is in progress, and, in
this paper, we provide insights on a first model covering: the main features of
the platform's sandboxing concept, and an analysis on the table of roles and
profiles by the users, that regulates the user access to the different
sandboxing capabilities.  The model is specified in VDM-SL and was validated
using the Overture tool combinatorial testing.  The analysis has uncovered some
potential issues, that have been shared with the platform implementation team,
who have used it to more clearly define the table of roles and profiles within
the system. Although useful, the work on implementing and securing the
sandboxing is still in its initial stages, and given the novelty of the
sandboxing concept, a definite answer regarding its security aspects is
unclear, but a challenging research question. 

\keywords{Sandbox  \and Collaborative Platforms \and Model Based Engineering.}
\end{abstract}

\section{Introduction}
A new collaborative platform for model based development (MBD) is emerging from
the HUBCAP project \cite{Larsen&20c,Larsen&20d}. Differently from the
traditional collaborative platforms, where users are given access to shared
documents or models (with an interface to perform changes), the HUBCAP project provides multiple virtual machines (VMs) that are fully
accessible via a browser web page. Users are given access to read and write 
to the operating system's file systems and can jointly edit the state of the sandbox components.  Whether the concept will
be widely adopted, or it suits only the needs of the MBD community is uncertain. What is known is that there is a need to perform a thorough analysis of the security implications of the design.

Conceptually, a HUBCAP sandbox is a group of virtual machines managed and
hosted by a trusted third-party cloud provider. The users of a sandbox are MBD
providers or adopters, which despite not being granted access to the source
code of the platform details, receive VMs prepared with all the required
dependencies to run and develop tools and models. 
The advantages of the concept are twofold. Firstly, the users can explore
and interact with a system without having to give permission to either their
local machine  OS facilities or data. This goes beyond remote assistance tools
that allow an external party to provide training or help configure the requesting users' local
machine. Secondly, the users are able to manipulate a whole system,
beyond a file or a view of the model data, which is interesting when the
collaboration involves the development of cyber-physical systems and its 
associated need to use multiple tools and configure various system parameters.  On
the down side, and beyond the obvious resource consumption, the concept is
new and the security requires trust between all the users and the cloud third
party provider. 

This paper is the first step into a full analysis of the security aspects of
the HUBCAP platform. The work lays the foundation for future extensions to the
Sandbox concept and evaluates whether it is possible to ensure security in
multiple scenarios. For example, if the third party cloud provider is
compromised, or proven to be untrustworthy. 

To study the security properties of the HUBCAP platform we are developing a Vienna Development Method (VDM)
model of the system components and functionality that is security critical. VDM
is a formal method prescribing the development of models of computer systems,
which are used to generate code for the system implementation, or to provide
insight to the implementation team. The models have a precise semantics and
have proven to be useful in the process of software development and security
analysis. 

Within this analysis we have used VDM-SL a specification language standardised by the International Standardisation Organisation \cite{ISOVDM96,Plat&92b}. It is a model-oriented formal specification language which is based around logic (e.g., for invariants, pre-conditions and post-conditions) and a number of abstract data types for different kinds of collections. It has been used industrially since the seventies where it was invented at IBM's laboratories in Vienna \cite{Larsen&96a,Kurita&09}. For the last two decades the Overture/VDM tool support has evolved after the VDMTools area \cite{2-Kulik:Larsen01,2-Kulik:Larsen&10a}. 

Our work provided feedback to the implementation team,
and led to the refinement of the platform user roles and profiles tables, which
functions as an ``access-control list'' to the features available to the
different users. Although we are far from being able to fully assert the security
properties of the platform, our work was able to elicit the dataflow and pinpoint several security checkpoints that are necessary to ensure the security of the
platform. We further intend to iterate our model continuously to align with potential updates to the roles and profiles table.

The rest of the paper is organized as follows, Section~\ref{sec:sandboxing} introduces the concepts of sandboxing and how it is utilised within the HUBCAP platform. It further introduces a table of profiles and roles governing access of clients to specific functionality within the platform. Section~\ref{sec:model} describes the formal model of the sandboxing platform created in VDM-SL with focus on the functionality of creation of sandboxes and features present in the table of profiles and roles. Section~\ref{sec:analysis} provides an overview of the formal analysis carried out to ensure that the platform is secure based on the table of profiles and roles. Section~\ref{sec:related} then defines related work in the area of using formal methods in security and sandboxing assurance. Finally Section~\ref{sec:conc} provides concluding remarks and introduces expected future work.

\section{Sandboxing Platform For Collaboration}
\label{sec:sandboxing}


\begin{figure*}[bt]\centering
\includegraphics[width=\textwidth]{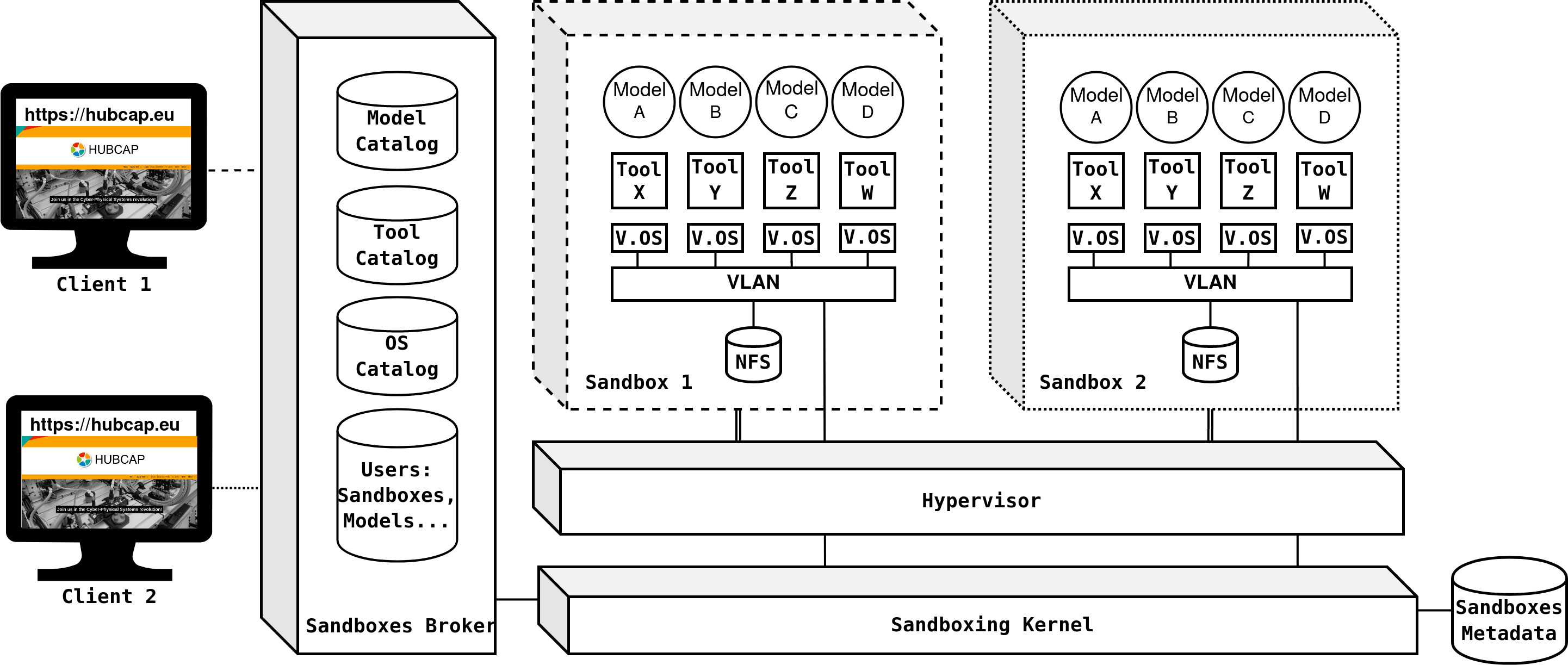}

\caption{\label{fig:sandbox}The HUBCAP Sandbox architecture (taken from  \cite{Larsen&20c})}
\end{figure*}

A sandbox is implemented as an isolated set of VMs (each one running a CPS tool) that interact with each other sharing a virtual dedicated subnet and a dedicated (Network File System) NFS storage service. No interaction is permitted between the VMs belonging to different sandboxes. The sandbox capability integrated with the web-platform is therefore a sort of private cloud service provider
plus the middleware to manage and mediate the access to those cloud services. In addition, as many cloud service providers offer the capability to select a combination of hardware and operating systems, the HUBCAP Platform offers you to select a combination of OS environments, tools, and models to run an experiment using the HUBCAP sandbox feature.

The sandbox service is outlined in Figure~\ref{fig:sandbox}. The web-platform is enhanced with a broker component (labelled as \emph{Sandboxes Broker} in the figure), which hosts a web application mediating the access of different users (\emph{Client 1} and \emph{Client 2}) to the sandboxes they requested (\emph{Sandbox 1} and \emph{Sandbox 2} respectively). All the users will use an Internet browser to access the tools in the sandbox and all the interactions are mediated by the broker.

The \emph{Sandbox Broker} has access to the catalogues of different models, tools, and pre-configured OSs that are available, so an end user can simply pick a valid combination to request a sandbox. In addition to those catalogues, the \emph{Sandbox Broker} keeps user information, such as the user's models (private copies of the model in the catalogue, which may have been modified by the user while using the sandbox) and the sandboxes the user created. This information is important to allow the creation of new sandboxes.

The operation of user requests and the sandboxing logic is provided by the \emph{Sandboxing Kernel}, which is a component that interacts with the system \emph{Hypervisor} to launch the different constituents of a sandbox, namely:
\begin{itemize}
	\item \textit{NFS} - Network File System providing storage in the form of shared folders where model files and tool outputs are placed.
	\item \textit{VLANs} - Virtual networks restricting the communications of the VMs inside a sandbox to the set of VMs composing it and those only.
	\item \textit{VOS} - Virtual machines running the OSs supporting a tool, a remote desktop protocol to provide the clients access to the tool display, and other monitor and interoperability tools to operate the VM inside the Kernel.
	\item \textit{Tools} - The tools running a model or a multi-model.
	\item \textit{Models} - A mathematical/formal description of a component.
\end{itemize}

The operation relies on a database of metadata about the different sandboxes (the \emph{Sandboxes Metadata} component in the figure). This component stores and keeps track of the sandboxes' states (running, suspended, \dots) and user ownership of the resources. It is worth highlighting that the Kernel has direct network connections to the Sandboxes' VLANs. We base our model on the building blocks of the sandboxing platform presented in this section, focusing on the components necessary for analysis of the client access based on roles and profiles, while abstracting away details such as VLAN addresses, specifics of the operating systems, file storage and tool properties.

\subsection{Designated Roles and Activities}
The sandbox platform as presented within this paper offers several features that are accessible by use of clients. Each feature is protected by requiring the client being of a specific role and profile. The full overview of features and their associated roles and profiles is shown in Table~\ref{tab:roles}. The profiles within the system are provider and consumer, where the provider can add new content to the sandboxing platform such as operating systems or tools, while the consumer could utilise the content already present within the platform. Furthermore the system defines two roles on the sandboxes themselves, where the owner role is assigned to a client creating the sandbox, while a guest role is assigned to clients that have been invited by the owner to the sandbox. Without the invitation the guest cannot access the sandbox.

\begin{table}[h!]
  \begin{center}
    \caption{Overview of profiles and roles}
    \label{tab:roles}
    \begin{tabular}{c|c c c c} 
      \multirow{2}{*}{\textbf{Feature}} & \textbf{Provider} & \textbf{Provider} & \textbf{Consumer} & \textbf{Consumer}\\
      & \textbf{Owner} & \textbf{Guest} & \textbf{Owner} & \textbf{Guest}\\
       \hline
      Access to remote viewer & X & X & X & X \\
      Upload archive & X & X & X & X \\
      Download archive & X &  & X &  \\
      Invite guests & X & & X & \\
      Destroy sandbox & X & & X & \\
      Select tool & X & & X & \\
      Select model & X & & X & \\
      Select operating system & X & & & \\
      Save tool & X & & & \\
      Upload new model & X & & & \\
      Delete repository tool & X & & &\\
      \hline
    \end{tabular}
  \end{center}
\end{table}


\section{Formal Model}
\label{sec:model}
The model consist of several sandbox building blocks, namely (1) Client, (2) Broker, (3) Gateway, (4) Sandbox, (5) Server and (6) System. Each of these components, excluding the gateway is represented as a VDM-SL file with varying level of detail. The model considers a single default module tracking the state of all of the components. The primary purpose of the model is to determine if the system functionality behaves according to the provided table of roles and profiles. This section provides detailed overview of the different model components.

\subsection{Client}
The client is the entry point into the sandbox. It is a component used by the user to access and interact with the system. The system can have multiple clients, where each client has an explicit identity, expressed as \lstinline[style=VDM]{ClientId = nat}. Furthermore the clients that can access the system must be considered valid, this is handled by ensuring that the client identity is present in a database of valid identities stored within the system. We express this check as a pre condition on all of the client calls as \lstinline[style=VDM]{pre cId in set validClients;}. The client can call several operations against the broker, namely starting a new sandbox, destroying a sandbox, selecting a model from a repository, selecting a tool from a repository, selecting an operating system from a repository, inviting a guest to a sandbox, removing own repository item, connecting to a sandbox and disconnecting from a sandbox. In order to create new sandbox the client needs to first select the tool, the operating system and optionally a model that shall be present within this sandbox. The selections of the client are recorded within a \lstinline[style=VDM]{ClientSt} record and are presented together with launching of new sandbox in Listing~\ref{lst:selections}. Once the sandbox is launched the client can access the sandbox as presented in Listing~\ref{lst:access}. The actual deployment of the sandbox and establishment of connections is handled by different components, namely, the broker and the system. 

\begin{vdmsl}[style=VDM, label={lst:selections}, caption={Sandbox elements selection and sandbox launch}]
SelectOS: ClientId * OSId ==> ()
SelectOS(cId, osId)== 
  clientst.selectedOS := SelectOperatingSystem(osId, cId)
pre cId in set validClients;

SelectToolFromRepository: ClientId * ToolId ==> ()
SelectToolFromRepository(cId, tId) == 
  clientst.selectedTool := SelectTool(tId, cId)
pre cId in set validClients;

SelectModelFromRepository: ClientId * ModelId ==> ()
SelectModelFromRepository(cId, mId) == 
  clientst.selectedModel := SelectModel(mId, cId)
pre cId in set validClients;

LaunchNewSandbox: ClientId ==> ()
LaunchNewSandbox(cId) == 
  StartNewSandbox(cId, clientst.selectedTool, 
    clientst.selectedModel, clientst.selectedOS)
pre cId in set validClients;
\end{vdmsl}

\begin{vdmsl}[style=VDM, label={lst:access}, caption={Client accessing a sandbox}]
-- Connect to a sandbox
AccessSandbox: ClientId * SandboxId ==> ()
AccessSandbox(cId, sId) == 
  if not BrokerInitiateSandboxAccess(cId, sId)  
  then GetError(cId, mk_token(1))
pre cId in set validClients;
\end{vdmsl}

\subsection{Broker}
The broker is a component of the overall system that the client interacts with, before being handed over a direct connection to a sandbox. This component is aware of currently active sandboxes, their respective ownership and records the items available within the repository, specifically the operating systems, tools and models. The broker also records ownership of these repository items. This is represented by mappings as for example ownership of a sandbox recorded in \lstinline[style=VDM]{Owners = map ClientId to set of SandboxId}, while the ownership of tools is recorded as \lstinline[style=VDM]{ToolOwners = map ClientId to set of ToolId}. Furthermore the broker records how specific tools, models and operating systems are utilised within different sandboxes, an example of what is a mapping \lstinline[style=VDM]{SandboxTools = map SandboxId to set of ToolId}. A client selecting for example an operating system calls a broker operation as shown in Listing~\ref{lst:brokerOS}, where the precondition states that the client either needs to have a profile of a provider or a consumer as presented in Table~\ref{tab:roles}.

\begin{vdmsl}[style=VDM, label={lst:brokerOS}, caption={Broker providing an operating system}]
SelectOperatingSystem: OSId * ClientId ==> [nat]
SelectOperatingSystem(osId, cId) == 
  return if osId in set brokerst.validOSs
  			then osId 
  			else nil
pre cId in set brokerst.providers or 
    cId in set brokerst.consumers;
\end{vdmsl}

The broker is also responsible for starting a new sandbox. This operation registers the sandbox within the system and it becomes available for connections. This operation further registers the requested operating system, tool and optionally a model within the sandbox and is presented in Listing~\ref{lst:startSandbox}. The precondition checks that the user (via the client) has made valid selections, while the post-condition ensures that the sandbox is registered correctly with a newly generated sandbox identity.
\begin{vdmsl}[style=VDM, label={lst:startSandbox}, caption={Broker starting a new sandbox within the system}]
StartNewSandbox : ClientId * ToolId * [ModelId] * OSId ==> ()
StartNewSandbox(cId, tId, mId, oId) == 
let sId = GenerateNewSandboxId()
in
  (brokerst.sandboxTools := brokerst.sandboxTools munion 
                            {sId |-> {tId}};
   brokerst.sandboxOSs := brokerst.sandboxOSs munion 
                          {sId |-> {oId}};
   if mId <> nil 
   then brokerst.sandboxModels := brokerst.sandboxModels munion 
                                  {sId |-> {mId}};
   systemSandboxes := systemSandboxes munion 
                      {sId |-> mk_Sandbox(sId, {1})};	
   if cId in set dom brokerst.owners 
   then brokerst.owners(cId) := brokerst.owners(cId) union {sId}
   else brokerst.owners := brokerst.owners munion 
                           {cId |-> {sId}})
pre tId in set brokerst.validTools
		and oId in set brokerst.validOSs
		and mId <> nil => mId in set brokerst.validModels
post card dom systemSandboxes = card dom systemSandboxes + 1
		and cId in set dom brokerst.owners
		and Max(systemSandboxes) in set brokerst.owners(cId);
\end{vdmsl}
\vspace{-4ex}
In order to provide connectivity to the sandbox the broker checks if the client is associated with the requested sandbox. This means that the client either needs to be an owner of the sandbox or a guest invited for access to this sandbox. This operation is shown in Listing~\ref{lst:provideAccess}. The precondition checks that the client actually has some role within the system.

\begin{vdmsl}[style=VDM, label={lst:provideAccess}, caption={Broker initiating access to a sandbox for a client}]
BrokerInitiateSandboxAccess: ClientId * SandboxId ==> bool
BrokerInitiateSandboxAccess(cId, sId) == 
	let sandboxes = GetSystemSandboxes(),
	    servers = dunion {s.sandboxServers 
	                     | s in set rng sandboxes 
	                     & s.sandboxId = sId} 
	in 
	  (for all x in set servers do
	     UpdateConnections(cId, x, sId, true);
	   return true)
pre not ClientIsNull(cId,brokerst.providers,brokerst.consumers, 
                     brokerst.owners, brokerst.guests) and 
    ((cId in set dom brokerst.owners and 
     sId in set brokerst.owners(cId)) or
    (cId in set dom brokerst.guests and 
     sId in set brokerst.guests(cId)));
\end{vdmsl}

The broker further handles operations such as assigning a guest to a sandbox invited by a sandbox owner updating the repository of items, i.e. by uploading new items or removing them, based on the wishes of the client and destroying the sandbox, thus removing it from the system.

\subsection{Gateway}
The gateway is a connection component providing connectivity between the sandbox and the client. More specifically the gateway opens a connection to every server that exists under a sandbox and registers which client holds the open connection to this sandbox. In the current model the focus is on analysing the specific roles of the clients against the actions they can take and hence the gateway connections are simply recorded under the system state as \lstinline[style=VDM]{GatewayConnections = map ClientId to set of ServerId} and also \lstinline[style=VDM]{GatewayConnectionsSandbox = map ClientId to set of} \lstinline[style=VDM]{SandboxId}. The client can connect to and disconnect from a sandbox- a functionality captured by an operation shown in Listing~\ref{lst:updatecons}. The listing further shows the broker keeping a set of active sandboxes (i.e. sandboxes with an open connection).

\begin{vdmsl}[style=VDM, label={lst:updatecons}, caption={Gateway recording the connections to a sandbox}]
UpdateConnections: ClientId * ServerId * SandboxId * bool ==> ()
UpdateConnections(cId, sId, sbId, connect)== 
  if connect 
  then (if cId in set dom gatewayConnections 
        then atomic
       (gatewayConnections(cId):= gatewayConnections(cId) 
                                  union {sId};
        gatewayConnectionsSandbox(cId):= 
             gatewayConnectionsSandbox(cId) union {sbId};
        brokerst.activeSandboxes:= brokerst.activeSandboxes 
                                   union {sbId}))
  elseif cId in set dom gatewayConnections 
	then atomic
	(gatewayConnections(cId):= gatewayConnections(cId) \ {sId};
	 gatewayConnectionsSandbox(cId):= 
	      gatewayConnectionsSandbox(cId) \ {sbId};
	 brokerst.activeSandboxes:= brokerst.activeSandboxes \ {sbId})
\end{vdmsl}

\subsection{Sandbox}
Sandbox is a secure container encapsulating servers, tools and a model that the user decided to work with. The nature of the sandbox is to provide a demo environment to users that want to try use of model based engineering within a hosted environment. The sandbox provides an isolation from other sandboxes and the wider system and users working with sandboxes follow functionality roles as shown in a Table~\ref{tab:roles}. The user can, via clients, create new sandboxes, connect to sandboxes, disconnect from sandboxes and destroy sandboxes. If a user creates a sandbox, this user can further invite another user as a guest to the sandbox, via the identity of the client that the guest uses. Each sandbox within the system could be understood as a collection of servers with specific tools installed and each sandbox carries an unique identity. The definition of the sandbox is shown in Listing~\ref{lst:sandbox}.

\begin{vdmsl}[style=VDM, label={lst:sandbox}, caption={Sandbox specification}]
types 
SandboxId = nat;
SandboxServers = set of ServerId;
Sandbox:: 
 sandboxId : SandboxId
 sandboxServers : SandboxServers
\end{vdmsl}

In the current iteration the model is only capable of creating sandboxes consisting of a single server, as this is sufficient for analysis of the access and functionality roles.

\subsection{Server}
Within the modelled system, the server is a machine running an operating system with a specific tool allowing a user to manipulate models with a possibility of a remote access. Each server has to be a part of a sandbox in order to be accessible and each server carries a unique identity as \lstinline[style=VDM]{ServerId = nat}. Within the model presented in this paper, the only considered functionality of the server is, the server being a part of a sandbox.

\subsection{System}
The System VDM file captures functionality of the system as a whole and facilitates operations not suitable for other components. The system for example generates new identities for new sandboxes as shown in Listing~\ref{lst:system}. Furthermore the system contains the state for the entire system and its components as shown in Listing~\ref{lst:state}. The system represents an abstract container for the system components with its own high level functionality.

\begin{vdmsl}[style=VDM, label={lst:system}, caption={Operations generating identity for new sandbox}]
pure Max: SystemSandboxes ==> nat
Max(ss) ==
  let max in set dom ss be st forall d in set dom ss & d <= max
  in
    return max
pre ss <> {|->};
GenerateNewSandboxId: () ==> nat
GenerateNewSandboxId() == 
  return Max(systemSandboxes) + 1;
\end{vdmsl}

\begin{vdmsl}[style=VDM, label={lst:state}, caption={System state}]
state SystemSt of
	gatewayConnections : GatewayConnections
	gatewayConnectionsSandbox : GatewayConnectionsSandbox
	systemSandboxes : SystemSandboxes
  toolOwners : ToolOwners
  modelOwners : ModelOwners
	brokerst : BrokerSt
	clientst: ClientSt
	validClients : ValidClients
inv ss == dom ss.gatewayConnections = 
          dom ss.gatewayConnectionsSandbox 	
init s == s = mk_SystemSt({|->},{|->},{|->},{|->},{|->},
     mk_BrokerSt({},{},{},{},{},{},{|->},{|->},[],{|->},{|->},{|->}),
     mk_ClientSt(nil,nil,nil),{})
end 
\end{vdmsl}

\section{Formal Analysis}
\label{sec:analysis}
This section specifies the traces that have been used to analyse several system features against the table of profiles and roles. The most important functionality analysed was the creation of new sandboxes and ensuring that uninvited clients do not have access to the sandboxes, an important aspect for the integrity of the platform.
\subsection{Traces under analysis}
To analyse the specific cyber security properties, we express several scenarios as traces and use the combinatorial testing feature of overture. The scenarios considered within this paper are creation of sandboxes, connection and disconnection to sandboxes and creation of sandboxes considering invitation of guests to the newly created sandboxes. These scenarios have been selected based on the communication with the implementation team in order to ensure that the table of roles and profiles provides the intended permissions in order to access the system. The analysis has uncovered a potential permissions issue within the system. In order to create a sandbox a trace as shown in Listing~\ref{lst:create} could consider several scenarios, where the user with a profile consumer, a user with a profile provider (profiles as specified in Table~\ref{tab:roles}) and a user without a specific profile attempt to create a new sandbox. To do this, the users select the operating system, the tool and the model that shall be used within the sandbox (only the operating system is mandatory as the tool and the model could be uploaded later to an existing sandbox). Once the selection is complete the users simply launch a new sandbox. The initial operation calls in the listing simply sets up the environment, ensuring that the system has the clients, the operating systems, the tools and the models registered.
\begin{vdmsl}[style=VDM, label={lst:create}, caption={A trace of different clients launching a sandbox}]
CreateSandboxMultipleClients:
SetupClients(1);
SetupClients(2);
SetupClients(3);
SetupProviders(1);
SetupConsumers(2);
SetupOSs(1);
SetupTools(1);
SetupModels(1);
let clientId in set {1, 2, 3}
in 
(SelectOS(clientId, 1);
 SelectToolFromRepository(clientId, 1);
 SelectModelFromRepository(clientId, 1);
 LaunchNewSandbox(clientId));
\end{vdmsl} 
\vspace{-3ex}
In order to initialize client within the system a \lstinline[style=VDM]{SetupClients} operation is called with further operations setting up the client as a consumer or a provider. This is shown in Listing~\ref{lst:setup}.
\begin{vdmsl}[style=VDM, label={lst:setup}, caption={Setting up a client within the system}]
SetupClients: ClientId ==>()
SetupClients(cId) == validClients:= validClients union {cId}

SetupProviders: ClientId ==> ()
SetupProviders(cId) == 
  brokerst.providers := brokerst.providers union {cId};
\end{vdmsl} 
This scenario has uncovered an issue where a user without a specific role and without a specific profile (but still a valid user within the system) attempts to launch a sandbox. While this user is allowed to launch a sandbox, it is only the users with profiles \textit{consumer} or a \textit{provider} that are allowed to select an operating system. This is captured as a precondition on the \lstinline[style=VDM]{SelectOS}. Since the user without an explicit profile can not select an operating system, it is not possible for this user to create a new sandbox. This led to a suggestion that users that are created on the platform but do not have assigned access rights should not be able to create a new sandbox.

Another trace considered was the trace specifying a scenario where a user (using a client) creates a sandbox and another user (using a separate client) attempts to connect to this sandbox without first receiving an invitation to this sandbox. This trace is shown in Listing~\ref{lst:uninvited}. The analysis of this trace has uncovered that the permissions model based on roles and profiles specified in Table~\ref{tab:roles}, expressed as a pre-condition for the operation \lstinline[style=VDM]{AccessSandbox} prevents this from happening, ensuring that only users with legitimate claim could access the sandbox, i.e. only the owners or invited guests can access the sandbox. 

\begin{vdmsl}[style=VDM, label={lst:uninvited}, caption={Uninvited client attempting connection to a sandbox}]
CreateSandboxAndUninvitedConnect:
SetupClients(1);
SetupClients(2);
SetupProviders(1);
SetupOSs(1);
SetupTools(1);
SetupModels(1);
let clientId in set {1}
in 
(SelectOS(clientId, 1);
 SelectToolFromRepository(clientId, 1);
 SelectModelFromRepository(clientId, 1);
 LaunchNewSandbox(clientId);
 AccessSandbox(2, 1));
\end{vdmsl} 

Several other traces have been analysed, covering the launching of the sandbox, where these traces have confirmed that the permission table does allow access as specified. One of these flows is shown in Listing~\ref{lst:createValid}, representing a simple scenario of a user with a provider profile launching a new sandbox.

\begin{vdmsl}[style=VDM, label={lst:createValid}, caption={Legitimate user creating a sandbox}]
CreateSandboxNoModelNoTool:
SetupClients(1);
SetupProviders(1);
SetupOSs(1);
let clientId in set {1}
in 
(SelectOS(clientId, 1);
 SelectToolFromRepository(clientId, 1);
 SelectModelFromRepository(clientId, 1);
 LaunchNewSandbox(clientId)
);
\end{vdmsl} 

\subsection{Results and suggestions}
The security analysis of the sandbox platform has uncovered potential improvements to the table of roles and profiles. The first suggestion is to explicitly state what roles and profiles are needed in order for the client to be able to create a new sandbox. Second suggestion is to carry out a consistency check on the roles and profiles after each update of the roles and profiles table, a task well suited for VDM as during the modelling work one such update has been provided by the implementation team with minimal amount of effort to update the model. This first update has been a result of inconsistencies uncovered during encoding of the table into the VDM model.

In total 11 traces have been created, most of them considering only a single value for an input parameter, while a single trace has considered a selection of three values for the input parameter. This lead to creation of 13 tests covering the actions of interest from the roles and profiles table as well as considering an important action of creating a new sandbox. The evaluation of traces has taken less than two seconds, providing a well performing solution for analysis of client permissions. Since the roles and profiles table is continuously updated it is expected that the VDM model pre- and post-conditions on different operation calls will be updated as well in order to carry out an analysis ensuring the security of the sandbox platform.

\section{Related Work}
\label{sec:related}
Given its popularity, there are several works covering formal verification of
cloud systems \cite{Kulik&20}. In fact, it is not unusual for major cloud
vendors to adopt, use, and have internal departments doing research in formal
methods and verification tools.  Our work is different from the usual in the
sense that cloud operators typically do not provide access to the same
sandboxed OSs to multiple users at the same time.  The work in \cite{Zeng&16}
is an example of the application of formal methods to secure the assets of
organisation that wish to use federated cloud systems (systems where several
providers and local clouds are combined), but are afraid of compromising the
security of their solution. The paper shows how the security of information
flow can be analysed.  Our work focus is also at the dataflow level, although
information security is not our current focus.  Our goal is to achieve a
solution similar to the one presented in \cite{Madi&18} and
\cite{Bleikertz&15}, but generalised beyond network checks and the cloud
machinery provisioning respectively.  We foresee the development of a security
middleware applying checks at the pain points elicited during the platform
modelling and auditing tasks. At project proposal phase we envisaged the
possibility to specialise and apply malware detection techniques as the one in
\cite{Macedo&13}, but the sandbox prototype based on sandboxed OS images is
much more suited to generalist and COTS malware checkers.

Sandboxing is a widely used to establish security.  Whether we think of the
popular browser tabs, an operating system kernel, or a C program memory
sandboxes, their security involves a sandbox. Formal methods have been widely
used to verify the good application of sandboxing. In \cite{Dongseok&12}, the
authors propose to extract a browser sandboxing kernel with leak-proof
security. Our work does not leverage on theorem proving, and it does not intend
to substitute the current implementation with a correct-by-construction one, but
that is still a possibility. Our current focus is the usage of lightweight
specification in the form of pre/post-conditions and invariants to reason about the system security. 
In \cite{Becker&16} the authors extract an executable COQ model of an
hypervisor, a ``real-world'' C++ implementation of a virtualisation
architecture. The executable model is then used to as an oracle for the
expected behaviour. Our work threads along the same lines, we expect to model
our sandboxing platform in VDM, from which we expect to derive an oracle and
expected traces exhibiting secure and insecure dataflows. 
In \cite{Besson&19}, a similar approach is used to program data memory. In all
the approaches implementations are expected to be secure, after modelling and
using tools in the security audit.

\section{Conclusion}
\label{sec:conc}

This paper reported on findings of use of VDM-SL in security analysis of online (potentially cloud based) collaboration platforms. The system architecture presented in this paper is based on a real life sandboxing platform, a part of a project aimed at collaborative use of model based engineering while preserving the information security and intellectual property. We created a model of the system, considering features provided within a table of roles and profiles by the implementation team. To model this feature set we have considered the mentioned actions and modelled the roles and profiles as pre conditions. We have then analysed several specific scenarios within the model by use of the combinatorial testing. The analysis has uncovered a potential issue, presenting an inconsistency within the intended sandbox creation functionality, specifically an issue within a hierarchy of operation calls, where an action can be requested without specifically defined roles, however it is dependent on another action requiring specific role. These results have been shared with the implementation team, that has used it to more clearly define the table of roles and profiles within the system. The analysis has in total considered 13 tests and could be carried out in seconds.

As a future work we intend to align with the progress of the implementation team and update the model to analyse every update to the table of roles and profiles. Furthermore we plan to extend our model to be able to analyse aspects of federated cloud, where some sandbox servers could exist within an on premises data-center while other servers would exist within a public cloud service. To this end we intend on proposing an access model, i.e. and updated list of roles and profiles, that has been modelled using VDM and share the findings with the implementation team. 


\subsubsection*{Acknowledgements.}
The work presented here is partially supported by the HUBCAP Innovation Action funded by the European Commission's Horizon 2020 Programme under Grant Agreement 872698. We would also like to express our thanks to the anonymous reviewers.

\clearpage

\endgroup

\begingroup
\hypersetup{
bookmarks=false,
breaklinks=true,
colorlinks=true,
linkcolor=black,
citecolor=black,
urlcolor=black,
pdfpagelayout=SinglePage,
pdfstartview=Fit}
\title{Visual Studio Code VDM Support}

\author{ Jonas Kjær Rask\inst{1}
\and Frederik Palludan Madsen\inst{1}
\and Nick Battle\inst{2} 
\and Hugo Daniel Macedo\inst{1}
\and Peter Gorm Larsen\inst{1} }
\authorrunning{J. K. Rask, F. P. Madsen, N. Battle, H. D. Macedo and P. G. Larsen}

\institute{
DIGIT, Aarhus University, Department of Engineering, \\
Finlandsgade 22, 8200 Aarhus N, Denmark\\
\email{\{201507306, 201504477\}@post.au.dk, \{hdm, pgl\}@eng.au.dk}
\and
Independent, \email{nick.battle@acm.org}
}
			
\maketitle
\setcounter{footnote}{0} 

\makeatletter
\def\input@path{{3-Rask/}}
\makeatother

\graphicspath{{3-Rask}}
\begin{abstract}
 How is it possible to significantly improve the Integrated Development Environment (IDE) for VDM from the existing Eclipse-based IDE? The proposal made in this paper is to use language-agnostic protocols such as the Language Server Protocol (LSP) and the Debug Adapter Protocol (DAP) connecting a general editor such as Visual Studio Code with core server functionality. This is demonstrated for editor related features, debugging, and Proof Obligation Generation and Combinatorial Testing support. We also believe that the extension of LSP and DAP will be useful for extending other IDEs for similar specification languages, since using such standard protocols will require less effort to upgrade to modern front-ends for their IDEs. 
\end{abstract}


\section{Introduction}
\label{2-Rask:sec:intro}


\section{Background}
\label{3-Rask:sec:background}
\todo{Tomas + Prasad + PGL 1p}

\begin{figure*}[bt]\centering
\includegraphics[width=\textwidth]{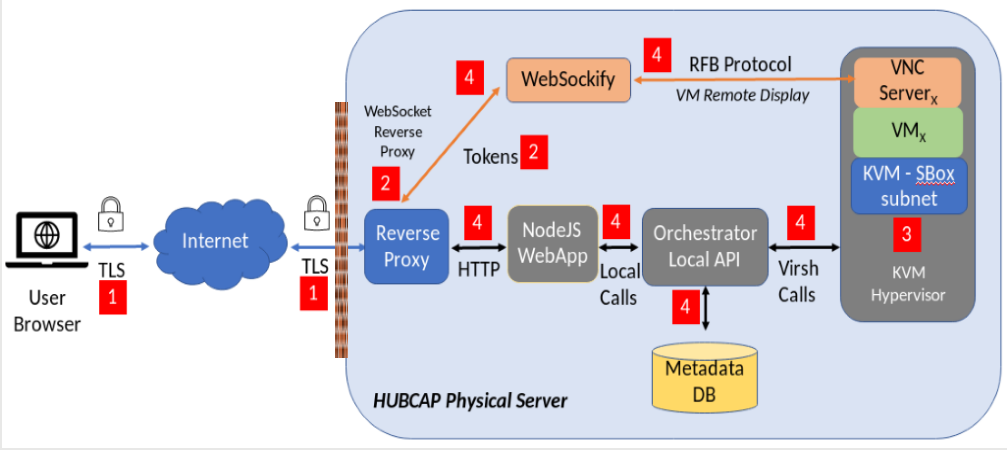}
\caption{\label{fig:sandbox-impl}The implementation of HUBCAP Sandbox system (taken from  \underline{sandbox user guide})}
\end{figure*}

\begin{itemize}
	\item explain Figure \ref{fig:sandbox-impl}
	\item VNC, VDM-SL
\end{itemize}

The Vienna Development Method (VDM Specification Language (VDM-SL) has been standardised by the International Standardisation Organisation \cite{ISOVDM96,Plat&92b}. It is a model-oriented formal specification language which is based around logic (e.g., for invariants, pre-conditions and post-conditions) and a number of abstract data types for different kinds of collections. It has been used industrially since the seventies where it was invented at IBM's laboratories in Vienna \cite{Larsen&96a,Kurita&09}. For the last two decades the Overture/VDM tool support have evolved after the VDMTools area \cite{2-Kulik:Larsen01,2-Kulik:Larsen&10a}.

\section{Implementing Language Features}
\label{sec:Implementation}

The support for \gls{vdm} in \gls{vscode} is accomplished using an extension that communicates with a language server based on the language support found in VDMJ, as illustrated in \Cref{fig:ExtensionArchitecture}.
The standard protocols \gls{lsp} and \gls{dap} are developed with programming languages in mind. 
This means that the specification language features that are common for these languages are not supported, \eg \gls{pog} and \gls{ct}, as discussed in \Cref{sec:ProtocolCoverage}.
To compensate for this we have developed the \gls{slsp} protocol, which is introduced in the section \Cref{sec:slspIntroduction}. 
Followed by a description of the components that the \gls{vscode} extensions\footnote{The extensions can be found at: \url{https://github.com/jonaskrask/vdm-vscode}} are comprised of and how they have been implemented.
Finally, we describe the work put into VDMJ to provide support for the protocols.

\begin{figure}[htb]
    \centering
        \includegraphics[width=1\textwidth]{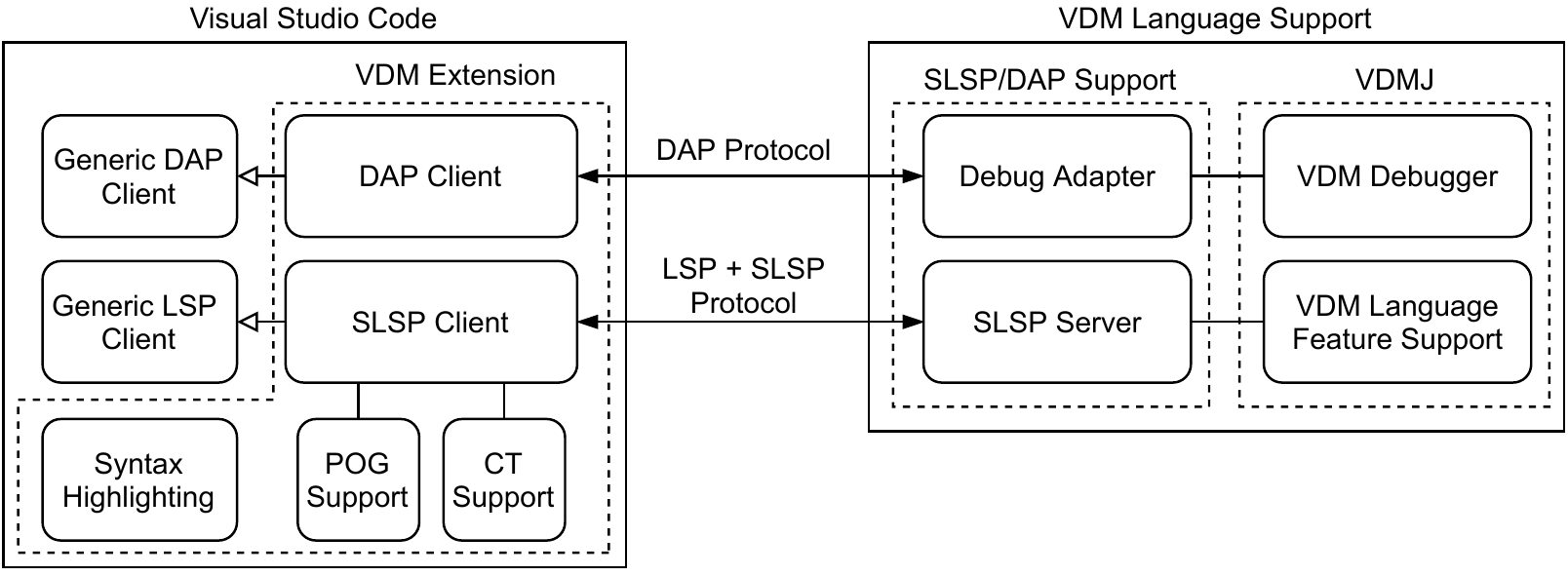}
    \caption{Overview for the suggested extension architecture}
    \label{fig:ExtensionArchitecture}
\end{figure}

\subsection{Specification Language Server Protocol}
\label{sec:slspIntroduction}
The purpose of the \gls{slsp} protocol is to facilitate support for specification language specific features using the same architecture as for \gls{lsp} and \gls{dap}.
The \gls{slsp} protocol is an extension to the \gls{lsp} protocol.
This means that it uses the same base protocol as \gls{lsp} and relies on functionality from \gls{lsp} such as the synchronisation between client and server.
The protocol is developed with the intention that it should also be usable by other languages than \gls{vdm} and possibly be included in the \gls{lsp} specification. 
Hence, the \gls{slsp} protocol uses language neutral data types, which allows the client to be language-agnostic enabling it to be used with any server that supports the \gls{slsp} protocol.
At the time of writing the \gls{slsp} protocol defines messages to support \gls{pog} and \gls{ct}, and it is used in the \gls{vscode} extension for those features. 

\subsection{Visual Studio Code Extension}
\Gls{vscode} operates with a rich extensibility model that enables users to include support for different programming languages, debuggers and various tools to support the development workflow, by downloading extensions from the \gls{vscode} extensions marketplace\footnote{See \url{https://marketplace.visualstudio.com/vscode}}.
This also enables developers to make extensions that they find missing from the marketplace, which is how \gls{vdm} is supported in \gls{vscode}; by creating an extension for each \gls{vdm} dialect.
The extensions include a \gls{slsp} client and a \gls{dap} client.
Besides communicating with the server the extension also has the responsibility of doing syntax highlighting as this is not supported in any of the protocols.

As illustrated in \Cref{fig:ExtensionArchitecture}, \gls{vscode} includes generic support for both the \gls{lsp} and \gls{dap} protocols which allows the client implementation to be carried out with little effort.
The generic protocol support provides the extension with handlers for all the \gls{lsp} and \gls{dap} messages which include code to synchronise the editor and the server, handling of the editor navigation and displaying messages from the server.
Thus, the extension provides configuration of the generic protocol support, launching and connecting to the server, syntax highlighting, and the \gls{gui} to support the \gls{slsp} features.
\Cref{fig:VSCodeSnippet_pog} depicts the \gls{vscode} environment with the VDM-SL extension active.

\begin{figure}[htb]
    \centering
        \includegraphics[width=1\textwidth]{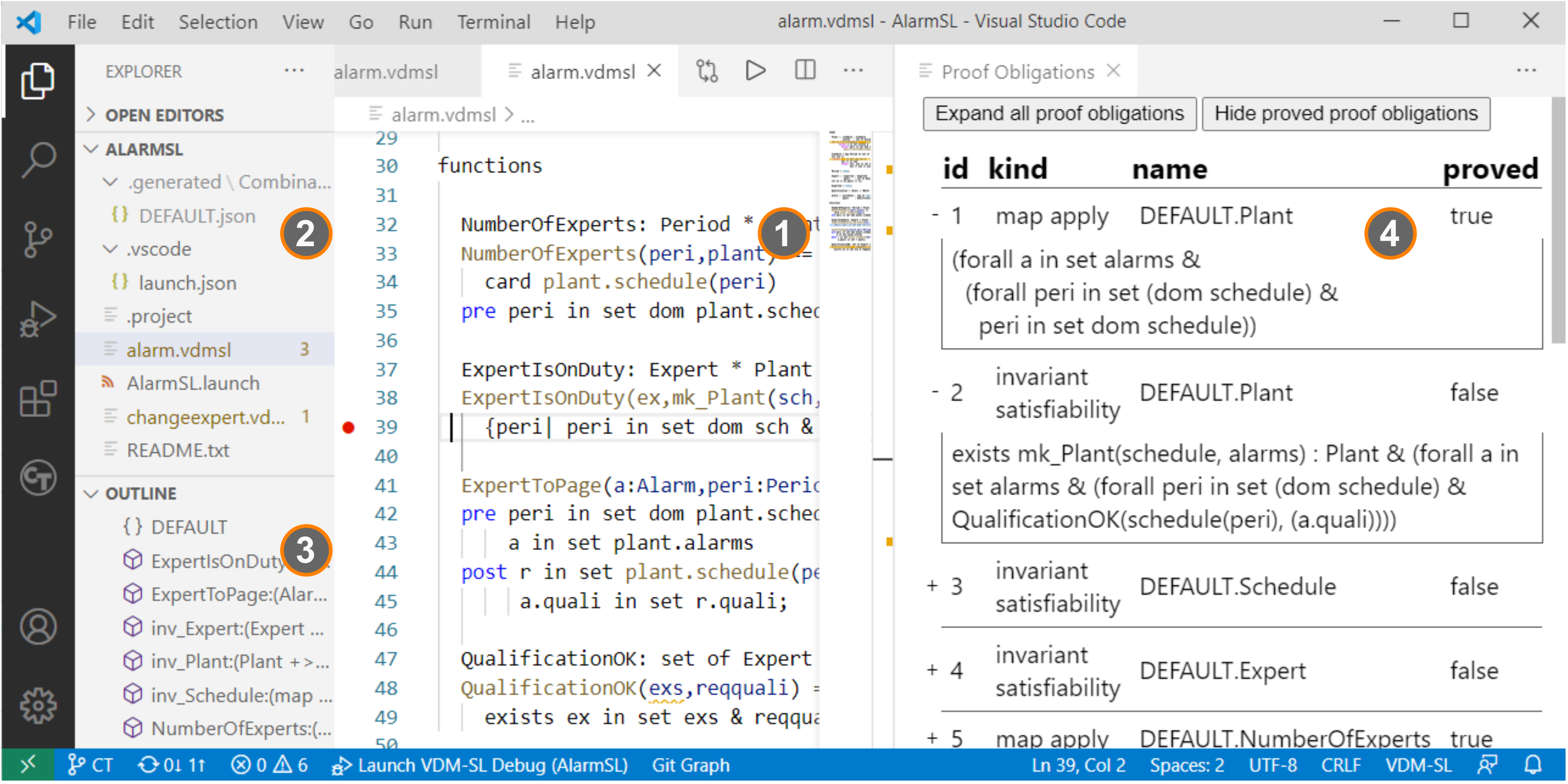}
    \caption{VDM-SL VS Code Extension: (1) Main Editor; (2) Project Explorer; (3) File Outline; (4) Proof Obligation View}
    \label{fig:VSCodeSnippet_pog}
\end{figure}

\subsubsection{Syntax Highlighting:}
Syntax highlighting in \gls{vscode} is handled on the client side as it is not supported by the protocols. 
Instead syntax highlighting is performed using TextMate grammars \cite{Gray07}. 
The TextMate grammars are structured collections of Oniguruma regular expressions\footnote{See \url{https://github.com/kkos/oniguruma}}, implemented using a JSON schema. 
The grammar file specifies a set of rules, that is used with pattern matching to transform the visible text into a list of tokens and colour these according to a colour scheme.

\subsubsection{Specification Language Server Protocol Client:}
As custom for language extensions the VDM extensions are activated when a file with a matching file format is opened, \eg opening a `\texttt{.vdmsl}' file actives the VDM-SL extension.
On activation the client launches the server and connects to it.
When connected, the client forwards all relevant information to the server using the \gls{lsp} protocol. 
Furthermore, any responses or notifications from the server triggers feedback to the user, which is handled by the generic \gls{lsp} client integration in \gls{vscode}.
In addition to setting up the support for \gls{lsp} the client also provides customised \gls{gui} views to support \gls{pog} and \gls{ct}. This is not directly supported by \gls{vscode} and \gls{lsp} since these features are not commonly used for traditional programming languages. 

The \gls{pog} view, illustrated in \Cref{fig:VSCodeSnippet_pog}, provides a list of the proof obligations for a specification, where expanding an element displays the actual proof obligation.
The view is implemented using the \gls{vscode} Webview API\footnote{See \url{https://code.visualstudio.com/api/extension-guides/webview}.}, which is customised using CSS, HTML and JavaScript.

The \gls{ct} view, illustrated in \Cref{fig:VSCodeSnippet_ct}, shows the tests that are generated based on the traces in the specification, these are structured in a tree view.
Traces can be fully or partially executed using either an execution filter or executing a test group at a time. 
A tests execution sequence and result is displayed in a separate tree view.
Tests can also be debugged, which is facilitated using the \gls{dap} protocol.
The \gls{ct} view is implemented using the \gls{vscode} Tree View API\footnote{See \url{https://code.visualstudio.com/api/extension-guides/tree-view}.}. which reduces the complexity of implementing a view.

\begin{figure}[htb]
    \centering
        \includegraphics[width=1\textwidth]{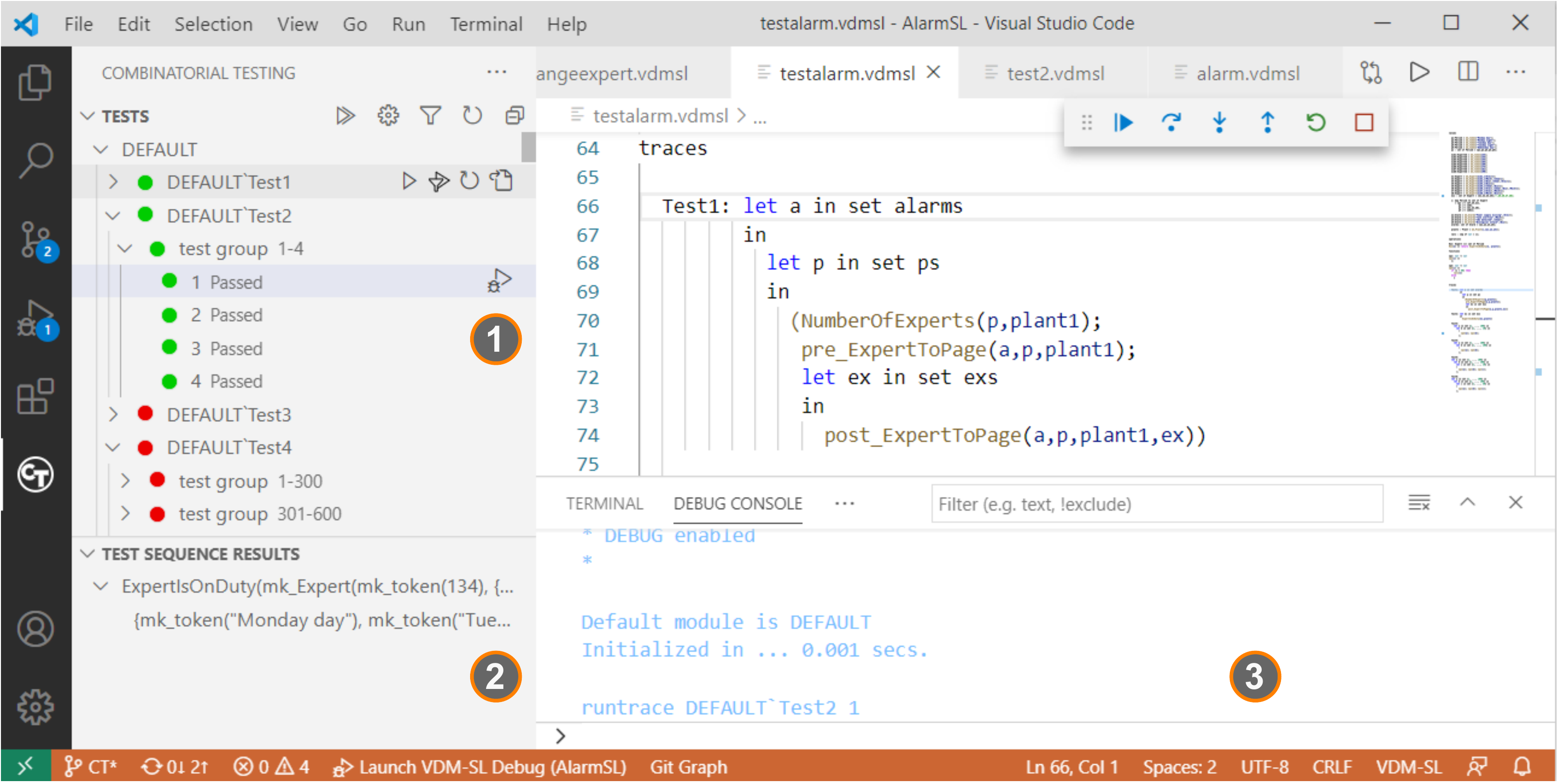}
    \caption{VDM-SL VS Code Extension: (1) Combinatorial Testing View; (2) Test Result View; (3) Debug Console}
    \label{fig:VSCodeSnippet_ct}
\end{figure}

\subsubsection{Debug Adapter Protocol Client:}
The \gls{dap} protocol support is added to the client by specifying a \texttt{DebugAdapterDescriptorFactory} which establishes the connection between the generic debugger client and the server upon starting a debug session.
This would normally include launching a \gls{dap} server and connect to it.
However, as the \gls{lsp} and \gls{dap} server is combined into one we simply connect to the port specified for \gls{dap} communication. 


\subsection{Enabling the use of LSP and DAP in VDMJ}

    


Internally, VDMJ provides a set of language services to enable VDM specifications to be processed, and by default the coordination of these services is handled by a command-line processor. So for example, a Parser service is used to analyse a set of VDM source files to produce an Abstract Syntax Tree (AST); if there are no syntax errors, that AST is further processed by a Type Checker; and if there are no type errors, the tree is further processed by an Interpreter (in a read-eval loop) or by a Proof Obligation Generator. If the Interpreter encounters a runtime problem or a debug breakpoint, it stops and interacts with the command-line to allow the user to examine the stack and variables, for example.

Although the command-line is the default way to interact with VDMJ, this is not assumed in the design. An alternative means to interact is provided using the DBGP protocol\footnote{See \url{https://xdebug.org/docs/dbgp}}. This allows VDMJ to interact with a more sophisticated IDE, such as Overture. DBGP messages are exchanged over a local TCP socket, allowing the IDE to coordinate the execution of expressions within the specification, and to debug using a richer user interface. The VDMJ end of the DBGP socket coordinates the same services as the command-line does, but with requests and responses being exchanged via the socket, using DBGP messages.

So adapting VDMJ to use the LSP and DAP protocols is a matter of creating a new handler to accept socket connections from a client, and process the JSON/RPC messages defined in the respective LSP and DAP standards. Since LSP and DAP are expressed in general language terms, there is generally a natural mapping from abstract concepts in these protocols to the specific case of a VDM specification.

Creating the LSP part of the VDMJ handler was relatively simple. The biggest difference to the existing connection handlers is that LSP is an ``editing'' protocol. That is, rather than being asked to process a fixed set of unchanging specification files, LSP requires the server to accept the creation and editing of files on the fly. So whereas normally, the Parser would only be called once to process a set of fixed specification files, in the LSP handler it can be called repeatedly as edits are passed from the client. Since the VDMJ parser is relatively efficient, it is responsive enough to re-parse the affected file as edits are made. A decision was taken to only type-check the specification when the client deliberately saves the work (\ie writes changes to disk). This is the same behaviour as Overture.

Execution and debugging via the DAP protocol required the creation of a new ``DebugLink'' subclass in VDMJ. The existing command-line and DBGP protocols each extend an abstract DebugLink class, which allows the Interpreter service to interact with an arbitrary client to coordinate a debugging session. In this case, the client is the DAP client. Since the DebugLink was designed to be extended, this was relatively straightforward -- in fact the DAPDebugLink is an extension of the command-line class with the protocol aspects changed from textual read/writes to JSON message exchanges. It is about 200 lines of Java. The only complexity in this area is in the case of multi-threaded VDM++ specifications. Special care must be taken to be sure that all of the separate threads and their data are coordinated correctly, with appropriate mutex protection of the (single) communication channel to the DAP client.

In addition to parsing, type-checking and executing a specification, the LSP protocol also enables the server to offer various language services that allow the client to build a more intelligent user interface. For example, the outline of a file (its contents, in terms of the VDM definitions within and their location) can be queried; any name symbol in a file can be used to navigate to that symbol's definition (\eg moving from a function application to its definition, possibly in a different file); and name-completion services are offered, where the first few characters of a name can be typed, and the server will offer possible completions. These services do not exist in the base VDMJ, and were therefore added as part of the LSP development. They are mostly implemented via visitors (using the VDMJ visitor framework) which process the AST.

The Proof Obligation Generation and Combinatorial Testing features of VDMJ are maked available with the SLSP protocol by adding new ``slsp'' RPC methods. Internally, the server routes these new requests to the VDMJ components that are able to perform the analyses.

Proof Obligations (POs) can be generated relatively quickly (usually in a fraction of a second, even for hundreds of obligations), so the ``slsp/POG/generate'' method takes a one-shot approach, generating and returning all the POs in the specification by default, optionally allowing the UI to limit the scope to a sub-folder of specification files or a single file. The returned obligations are represented as VDM source, as a stack of contexts with a primitive obligation at the end. This allows the UI to decide how to represent the indentation of the full obligation without requiring it to understand the VDM AST.

Combinatorial Testing is a more complex problem, requiring multiple interactions between the UI and the server as new ``slsp/CT'' methods. Unlike Proof Obligations, Combinatorial Tests can expand to millions of test cases and their execution can take many hours. This in turn means that the UI must have some indication of the progress of the execution, and it must also have the ability to cleanly terminate the execution in the event that the test run is taking too long. The base LSP protocol allows for the asynchronous cancellation of an action, and this is implemented in the LSP Server by running CT in a separate thread, allowing the main thread to listen for more interactions from the UI. This complication means that a check has to be added to prevent the user from trying to launch more than one test execution at the same time (which the runtime could not easily support). The server also has to guard against changes to the specification between the expansion of the tests and their subsequent execution. Once a test run is complete, individual tests can be sent to the Interpreter for debugging. This is enabled via the standard DAP protocol, with the addition of a new ``runtrace'' launch command. So the runtime thinks that the user has started a normal interactive debugging session, but instead of executing something like ``print fac(10)'', they are executing ``runtrace A'Test 1234'', which will debug test number 1234 of the expansion of the A'Test trace.

In terms of packaging, the SLSP/DAP combination of server functionality is in its own jar, separate from (and dependent on) the standard VDMJ jar. The SLSP/DAP jar is about 174Kb. VDMJ itself is about 2.4Mb.

\section{Results and Discussion}
\label{sec:Evaluation}

In this section we discuss the implementation effort carried out in order to support the \gls{lsp}, \gls{slsp} and \gls{dap} protocols for VDMJ and the efforts needed on the client side to provide \gls{vdm} support.
This is compared to the efforts necessary to migrate the Overture language features to a different \gls{ide} without using the protocols as performed in \cite{Tran&19}.
Finally, we discuss to which extent it is currently possible to decouple specification language features from development tools using standardised protocols and what is required to have fully decoupled language support using protocols.

\subsection{Assessing the Implementation Effort}

The implementation effort is quantified by counting the Lines of Code (LoC) used for implementing support for a given protocol and the features it enables.
Providing support the protocols for VDMJ requires 7855 LoC, whereas support for \gls{vdm} in the \gls{vscode} extension only requires 1880 LoC where most are used for the new \gls{slsp} features.
The few lines of code for the extension can mainly be attributed to the generic \gls{lsp} and \gls{dap} protocol implementations exposed by the \gls{vscode} API\footnote{See \url{https://code.visualstudio.com/api/references/vscode-api}}.

A detailed overview of the lines needed to implement the language support is shown in \Cref{tab:LoCOverview}. 
In the table the `json' and `rpc' directory contain a basic JSON/RPC system; `dap' and `lsp' are the protocol handlers for each; `lsp/lspx' is the handler for SLSP; `vdmj' is the DebugLink and visitors to implement the various features (i.e.\ these link to the VDMJ jar); and `workspace' contains the code to maintain the collection of files. 
For the \gls{vscode} extension `LSP client' and `DAP client' is the setup of each client; `POG' and `CT' is the protocol handlers and GUI support for each feature; `syntax highlighting' contains the TextMate grammars for \gls{vdm}; and `SLSP Protocol' is the definition of the \gls{slsp} protocol and expansion of the LSP client.

\begin{table}[htbp]
\centering
\caption{LoC measures for the files and directories used to support VDM in VS Code\protect\footnotemark}
\label{tab:LoCOverview}
\begin{tabular}{|l|r|r|r|r|lllrrrr}
\cline{1-5} \cline{8-12}
\multicolumn{5}{|c|}{\cellcolor[HTML]{C0C0C0}SLSP/DAP support for VDMJ}                 & \textbf{} & \multicolumn{1}{l|}{\textbf{}} & \multicolumn{5}{c|}{\cellcolor[HTML]{C0C0C0}VS Code Extension}                                                                                                                                 \\ \cline{1-5} \cline{8-12} 
\textbf{Directory} & \textbf{Code} & \textbf{Comment} & \textbf{Blank} & \textbf{Total} &           & \multicolumn{1}{l|}{}          & \multicolumn{1}{l|}{\textbf{Feature}} & \multicolumn{1}{r|}{\textbf{Code}} & \multicolumn{1}{r|}{\textbf{Comment}} & \multicolumn{1}{r|}{\textbf{Blank}} & \multicolumn{1}{r|}{\textbf{Total}} \\ \cline{1-5} \cline{8-12} 
dap                & 699           & 450              & 185            & 1334           &           & \multicolumn{1}{l|}{}          & \multicolumn{1}{l|}{LSP client}       & \multicolumn{1}{r|}{156}           & \multicolumn{1}{r|}{16}               & \multicolumn{1}{r|}{29}             & \multicolumn{1}{r|}{201}            \\ \cline{1-5} \cline{8-12} 
json               & 635           & 184              & 139            & 958            &           & \multicolumn{1}{l|}{}          & \multicolumn{1}{l|}{DAP client}       & \multicolumn{1}{r|}{54}            & \multicolumn{1}{r|}{10}               & \multicolumn{1}{r|}{15}             & \multicolumn{1}{r|}{79}             \\ \cline{1-5} \cline{8-12} 
lsp                & 1653          & 625              & 337            & 2615           &           & \multicolumn{1}{l|}{}          & \multicolumn{1}{l|}{POG}              & \multicolumn{1}{r|}{460}           & \multicolumn{1}{r|}{51}               & \multicolumn{1}{r|}{96}             & \multicolumn{1}{r|}{607}            \\ \cline{1-5} \cline{8-12} 
lsp/lspx           & 169           & 44               & 30             & 243            &           & \multicolumn{1}{l|}{}          & \multicolumn{1}{l|}{CT}               & \multicolumn{1}{r|}{732}           & \multicolumn{1}{r|}{105}              & \multicolumn{1}{r|}{165}            & \multicolumn{1}{r|}{1002}           \\ \cline{1-5} \cline{8-12} 
rpc                & 187           & 138              & 56             & 381            &           & \multicolumn{1}{l|}{}          & \multicolumn{1}{l|}{Syntax highlight} & \multicolumn{1}{r|}{374}           & \multicolumn{1}{r|}{0}                & \multicolumn{1}{r|}{0}              & \multicolumn{1}{r|}{374}            \\ \cline{1-5} \cline{8-12} 
vdmj               & 2026          & 451              & 392            & 2869           &           & \multicolumn{1}{l|}{}          & \multicolumn{1}{l|}{SLSP protocol}    & \multicolumn{1}{r|}{104}           & \multicolumn{1}{r|}{167}              & \multicolumn{1}{r|}{27}             & \multicolumn{1}{r|}{298}            \\ \cline{1-5} \cline{8-12} 
workspace          & 2655          & 641              & 545            & 3841           &           & \multicolumn{1}{l|}{}          & \multicolumn{1}{l|}{\textbf{Sum}}     & \multicolumn{1}{r|}{\textbf{1880}} & \multicolumn{1}{r|}{\textbf{349}}     & \multicolumn{1}{r|}{\textbf{332}}   & \multicolumn{1}{r|}{\textbf{2561}}  \\ \cline{1-5} \cline{8-12} 
\textbf{Sum}       & \textbf{7855} & \textbf{2489}    & \textbf{1654}  & \textbf{11998} &           &                                &                                       &                                    &                                       &                                     &                                     \\ \cline{1-5}
\end{tabular}
\end{table}
\footnotetext{The LoC is counted using the ``VS Code Counter'' extension found at: \url{https://marketplace.visualstudio.com/items?itemName=uctakeoff.vscode-counter}}


A comparison between LoC for the \gls{vscode} extension, migrating Overture to Emacs (as in \cite{Tran&19}) and the Overture IDE (version 3.0.1) not counting the core is seen in \Cref{tab:LoCCompare}. 
From the comparison we find that the \gls{vscode} extension requires more code than the Emacs migration but Overture consists of far more LoC than both.
However, if we only look at the \gls{lsp} and \gls{dap} client support that covers all the normal programming language features they only require 210 LoC.
The \gls{slsp} parts of the extension are made almost generic, hence these parts can be reused for other specification languages to enable support for the same features in \gls{vscode}.

For VDMJ to support the protocols it requires considerably more LoC compared to the Emacs approach from \cite{Tran&19}, which weights against the protocol approach for enabling \gls{vdm} feature support.
However, the protocol solution is a standardisation of the decoupling, hence the server supporting the language features (in this case VDMJ) only has to be implemented once to be used with any development tool supporting the protocols. 
In comparison, the \gls{vscode} extension and migration of the Overture language core to another development tool, as done with Emacs, must be performed for each \gls{ide} that is to be supported.
Furthermore, the Overture migration depends on Emacs packages, which may not exist in other \glspl{ide} or have a different interface.
Thus, the analysis of finding suitable packages for another \gls{ide} will have to be repeated, possibly making the server solution faster to implement in a new \gls{ide}.

\begin{table}[htbp]
\centering
\caption{LoC measure for all IDE related files in the VS Code extension, the Emacs extension and Overture}
\label{tab:LoCCompare}
\begin{tabular}{|l|c|c|c|}
\hline
Project       & VS Code (+VDMJ)                 & Emacs                    & Overture                               \\ \hline
Lines of Code & \multicolumn{1}{r|}{1880 (+7855)} & \multicolumn{1}{r|}{349} & \multicolumn{1}{r|}{\textgreater{}63K} \\ \hline
\end{tabular}
\end{table}

To provide a fair comparison between the three implementation efforts, we compare the features that are available in each.
The comparison is done by the same set of features as in \cite{Tran&19}, this is found in \Cref{tab:featureCompare}.
As illustrated, the Overture IDE and the Emacs migration both support more features than the \gls{vscode} extension.
However, some of the features provided by the Emacs migration is only available through a command-line interface and not by a \gls{gui}.
\gls{vscode} does support implementation of integrated command-line interfaces like the Emacs migration, however the implementation efforts related to this are unknown.
Providing this kind of command-line interface has intentionally been left out of the \gls{vscode} extension, as we wanted to replicate the workflow from Overture.

\begin{table}[htbp]
\centering
\caption{Feature comparison between the VS Code extension, Overture and Emacs. X indicates that a feature is supported. (X) indicates that the feature is only available through a command-line interface, meaning that it is not supported by a graphical user interface.}
\label{tab:featureCompare}
\begin{tabular}{|l|c|c|c|}
\hline
\textbf{Features}         & \textbf{VS Code} & \textbf{Overture} & \textbf{Emacs} \\ \hline
Syntax highlighting       & X                & X                 & X              \\ \hline
Symbol prettyfication     &                  &                   & X              \\ \hline
Syntax validation         & X                & X                 & X              \\ \hline
Evaluation                & X                & X                 & (X)            \\ \hline
Debugging                 & X                & X                 & (X)            \\ \hline
POG                       & X                & X                 & (X)            \\ \hline
LaTeX report generation   &                  & X                 & (X)            \\ \hline
Combinatorial testing     & X                & X                 & (X)            \\ \hline
Code generation           &                  & X                 &                \\ \hline
Auto completion (limited) & X                & X                 & X              \\ \hline
Template expansion        &                  & X                 & X              \\ \hline
Standard library import   &                  & X                 &                \\ \hline
\end{tabular}
\end{table}

\subsection{Protocol Coverage of Language Features}
\label{sec:ProtocolCoverage}

To get an overview of the specification language features that are covered by the protocols, we have examined and grouped the features that are sought after to support specification languages such as \gls{vdm}, B and Z.
This is illustrated in \Cref{fig:ProtocolCoverage}, where the language features have been divided into four categories:
\begin{enumerate}
    \item \textbf{Editor}: features that support writing a given specification, e.g.\ type- and syntax-checking.
    \item \textbf{Translation}: features that support translating a specification to other formats, e.g.\ executable code and LaTeX.
    \item \textbf{Validation}: features that support validation of a specification.
    \item \textbf{Verification}: features that support verification of a specification.
\end{enumerate}

\begin{figure}[htbp]
    \centering
        \includegraphics[width=0.53\textwidth]{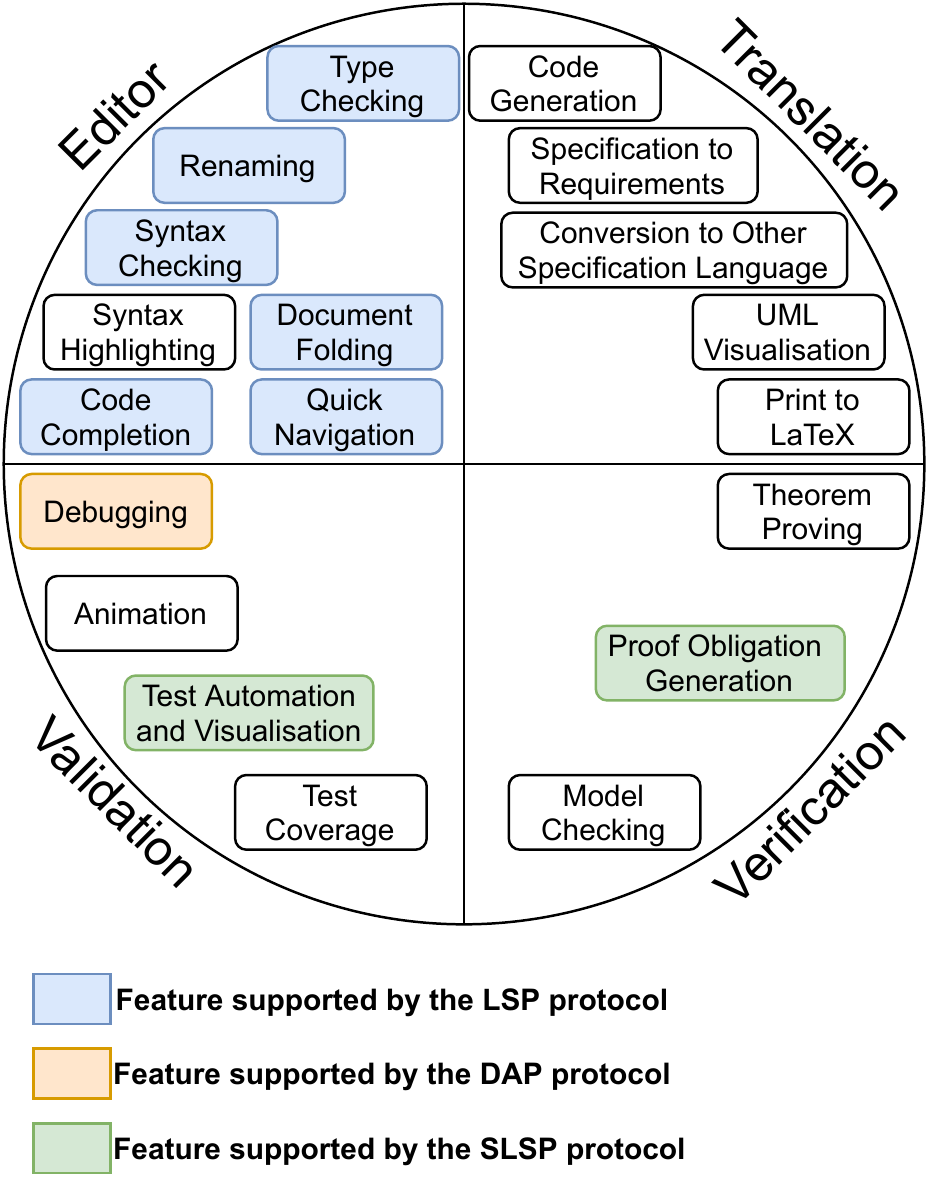}
    \caption{Specification language features covered by existing protocols}
    \label{fig:ProtocolCoverage}
\end{figure}

As a result of the initial implementation and by comparing the features supported by the \gls{lsp} protocol with the features from Overture, it is found that the protocol is able to support all of the editor related features except syntax highlighting.
Thus, specification languages that benefits from the editor features found in \Cref{fig:ProtocolCoverage}, can use a direct implementation of the \gls{lsp} protocol to support this feature set.


The \gls{dap} protocol can be used for decoupling the debugging feature as it supports common debug functionality such as different types of breakpoints, variable values and more.
This is useful for specification languages that allow execution of specifications.

Additionally, the \gls{slsp} extension to \gls{lsp} facilitates support for both \gls{pog} and \gls{ct}. 
As found in \Cref{fig:ProtocolCoverage} there are still many features related to specification languages that are not supported by protocols.
Thus, in order to achieve fully decoupled language server support for \gls{vdm} one or more protocols must be developed that support the remaining features.
From our development of \gls{slsp} we believe that it is possible to develop a language-agnostic protocol to support all of these features, which should make it possible to apply the protocol to multiple specification languages and potentially increase the industrial uptake of specification languages.

\section{Related Work}
\label{sec:relatedWork}



Support for multiple editors has become common for programming languages, and similar work is now being conducted to achieve the same for modelling and specification languages. 
Previous work has been focused on providing VDM support for specific platforms, such as Emacs \cite{Tran&19}, VDMPad \cite{Oda&15} and Overture Web IDE \cite{Reimer&16}.
For programming languages it is common to provide language support using standardised protocols to contain the language support in a language-specific server that can be used with multiple language-agnostic clients.
This tendency is also becoming apparent for other computer-based languages, such as domain-specific languages (DSL) \cite{Bunder19b}, graphical modelling \cite{Rodriguez&18} and formal specification languages such as PVS \cite{Masci&19}, Dafny \cite{Hess&19} and our work for VDM.

Tran and Kulik \cite{Tran&19} is able to migrate the Overture core from the Eclipse IDE to Emacs. 
This is carried out using as many Emacs packages as possible which allows them to provide VDM support with very few additional LoC. 
This is interesting as it provides knowledge about how few LoC is necessary, which reduces the potential gain from a generalised solution, \ie if the generalised solution requires a lot of effort you could argue that you might as well have several small dedicated solutions.

Both Oda et al. \cite{Oda&15} and Reimer and Saaby \cite{Reimer&16} presents a web IDE for VDM, based on VDMJ and Overture respectively. 
Even though both are web based they do not utilise the same language core, thus not providing the same set of language features.
Both solutions can probably benefit from unifying the efforts towards shared language support, this could be facilitated using the LSP, DAP and SLSP protocols.
Also, \gls{vscode} is build using HTML, CSS, and JavaScript which allows the IDE to be launched in a web environment, \eg Coder\footnote{See \url{https://coder.com/}.}. Thus, adding to the number of web IDEs for VDM.



Masci and Muñoz \cite{Masci&19} provides support for PVS using the LSP protocol and VS Code. They too use the extensibility of the protocol to provide client-server support for non-LSP features, such as \gls{pog} and theorem proving.
These features are facilitated using a PVS-specific protocol.
It would be interesting to investigate if the SLSP protocol could also be used in their case, and what changes may have to be made. 
This could make way for an addition to the LSP protocol that can be used for a variety of specification languages and possibly be incorporated in the official LSP specification.

\section{Concluding Remarks and Future Work}
\label{3-Rask:sec:conclusion}

\subsubsection*{Acknowledgments}
We would like to thank Futa Hirakoba for providing information about using VS Code and LSP to support VDM and allowing us to use his implementation of a VS Code extension as a basis for our extensions. We would also like to express our thanks to the anonymous reviewers.

\bibliographystyle{splncs03}

\clearpage
\endgroup

\begingroup
\colorlet{punct}{red!60!black}
\definecolor{background}{HTML}{EEEEEE}
\definecolor{delim}{RGB}{20,105,176}
\colorlet{numb}{magenta!60!black}

\lstdefinelanguage{json}{
	basicstyle=\scriptsize\ttfamily,
	numbers=left,
	numberstyle=\scriptsize,
	stepnumber=1,
	numbersep=8pt,
	showstringspaces=false,
	breaklines=true,
	frame=lines,
	backgroundcolor=\color{white},
	literate=
	*{0}{{{\color{numb}0}}}{1}
	{1}{{{\color{numb}1}}}{1}
	{2}{{{\color{numb}2}}}{1}
	{3}{{{\color{numb}3}}}{1}
	{4}{{{\color{numb}4}}}{1}
	{5}{{{\color{numb}5}}}{1}
	{6}{{{\color{numb}6}}}{1}
	{7}{{{\color{numb}7}}}{1}
	{8}{{{\color{numb}8}}}{1}
	{9}{{{\color{numb}9}}}{1}
	{:}{{{\color{punct}{:}}}}{1}
	{,}{{{\color{punct}{,}}}}{1}
	{\{}{{{\color{delim}{\{}}}}{1}
	{\}}{{{\color{delim}{\}}}}}{1}
	{[}{{{\color{delim}{[}}}}{1}
	{]}{{{\color{delim}{]}}}}{1},
}

\lstdefinelanguage{VDM++}
  {morekeywords={act, active, fin, req, waiting, abs, all, allsuper, always, and, answer, 
     assumption, async, atomic, be, bool, by, card, cases, char, class, comp, compose, conc, cycles,
     dcl, def, definitions, del, dinter, div, dlmodule, do, dom, dunion, duration, effect, elems, else, elseif, end,
     error, errs, exists, exists1, exit, exports, ext, floor, for, forall, from, functions, 
     general, hd, if, imports, in, inds, infer, init, inmap, input, instance, int, inter, inv, inverse, iota, is, 
     isofbaseclass, isofclass, inv, inverse, lambda, len, let, map, measure, mu,
     mutex, mod, module, nat, nat1, new, merge, 
     munion, not, of, operations, or, others, per, periodic, post, power, pre, pref, 
     private, protected, public, qsync, rd, responsibility, return, reverse,  
     sameclass, parameters, psubset, pure, rem, renamed, rng, sel, self, seq, seq1, set, skip, specified, st, 
     start, startlist, state, static, stop, stoplist, sporadic, subclass, subset, subtrace, sync, system, then, thread, 
     threadid, time, tixe, tl, to, token, traces, trap, types, undefined,
     union, uselib, using, values, 
     variables, while, with, wr, yet, RESULT, false, true, nil, periodic pref, rat, real},
   sensitive,
   frame=trBL,
   morecomment=[l]--,
   morestring=[b]",
   morestring=[b]',
   frameround=fttt
  }[keywords,comments,strings]
\lstdefinelanguage{JavaCC}
  {morekeywords={options, PARSER\_BEGIN, PARSER\_END, SKIP, TOKEN},
   sensitive=false,
  }[keywords]

\lstset{basicstyle=\scriptsize,tabsize=2,frame=trBL,frameround=fttt}
\keepvalues
\setlength\textfloatsep{8.0pt plus 2.0pt minus 2.0pt}
\setlength\intextsep{4.0pt plus 2.0pt minus 2.0pt}
\setlength\floatsep{4.0pt plus 2.0pt minus 2.0pt}
\setlength\abovecaptionskip{4.0pt plus 2.0pt minus 2.0pt}
\setlength\belowcaptionskip{0pt}

\title{Tuning Robotti: the Machine-assisted Exploration of Parameter Spaces in Multi-Models of a Cyber-Physical System}
\titlerunning{Tuning Robotti: the Machine-assisted Exp. of Par. Spaces in Multi-Models of a CPS}

\author{Sergiy Bogomolov\inst{1} \and
  John Fitzgerald\inst{1} \and
  Frederik F. Foldager\inst{2} \and
  Carl Gamble\inst{1} \and
  Peter Gorm Larsen\inst{3} \and
  Kenneth Pierce\inst{1} \and
  Paulius Stankaitis\inst{1} \and
  Ben Wooding\inst{1}
  }
\authorrunning{S. Bogomolov, J. Fitzgerald, F. Foldager, P. G. Larsen, K. Pierce,P. Stankaitis and B. Woodin}

\institute{School of Computing, Newcastle University, United Kingdom \\\email{Sergiy.Bogomolov,John.Fitzgerald,Carl.Gamble@ncl.ac.uk}
\\\email{Ken.Pierce,P.Stankaitis,B.Wooding1@ncl.ac.uk}
\and
Agro Intelligence, Agro Food Park 13, 8200 Aarhus N, Denmark, \\\email{ffo@agrointelli.com}
\and
DIGIT, Department of Engineering, Aarhus University, Aarhus, Denmark\\
\email{pgl@eng.au.dk}
}

\maketitle
\setcounter{footnote}{0}

\makeatletter
\def\input@path{{4-Bogomolov/}}
\makeatother

\graphicspath{{4-Bogomolov}}

\begin{abstract}
We describe a pilot study that aims to evaluate a systematic approach to tuning the design parameters of Cyber-Physical Systems by using machine-assisted Design Space Exploration supported by the INTO-CPS Toolchain. The study focuses on the design of a modified controller for a semi-autonomous agricultural robot. Beginning from a candidate multi-model built from VDM and 20-sim, the study uses data derived from test runs of the robot to tune selected parameters of a Continuous-Time model of the robot's physics in order to improve fidelity.
The study raises questions around support for the selection of parameters for tuning, selection of test scenarios, and the ranking of results. We identify further work to complete the design loop by assessing the consequences of introducing a modified controller and deploying it in further field tests.
\end{abstract}

\section{Introduction}
\label{4-Bogomolov:sec:intro}

Cyber-Physical Systems (CPSs) integrate computational and physical processes with data and network technologies, creating opportunities for new smart products and services and the digitalisation of industrial processes~\cite{Reimann&17}. CPS design is inherently multidisciplinary: a developer must bring together the products of diverse disciplines, often from diverse suppliers, with the attendant risks of incompatibilities and misunderstandings which may only come to light in prototyping. Model-based methods of \emph{virtual engineering}~\cite{VanderAuweraer&13} have been proposed as a means of addressing these challenges. Among the proposed advantages are the ability to use design models as a basis for optimising parameters by performing systematic Design Space Exploration~(DSE) with the goal of reducing the number of prototyping cycles required to converge on an optimal design.

A model-based approach to multi-disciplinary CPS design entails the federation of semantically heterogeneous models. We call such federations \emph{multi-models} and the coordinated execution of multi-models we term \emph{co-simulation}~\cite{Gomes&18}. Co-simulation allows the exploration of CPS-level behaviours that arise from the interaction between cyber and physical elements. In practice, this means that open tool chains should allow for the collaborative construction and co-simulation of multi-models. However, there is only limited experience with DSE over such multi-models within CPS product development.

A practical challenge in CPS design is that of estimating the properties of the physical product, and hence ensuring that cyber elements such as control and communication are matched to physical system elements' properties, to deliver the desired CPS-level performance. In this paper, we report progress in a study that aims to pilot the use of DSE to address this challenge. The underlying co-simulation technology is that of the ``Integrated Tool Chain for Model-based Design of Cyber-Physical Systems'' (INTO-CPS) project \cite{Fitzgerald&15}. This makes use of the mature Functional Mock-up Interface (FMI) 2.0 standard for co-simulation \cite{Blochwitz&11}\footnote{{\url{http://fmi-standard.org}}}. The core FMI-compliant tool of the INTO-CPS tool chain is a Co-simulation Orchestration Engine supporting both fixed and variable step-size simulations~\cite{Thule&19}.

The ultimate goal of the study is to evaluate the use of DSE over a multi-model to determine optimal control parameters under different operating conditions for an autonomous agricultural field robot, Robotti~\cite{Foldager&18}. The robot is a tool-carrier designed for accommodating a wide range of field operations. By combining the robot with several tools, a large number of load cases emerge. To reduce time to market and the cost of testing diverse configurations in the field, a structured simulation-based approach is needed to address the increasing number of possible configurations in a virtual setting.

The intention is to use DSE to optimise the steering controller, minimising cross-track error by co-simulation of a simple model representing the dynamics of the robot in a continuous-time (CT) formalism and a model representing the steering controller in a discrete-event (DE) formalism~\cite{Christiansen&15b}.  In the work performed to date and reported in this paper, the model of the machine dynamics is tuned experimentally to ensure that it satisfactorily captures the motion of the robot over a series of scenarios~(which are intended to exercise the controller rather than reflecting real field-work). Model calibration and fidelity are critical to applicability~\cite{Gomes&17,Gomes&18}.

We briefly discuss related work on DSE in Section~\ref{sec:related_work}. The relevant elements of the INTO-CPS tool chain and its underlying co-simulation are summarised in Section~\ref{sec:into}. The INTO-CPS support for DSE technology is explained in Section~\ref{sec:DSE}. The pilot study of the Robotti case is presented in Section~\ref{sec:robottimodel} and the DSE analysis of it in Section~\ref{sec:RobottiDSE}. Finally, Section~\ref{sec:discussion} discusses our results so far and future work.

\section{Related Work}
\label{sec:related_work}

The techniques of automated DSE originated in the embedded systems domain, focusing on optimum placement of software functionality on constrained hardware. The need for efficient DSE was noted early~\cite{Chou&95} and applied to integrated circuits (e.g. \cite{Apvrille&06}). The aerospace domain pioneered adoption of modelling and DSE at the larger-scale systems level, e.g. co-design of the Chinook helicopter's integrated circuit and controller~\cite{Chou95b} and to multi-objective energy-based optimisation~\cite{Ahmadi&12}. As in agriculture, the component or components being optimised in aerospace are often from a single domain (e.g.\ a turbine~\cite{Barzegar&11} or exhaust units~\cite{Karaali&15}), although the objectives being optimised may be from multiple domains. Where DSE is applied with components from different domains, these models still tend to represent only the physical elements of the system (e.g.\ combined heat and power~\cite{Sanaye&14} and hybrid energy systems cells~\cite{Mamaghani&15}).

A key challenge in DSE
is that the size of the design space increases exponentially as the number and range of parameters is increased~\cite{Kang&11}. Large design spaces can be managed by increasing simulation resources, or by reducing the number of simulations required. Guidelines on experiment design can help engineers identify important factors to focus their searches, for example using the Taguchi Method~\cite{Gamble&14}. This can be supported using techniques such as  computational data clustering algorithms~\cite{Li&13} and  global sensitivity analysis~\cite{Kucherenko&11}.

\section{The INTO-CPS Toolchain}
\label{sec:into}
\label{sec:intocps}

INTO-CPS is an open tool chain that facilitates collaborative development and co-simulation for multi-models\footnote{\url{www.into-cps.org}}.
Although it supports traceable systems engineering activities from requirements analysis to test, the tool chain's central feature is support for construction and co-simulation of multi-models using FMI.
%

The tool chain allows requirements to be described using SysML~\cite{SysML17} using a profile for CPSs which has been implemented in the Modelio tool~\cite{Bagnato&17}. This profile allows the architecture of a CPS to be described, including both computational and physical elements. Based on a profile-conformant model, one can automatically generate FMI descriptions for each constituent model. It is also possible to generate the connections between different Functional Mockup Units (FMUs) for a co-simulation.

The FMI model descriptions can be imported into diverse modelling and simulation tools. Each of these can produce detailed models that, exported as FMUs, can be incorporated into an overall co-simulation as independent simulation units. The element models can either be in the form of DE models or in the form of CT models combined in different ways. Thus, heterogeneous constituent models can then be built around this FMI interface, using the initial model descriptions as a starting point. A Co-simulation Orchestration Engine~(COE) allows these constituent models to be evaluated through co-simulation~\cite{Thule&19}. The COE also allows real software and physical elements to participate in co-simulation alongside models, enabling both Hardware-in-the-Loop (HiL) and Software-in-the-Loop (SiL) simulation.

A web-based INTO-CPS Application has been produced to allow engineers to manage the multi-modelling and co-simulation processes~\cite{Macedo&20}. This can be used to launch the COE, enabling multiple co-simulations to be defined and executed, and the results collated and presented automatically. The tool chain will allow these multiple co-simulations to be defined via DSE.


\section{Design Space Exploration in INTO-CPS}
\label{sec:DSE}

DSE is the activity of evaluating multi-models from a collection of alternatives in order to reach a basis for subsequent more detailed design~\cite{Gamble&14}.
The set of multi-models under consideration is termed a \emph{design space}. The multi-models within a design space may differ from one another solely in terms of the values of specific \emph{design parameters}, or may be more fundamentally different, for example in terms of structure or component elements. The purpose of DSE is to evaluate these multi-models under a series of co-simulations, in order to arrive at a \emph{ranking} against an \emph{objective}~(or \emph{cost function}), i.e., a set of criteria or constraints that are important to the developer, such as cost or performance. A multi-model or group of multi-models selected from a ranking might then form a basis for further design steps. The value of DSE is in prioritising cyber-physical prototypes which may have to be produced and on which expensive tests may have to be performed.

Design spaces for CPSs may be extremely large, and so different DSE techniques aim to explore the space of alternatives rapidly. These include both \emph{open} and \emph{closed}-loop approaches. In this kind of assessment there can be tradeoffs over multiple potentially competing objectives (e.g., speed and energy consumption). To enable the engineer to select optimal configurations it is necessary to have a convincing way to present the results, for example in the form of a \emph{pareto-front}.

Since system-level properties of CPSs are often the result of interactions between computational and physical processes, it is worth noting that tradeoffs can be made across these domains. For example, one might want to examine any tradeoff between different error detection and recovery strategies in supervisory controller code and, say, the performance and cost of sensors. The ability to do this over a multi-model is one of the advantages of taking a multi-disciplinary co-simulation approach.

The INTO-CPS tool chain supports DSE by running co-simulations on selected points in the design space of interest. Both open loop (exhaustive search) and closed loop (generic search or simulated annealing) approaches can be supported in exploring alternative solutions~\cite{Fitzgerald&16a}.

Given system requirements (potentially including constraints for different aspects of a CPS) and a multi-model, a desired multi-model configuration can be create for the INTO-CPS Application. Using scripts implemented in a combination of Java and Python, the DSE analysis can be performed from the INTO-CPS Application, and the ranking of results can be displayed after individual simulations have been conducted. This workflow was followed in the process illustrated later in Figure~\ref{fig:method}.


INTO-CPS supports cloud deployment of DSE through the open-source CondorHTC framework, and genetic algorithms for reducing simulations~\cite{Fitzgerald&17a}. Existing guidelines on experimental design for DSE over multi-models, for example using Taguchi Methods, can also help engineers using INTO-CPS~\cite{Gamble&14}. 

\section{Robotti Pilot Study: Overview and Experimental Setup}
\label{sec:robottimodel}

%

\subsection{Goal of the Study}
\label{sec:pilot}

The purpose of this pilot study is to assess the viability of DSE as a tool in  improving the fidelity of a multi-model, based on real-world data observed during field trials of a prototype system. The study is based on the multi-model of the Robotti unmanned platform developed by Agro Intelligence (Agrointelli) for field operations in commercial agriculture~(see Figure~\ref{fig:RealRobotti}).

\begin{figure}
\centering
\includegraphics[width=0.75\columnwidth,trim=0 150 0 130, clip]{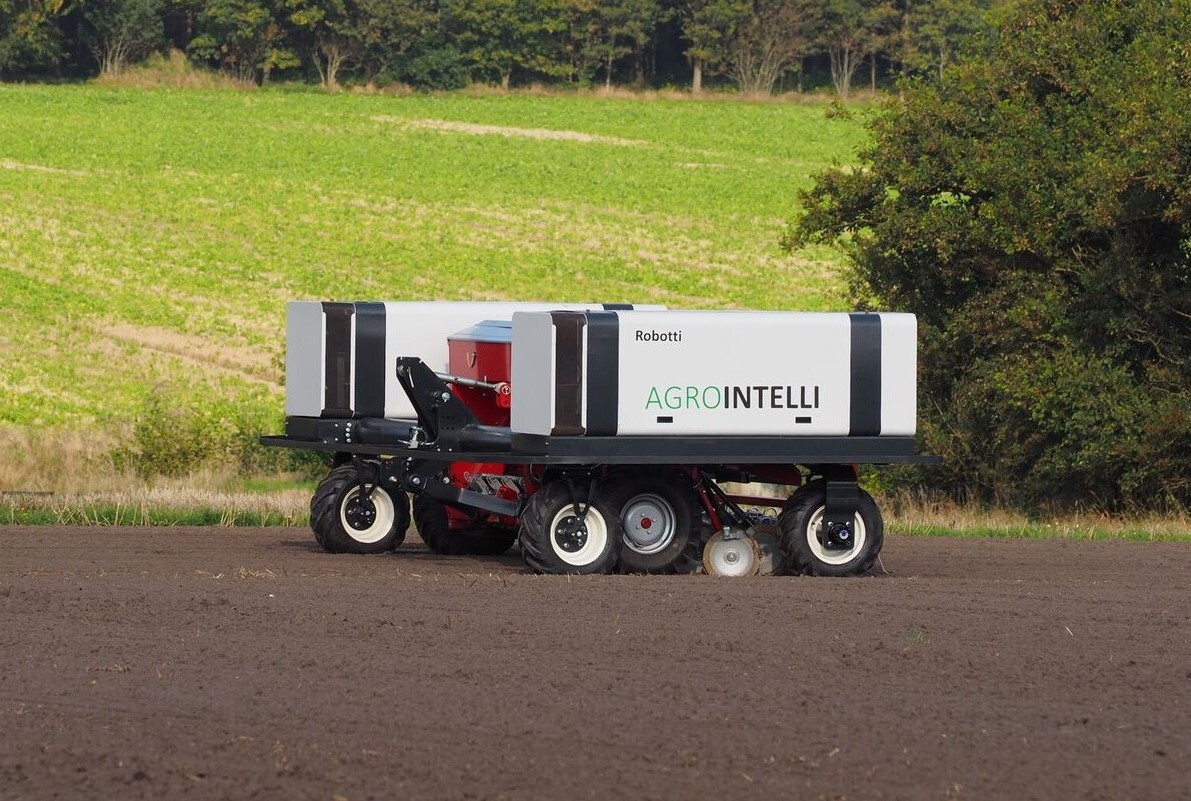}
\caption{\label{fig:RealRobotti}The Robotti unmanned platform}
\end{figure}

The challenge we address is that some physical properties can be difficult to estimate during design. These include those affected by changing conditions, such as surface friction and tyre stiffness. In order to develop a useful (multi-)model, we aim to approximate these values by measuring the overall performance of a prototype design and tuning the physical parameters of a simple model such that the modelled behaviour of the application matches the measurements. A number of field tests are carried out in order to characterise the actual motion of the robot as a function of given input trajectories. We then vary the physical parameters of the multi-model via DSE and identify the inputs that result in the closest correlation between the modelled and measured trajectories in corresponding scenarios. Note that we do not undertake modelling of the terramechanics including soil characteristics~\cite{WONG196781}.

The goal is to be able to make a simple, low degree of freedom model with only a few parameters that, to an acceptable degree, describes the motion of the robot. The fundamentals of the CT model are also described in \cite{Legaard2020}. This allows us to tune the model without knowing and modelling in detail the entire physical system by simply finding the best possible set of parameters to describe this motion. In the end, if we can achieve a validated multi-model of both the CT model describing the dynamics and the DE model describing the controller, we can use this knowledge to provide an estimate of how to adjust the control parameters for the best possible performance based on a future round of DSE.



\subsection{Field Tests and Scenarios}
\label{sec:fieldtests}

The field tests comprised a total of 27 runs across four scenarios. Each scenario is a pattern of control outputs designed to assess the dynamical response of Robotti to different motion profiles. Within each test run, the controller was set to ``open loop'' mode, where the control outputs change but there is no active control based on feedback. The control outputs and GPS position were recorded for each run. The position of the Robotti in a selection of test runs is shown in Figure~\ref{fig:scenarios}. The four scenarios are:

\begin{description}
  \item[Speed step:] The speed of each wheel is increased incrementally in lock step, resulting in the Robotti driving straight as in Figure~\ref{fig:path_speed_ramp1}. This test was repeated eight times.
  \item[Speed ramp:] The speed of each wheel is increased smoothly. This also results in a straight path as in the speed step tests. This test was repeated three times.
  \item[Turn ramp:] The speed of the right wheel is increased smoothly, while the left wheel is kept constant. This results in the turning motion seen in Figure~\ref{fig:path_turn_ramp1}. This test was repeated eight times.
  \item[Sin:] The speed of the left and right wheel is increased and decreased to produce the sinusoidal motion seen in Figure~\ref{fig:path_sin1}. This test was repeated eight times.
\end{description}

\begin{figure}
\centering
\subcaptionbox{Speed step / speed ramp\label{fig:path_speed_ramp1}}
{\includegraphics[width=0.32\textwidth]{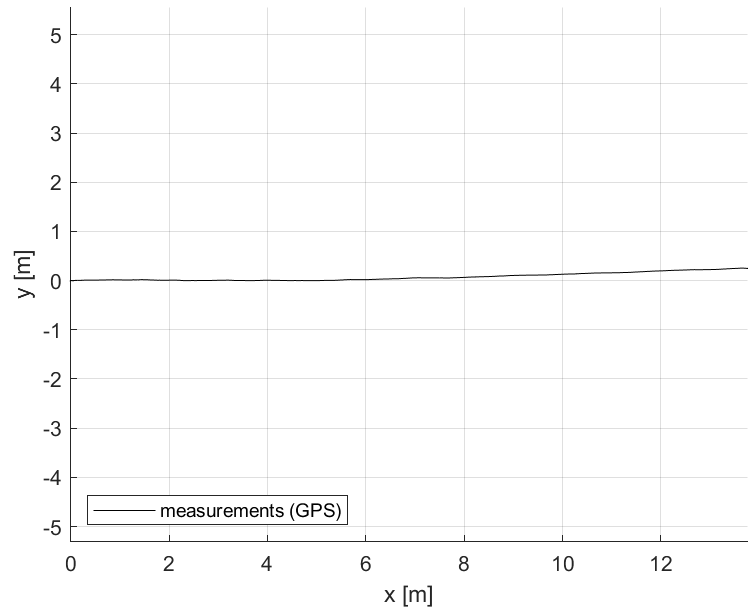}}
\subcaptionbox{Turn ramp\label{fig:path_turn_ramp1}}
{\includegraphics[width=0.32\textwidth]{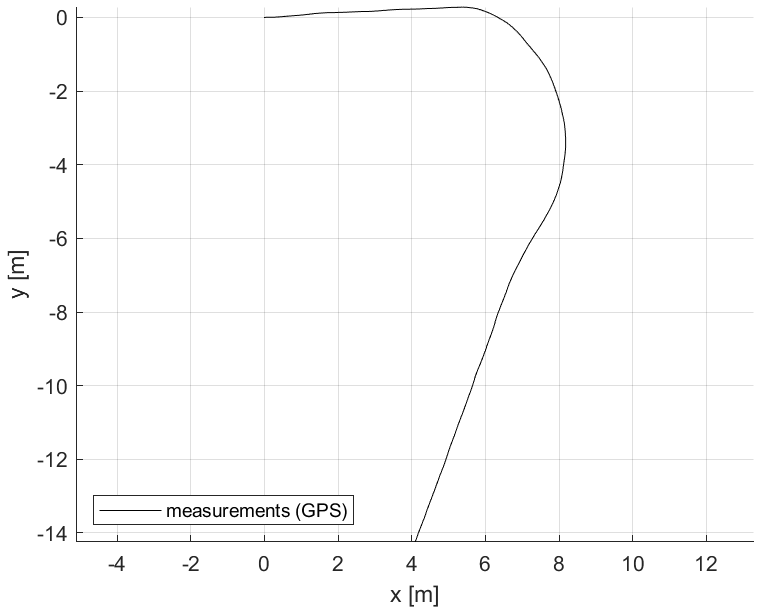}}
\subcaptionbox{Sin\label{fig:path_sin1}}
{\includegraphics[width=0.32\textwidth]{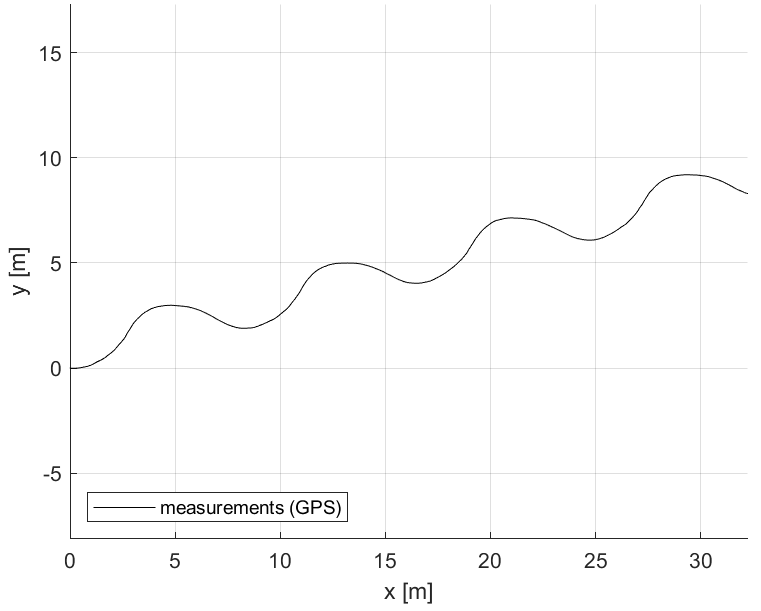}}
\caption{Path taken by the Robotti during field tests across various scenarios.}
\label{fig:scenarios}
\end{figure}

\subsection{Baseline Robotti Multi-model}
\label{sec:robottibaseline}

The baseline multi-model used for the DSE experiments comprises two FMUs: a CT model of the vehicle realised in 20-sim and a DE model of the open-loop controller realised in VDM-RT and Overture. Figure~\ref{fig:multi-model} shows the architecture of the multi-model and the connections between FMUs. The vehicle FMU receives two control signals from the controller FMU, \emph{velocity} and \emph{delta\_f}. The former controls the overall velocity of the vehicle and the latter is the steering angle where a value 0 corresponds to driving in a straight line, and positive or negative values indicate right or left turns respectively. Higher absolute values of \emph{delta\_f} result in a sharper turn. The vehicle FMU reports the current position and rotation of the Robotti through output signals \emph{x}, \emph{y}, and \emph{theta}.

\begin{figure}[tb]
\centering
\subcaptionbox{Architecture Diagram\label{fig:architecture_diagram_rob}}
{\includegraphics[width=0.45\textwidth]{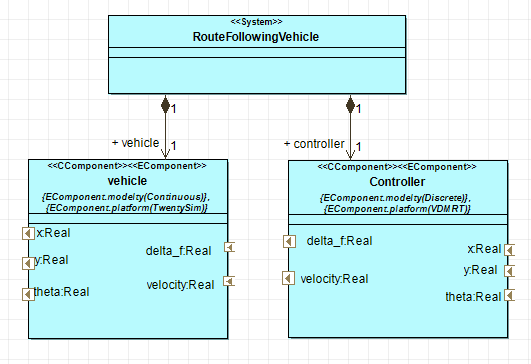}}
\subcaptionbox{Connection Diagram\label{fig:connection_diagram_rob}}
{\includegraphics[width=0.45\textwidth]{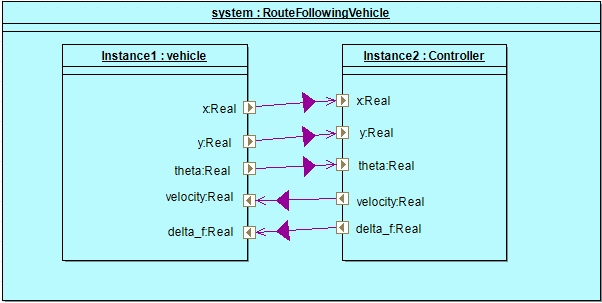}}
\caption{SysML diagrams showing the baseline multi-model of the Robotti.}
\label{fig:multi-model}
\end{figure}

In turn, the controller FMU sets the control signals as outputs and receives the position and rotation signals from the vehicle as inputs. In this pilot study, these inputs are not used. However this interface means that a closed-loop FMU containing a feedback controller can be easily substituted in future with a closed loop controller. The control outputs for the open-loop field tests were recorded in Comma-Separated Values (CSV) format. In order to perform the same experiments as carried out in the field in a co-simulation, it is necessary to ``replay'' these control outputs at the appropriate time during co-simulation. The CSV library included with the Overture tool is used to load the data, then each data row is compared with the wall clock of the VDM-RT interpreter and the output set appropriately.

\section{Robotti Pilot Study: Design Space Exploration}
\label{sec:RobottiDSE}

The objective of the DSE was to calibrate Robotti model parameters using results obtained from a physical robot trial. In this section we describe the DSE methodology applied (Section~\ref{sec:DSEoverview}), the setup (Section~\ref{sec:DSEsetup}) and results obtained so far~(Section~\ref{sec:DSEresults}).

\subsection{Robotti Pilot Study: DSE Analysis Overview}
\label{sec:DSEoverview}

The INTO-CPS toolchain provides a systematic and flexible method for DSE. Flexibility is achieved by allowing the user to configure DSE analysis with a configuration file ($\mathsf{.json}$ format) which defines and orchestrates the exploration process. The methodology also provides an infrastructure for developing and importing external objective functions (as Python scripts) which are used to quantify and rank designs.

Figure~\ref{fig:method} shows the workflow and key components of the Robotti DSE. The region in the dashed box represents an existing INTO-CPS DSE infrastructure which contains (and co-simulates) the Robotti multi-model ($\mathsf{MM}$) and a series of manoeuvre scenarios ($\mathsf{SN}$) with corresponding steering inputs ($\mathsf{steering\_inputs[s_n].csv}$) and recorded positions from the field trial ($\mathsf{gps\_position[s_n].csv}$) for a scenario $\mathsf{s_n}$. Computing a cost of the specific design is performed by objective function script~(Algorithm~\ref{alg:1}) which compares simulated ($\mathsf{results[s_n].csv}$) and recorded positions of the Robotti. The whole DSE process is configured and orchestrated by configuration file ($\mathsf{dse\_configuration.json}$) which specifies process parameters such as search algorithm type, multi-model parameters ($\mathsf{[cAlpha, mass, mu]}$ which represent total mass of the vehicle, tyre stiffness and surface friction respectively) and their value ranges as well as objective function script file. Once the DSE experiment has been configured, an internal search algorithm simulates the Robotti multi-model with specific parameters and the objective function algorithm computes a cost value of the design. The result of this DSE stage is a $\mathsf{dse\_results.csv}$ file which maps different simulated multi-model scenarios to a cost value.


\begin{figure}[h!]
	\centering
	\begin{tikzpicture}[every node/.append style={draw=black!60, inner sep=5pt, minimum size=50pt, fill=gray!5}, scale=\textwidth]
	
	\node[draw=black, fill=gray!5, minimum size = 40pt] (centre) at (0,0) {$\mathsf{MM}$};
	
	\node[draw=none, fill=none, minimum size = 10pt, right = 1 cm of centre] (param) {$\mathsf{[cAlpha, mass, mu]}$};
	
	\node[draw=black, fill=gray!5, minimum size = 40pt, above = 1.5 cm of centre] (centre2) {$\mathsf{SN}$};
	\node[draw=none, fill=none, minimum size = 10pt, left = 1 cm of centre2] (fakecentre2) {};
	
	\node[draw=black, circle,  fill=gray!5, minimum size = 10pt, below = 1.5 cm of centre] (sum)  {+};
	\node[draw=none, fill=none, minimum size = 10pt, left = 1 cm of sum] (mm) {};
	
	\draw[->] (centre) -- (sum) node[draw=none, fill=none,midway, minimum size = 5pt, right] (ref) {\footnotesize{$\mathsf{results[s_n].csv}$}};
	\draw[<-] (centre) -- (centre2) node[draw=none, fill=none,midway, minimum size = 5pt, right] (ref) {\footnotesize{$\mathsf{steering\_inputs[s_n].csv}$}};
	
	\draw[] (centre2.west) -- (fakecentre2.east) -- (mm.west) node[draw=none, fill=none,midway, minimum size = 5pt, left] (ref) {\footnotesize{$\mathsf{gps\_position[s_n].csv}$}};
	
	\draw[->] (mm.west) -- (sum.west);
	\draw[->] (param) -- (centre);

	\node[draw=none, fill=none, minimum size = 40pt, left = 4 cm of centre2, yshift=-7cm] (corner1) {};
	\node[draw=none, fill=none, minimum size = 40pt, left = 4 cm of centre2, yshift=0.5cm] (corner11) {};
	\node[draw=none, fill=none, minimum size = 40pt, right = 4 cm of centre2, yshift=-7cm] (corner2) {};
	\node[draw=none, fill=none, minimum size = 40pt, right = 4 cm of centre2, yshift=0.5cm] (corner21) {};
	
	\draw[dashed] (corner21.north) -- (corner11.north)-- (corner1.north) -- (corner2.north) -- (corner21.north);

	\node[draw=none, fill=none, rotate=0, right = .1 cm of sum] {$\mathsf{Algorithm \; 1}$};
	
	\node[draw=black, fill=gray!5, minimum size = 40pt, below = 1 cm of sum] (op) {$\mathsf{OP}$};
	\node[draw=none, fill=none, minimum size = 0pt, above = .55 cm of op] (op2) {};
	\draw[->] (op2) -- (op) node[draw=none, fill=none,midway, minimum size = 5pt, right] (ref) {\footnotesize{$\mathsf{dse\_results.csv}$}};
	
	\node[draw=none, fill=none, rotate=0, right = .1 cm of op] {$\mathsf{Algorithm \; 2}$};
	
	\node[draw=none, fill=none, minimum size = 10pt, left = .8 cm of op] (param2) {$\mathsf{[cAlpha, mass, mu]}$};
	
	\draw[->] (op) -- (param2);
	
	\node[draw=none, fill=none, rotate=0, left = .7 cm of centre2, yshift=5mm] {$\mathsf{dse\_configuration.json}$};
	
	\node[draw=none, fill=none, rotate=0, right = 1.2 cm of centre2, yshift=5mm] {$\mathsf{INTO-CPS \; Tool}$};
	
	\end{tikzpicture}
	\caption{The DSE analysis workflow overview of the Robotti pilot study}
	\label{fig:method}
\end{figure}
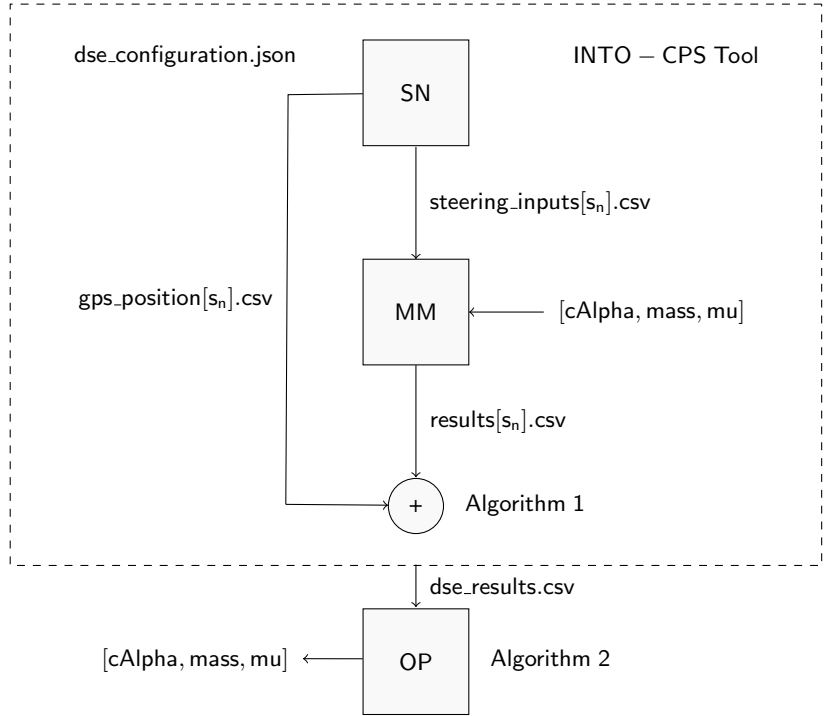

The second stage of the Robotti DSE analysis was concerned with finding Robotti multi-model parameters ($\mathsf{[cAlpha, mass, mu]}$) from a generated $\mathsf{dse\_results.csv}$ data file that would universally fit all simulated scenarios. Finding the best parameter values for all scenarios can be translated to a minimisation problem where an optimisation algorithm attempts to minimise a sum of cost values for different scenarios (with identical $\mathsf{[cAlpha, mass, mu]}$ parameters). In this study the optimisation algorithm ($\mathsf{OP}$) has been implemented as an external Python script, which takes $\mathsf{dse\_results.csv}$ data file and returns best fitting parameter values for all scenarios.

\subsection{Robotti Pilot Study: DSE Configuration and Objective Function}
\label{sec:DSEsetup}
\def\Plus{\texttt{+}}

Performing DSE entails defining an objective function (Python script) and configuration file which will guide the exploration and ranking processes. The configuration file for our pilot study is shown in Listing~\ref{listing1}. It first defines the DSE search algorithm (in our study exhaustive). Next it defines an objective function (external script $\mathsf{robottiCrossTrack.py}$) and simulation outputs used to produce an objective cost value. It then (line 10) specifies the ranges of design parameters which will be used by the exhaustive algorithm to generate different designs. Finally it gives the names of the scenarios to be used.


\renewcommand\thelstlisting{1}
\begin{lstlisting}[language=json,xleftmargin=3em, caption = {Robotti pilot study DSE configuration file}, label={listing1}]
{
"algorithm": {"type":"exhaustive"},

"objectiveDefinitions": {"externalScripts":{ "robottiCrossTrack":
{"scriptFile":"robottiCrossTrack.py", "scriptParameters":{
"1":"time",
"2":"{Robotti}.RobottiInstance.y3_out",
"3":"{Robotti}.RobottiInstance.y4_out"}
},
"parameterConstraints":[],					
"parameters":{
"{Robotti}.RobottiInstance.cAlphaF":[20k,24.5k,29k,33.5k,38k],
"{Robotti}.RobottiInstance.mu":[0.3,0.4,0.5,0.6,0.7],
"{Robotti}.RobottiInstance.m_robot":[1k,1.5k,2k,2.5k,3k]
},
"scenarios":["sin1, sin2, sin3, turn_ramp1, turn_ramp2, turn_ramp3, speed_ramp1, speed_ramp2, speed_step1, speed_step2, speed_step3"]
}
\end{lstlisting}

Given that the purpose here is calibration, we aim for minimal distance between measured and simulated results. Hence a root mean squared quadratic error is considered an appropriate cost function to evaluate. It was decided to base the tuning of the Robotti model parameters on the mean deviation between recorded and simulated robot positions for a specific manoeuvre. The deviation measure has been implemented with a root-mean-squared-error metric which measures mean a distance between two series of coordinates and can be expressed as Equation~\ref{eqn:1}.

\begin{equation}
\mathsf{rmse} = \dfrac{\sum\limits_{i=1}^{n} \sqrt{(x_{i2} - x_{i1})^{2} + (y_{i2} - y_{i1})^{2}}}{n}
\label{eqn:1}
\end{equation}

The measure was implemented as a Python script (described in Algorithm~\ref{alg:1}) which imports $\mathsf{gps\_pos.csv}$ and simulation result $\mathsf{results.csv}$ files. The $\mathsf{cross\_track\_error}$ function then firstly applies a numerator part of the root-mean-squared-error measure for corresponding rows of imported files and once every position entry of the  $\mathsf{gps\_pos.csv}$ has been compared the function computes and returns a mean error value. The measured objective value with the name of objective name are then stored in the $\mathsf{objective.json}$ file with an internal tool function $\mathsf{writeObjectiveToOutfile}$.

\begin{algorithm}[h]
	\caption{Root mean squared error objective function algorithm script}
	\begin{algorithmic}[1]
		\item $\mathsf{[Skeleton \: Code]}$
		\item
		\item $\mathsf{import \; gps\_pos.csv \; as \; gps}$
		\item $\mathsf{import \; results.csv \; as \; results}$
		\item
		\item \textbf{def} $\mathsf{cross\_track\_error(gps, results):}$
		\item \quad $\mathsf{result = [\: ]}$
		\item \quad \textbf{for} $\mathsf{p}$ \textbf{in} \textbf{range}$[0, \; \mathbf{len}(\mathsf{gps})]$
		\item \quad \quad $\mathsf{result}$.$\mathbf{append[}$$\mathsf{sqrt((gps[x_p] - results[x_p])^{2}  + (gps[y_p] - results[y_p])^{2}) }$]	
		\item \quad $\mathsf{mean\_error}$ =  $\mathbf{sum}(\mathsf{result}) / \mathbf{len}(\mathsf{gps})$
		\item \quad \textbf{return} $\mathsf{mean\_error}$
		\item
		\item  $\mathsf{objective\_result}$ = $\mathbf{cross\_track\_error}$$\mathsf{(gps, results)}$
		\item \textbf{writeObjectiveToOutfile}$(\mathsf{objectives.json}, \mathsf{objectiveName}, \mathsf{objective\_result})$
	\end{algorithmic}
\label{alg:1}
\end{algorithm}

Once a cost value has been computed for all permutations of parameters and scenarios, all data is collated and a $\mathsf{dse\_results.csv}$ file is produced, which in the Robotti DSE study contained the mapping between different simulated multi-model scenarios and a cost value.

\subsection{Robotti Pilot Study: Optimum Parameter Search and Results}
\label{sec:DSEresults}

The main objective of this exercise is finding the best fitting parameter value of the model given the different simulated and measured scenarios. The problem of finding the best fitting parameters can be considered as optimisation (function minimization) problem where a function in Robotti pilot study to be minimised maps parameter values to a mean cross track value error computed by Algorithm 1.

The function we intend to minimise is given in Equation 1 where $\mathcal{E}_{s}$ is a set of mean track error functions constructed from the simulated scenarios and parameterised by a scenario ${s \in \{sin1, sin2, sin3 \cdots \}}$ and constrained domain $dom(\mathcal{E})$. The domain of the function $\mathcal{E}_{s}$ is identical to parameter ranges used by the exhaustive search algorithm (see lines 12-14 in Listing \ref{listing1})

\begin{equation}
f^{-} = arg \; \min \sum_{s \in S} \mathcal{E}_{s}(\mathsf{cAlpha, mass, mu})
\end{equation}

The minimization function has been implemented as an external Python script and is described in Algorithm~\ref{alg:2}. The main function of the algorithm $\mathsf{searchParameters}$ explores all permutations of parameter values and for each permutation calls $\mathsf{getResults}$ function which takes specific parameter values and returns a sum of mean squared error values of all scenarios with given parameters. If the returned $\mathsf{total\_error\_value}$ is smaller than the current value of variable $\mathsf{minimum}$ then a new minimum value is assigned and optimal parameter values $\mathsf{params}$ are updated. The algorithm is exhaustive and will terminate once all parameter value permutations have been explored.

\begin{algorithm}[h]
	\caption{Total mean track error value minimisation algorithm}
	\begin{algorithmic}[1]
		\item $\mathbf{def}$ $\mathsf{getResults}(cAlpha, mass, mu):$
		\item \quad \quad $\mathsf{total\_error\_value \: = 0.0}$
		\item \quad \quad \quad $\mathbf{for}$ $\mathsf{s}$ $\mathbf{in}$ $\mathsf{[s_0, s_1, \cdots s_n]}$
		\item \quad \quad \quad \quad$\mathsf{total\_error\_value \: \Plus=  \mathcal{E}_s(cAlpha, mass, mu)}$
		\item \quad \quad $\mathbf{return}$ $\mathsf{total\_error\_value}$
		\item
		\item $\mathbf{def}$ $\mathsf{searchParameters}():$
		\item \quad \quad $\mathsf{minimum}$ = $\mathsf{init}$
		\item \quad \quad$\mathsf{cAlphaF}$ = $[20000, 24500, 29000, 33500, 38000]$
		\item \quad \quad$\mathsf{mass}$ = $[1000,1500,2000,2500, 3000]$
		\item \quad \quad $\mathsf{mu}$ = $[0.3, 0.4, 0.5, 0.6, 0.7]$
		\item
		\item \quad \quad $\mathbf{for}$ $i_0$ $\mathbf{in}$ $\mathbf{range}[\mathsf{0, len(cAlphaF)}]$
		\item \quad \quad \quad \quad $\mathbf{for}$ $i_1$ $\mathbf{in}$ $\mathbf{range}[\mathsf{0, len(mass)}]$
		\item \quad \quad \quad \quad \quad \quad $\mathbf{for}$ $i_2$ $\mathbf{in}$ $\mathbf{range}[\mathsf{0, len(mu)}]$
		\item \quad \quad \quad \quad \quad \quad \quad \quad$\mathbf{if} (\mathsf{getResults(cAlphaF[i_0], mass[i_1], mu[i_2])} < \mathsf{minimum}): $
		\item  \quad \quad \quad \quad \quad \quad \quad \quad  \quad \quad $\mathsf{minimum}$ = $\mathsf{getResults(cAlphaF[i_0], mass[i_1], mu[i_2])}$
		\item  \quad \quad \quad \quad \quad \quad \quad \quad  \quad \quad $\mathsf{parameters}$ = $\mathsf{cAlphaF[i_0], mass[i_1], mu[i_2]}$
		\item
		\item \quad \quad $\mathbf{return}$ $\mathsf{(minimum, parameters)}$
	\end{algorithmic}
\label{alg:2}
\end{algorithm}

The results of the Robotti DSE exercise are summarised in Table~\ref{tbl:1} and Figures~\ref{fig:res_scenarios} and~\ref{fig:res_scenario_groups}. In total 12 different Robotti manoeuvering scenarios have been considered where each scenario was simulated with 125 parameter $\mathsf{[cAlpha, mass, mu]}$ value variations giving us 1500 data points in $\mathsf{dse\_results.csv}$ file. The optimisation algorithm found the best fitting parameter values Robotti multi-model are $\mathsf{CAlphaF}$ = 38000, $\mathsf{mass}$ = 1000,  $\mathsf{\mu}$ = 0.3 resulting in a total mean cross track error of 17.865.

In Table~\ref{tbl:1} for each scenario we provide the smallest and largest possible error values as well as average error over 125 parameter value permutations. The computed parameters for 4 scenarios out of 12 give a below average mean squared error value and 5 slightly above. The interesting case was a $\mathsf{speed\_step}$ scenario (accelerating and braking in a straight line) where, as expected, a mean squared error value was unaffected by parameter variations and therefore had no effect on tuning Robotti parameters. From Figure~\ref{fig:res_scenarios} we can also notice how different parameter values were affecting four different groups of scenarios and we can notice that $\mathsf{sin1}$ and $\mathsf{turn\_ramp1}$ manoeuvres correlate similarly with changing parameter values, whereas $\mathsf{speed\_ramp}$ has the lowest mean squared error when $\mathsf{mu}$ is 0.3 and only very slightly was affected by varying other parameters. Figure~\ref{fig:res_scenario_groups} illustrates another correlation, which includes the scenario groups
$\mathsf{sin, speed\_ramp, turn\_ramp}$ and sub-scenarios (e.g. $\mathsf{sin1, sin2}$) in the individual sub-plots. With this figure we illustrate that the search space for the globally best parameters, where optimised parameters would be universally best was very small and for most scenario groups parameters affect mean squared error unevenly.

\begin{table}[h!]
\begin{center}
		\begin{tabular}{ c | c | c | c | c}
			$\mathsf{scenario}$ \hspace{5mm} & \hspace{2mm} $\mathsf{min\_error}$ \hspace{2mm} & \hspace{2mm} $\mathsf{max\_error}$ \hspace{2mm} & \hspace{2mm} $\mathsf{average\_error}$ \hspace{2mm} & \hspace{5mm} $\mathsf{computed}$ \hspace{5mm} \\
			\hline
			 $\mathsf{sin1}$ \hspace{2mm} & 1.188 & 1.371 & 1.323 & 1.318 (\textcolor{green}{-5E-3}) \\
			 $\mathsf{sin2}$ \hspace{2mm} & 1.680 & 1.960 & 1.884 & 1.891 (\textcolor{red}{+7e-3}) \\
			 $\mathsf{sin3}$ \hspace{2mm} & 2.139 & 9.218& 4.561 & 5.486 (\textcolor{red}{+9.25e-1}) \\
			 $\mathsf{speed\_step1}$ \hspace{2mm} & 0.742 & 0.742 & 0.742 & 0.742 (0)  \\
			 $\mathsf{speed\_step2}$ \hspace{2mm} & 0.578 & 0.578 & 0.578 & 0.578 (0)\\
			 $\mathsf{speed\_step3}$ \hspace{2mm} & 0.482 & 0.482 & 0.482 & 0.482  (0)\\
			 $\mathsf{turn\_ramp}$ \hspace{2mm} & 2.056 & 2.222 & 2.188 & 2.194 (\textcolor{red}{+6e-3})\\
			 $\mathsf{turn\_ramp2}$ \hspace{2mm} & 1.408 & 1.560 & 1.525 & 1.542 (\textcolor{red}{+1.7e-2}) \\
			 $\mathsf{turn\_ramp3}$ \hspace{2mm} & 2.233 & 2.497 & 2.431 & 2.456  (\textcolor{red}{+2.5e-2})\\
			 $\mathsf{speed\_ramp1}$ \hspace{2mm} & 0.307 & 5.120 & 3.244 & 0.312 (\textcolor{green}{-2.932}) \\
			 $\mathsf{speed\_ramp2}$ \hspace{2mm} & 0.367 & 3.965 & 3.328 & 0.430 (\textcolor{green}{-2.898})\\
			 $\mathsf{speed\_ramp3}$ \hspace{2mm} & 0.431 & 1.460 & 1.214 & 0.431 (\textcolor{green}{-1.1709})
		\end{tabular}
\end{center}
\caption{The summary of optimisation results and best fitting parameters: $\mathsf{mean\_cross\_track\_error_{total}}$ = 17.865 with $\mathsf{CAlphaF}$ = 38000, $\mathsf{mass}$ = 1000,  $\mathsf{\mu}$ = 0.3}
\label{tbl:1}
\end{table}

\begin{figure}[h!]
	\centering
	\subcaptionbox{$\mathsf{sin1}$ Scenario\label{fig:res_sin1}}
	{\includegraphics[width=0.48\textwidth]{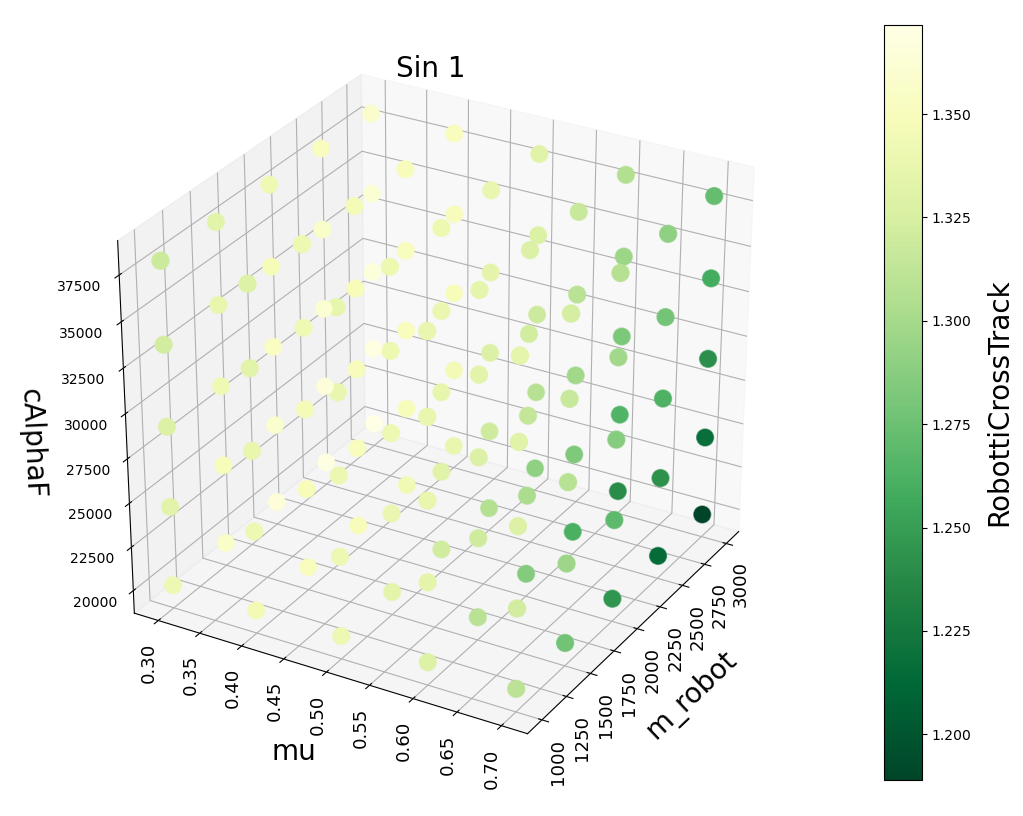}}
	\subcaptionbox{$\mathsf{turn\_ramp1}$ Scenario\label{fig:res_turn_ramp1}}
	{\includegraphics[width=0.48\textwidth]{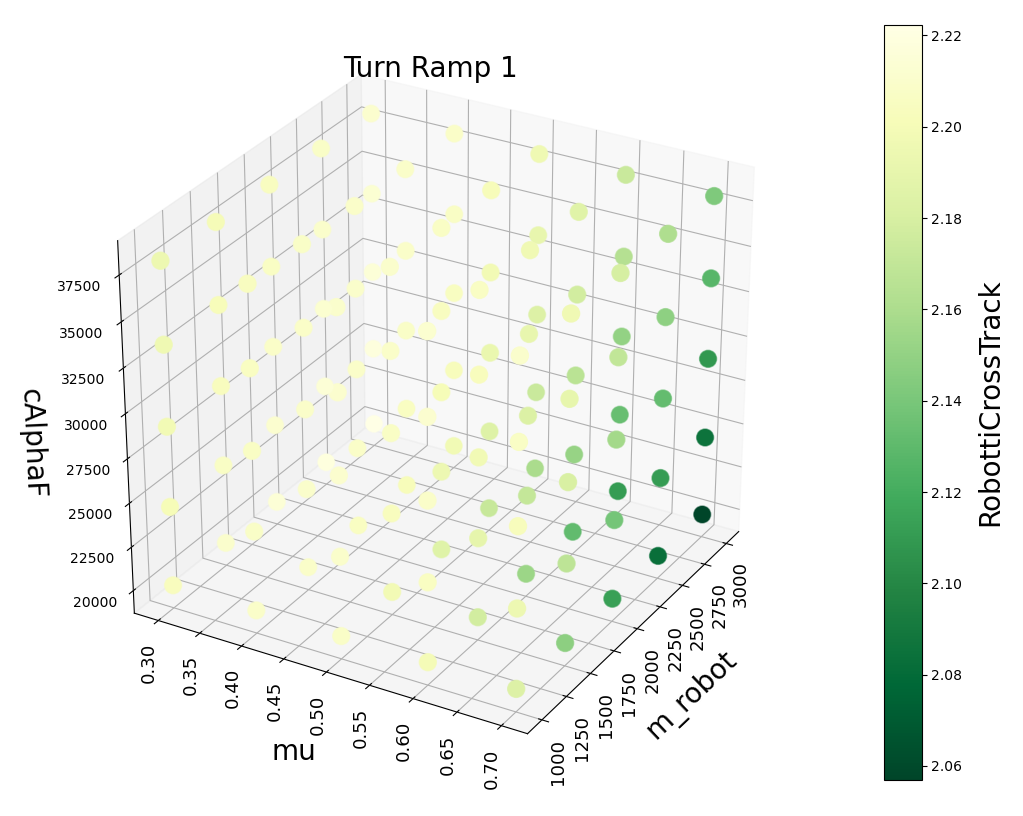}}\\
	\subcaptionbox{$\mathsf{speed\_ramp1}$ Scenario\label{fig:res_speed_ramp1}}
	{\includegraphics[width=0.48\textwidth]{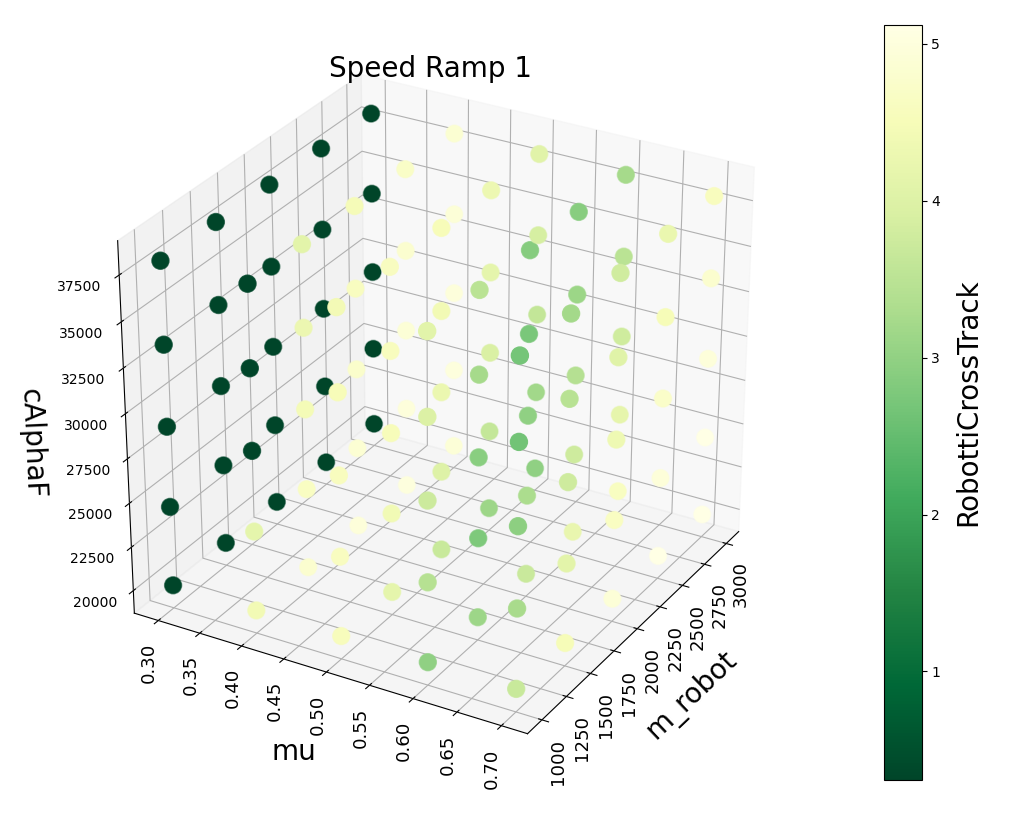}}
	\subcaptionbox{$\mathsf{speed\_step1}$ Scenario\label{fig:res_speed_step1}}
	{\includegraphics[width=0.48\textwidth]{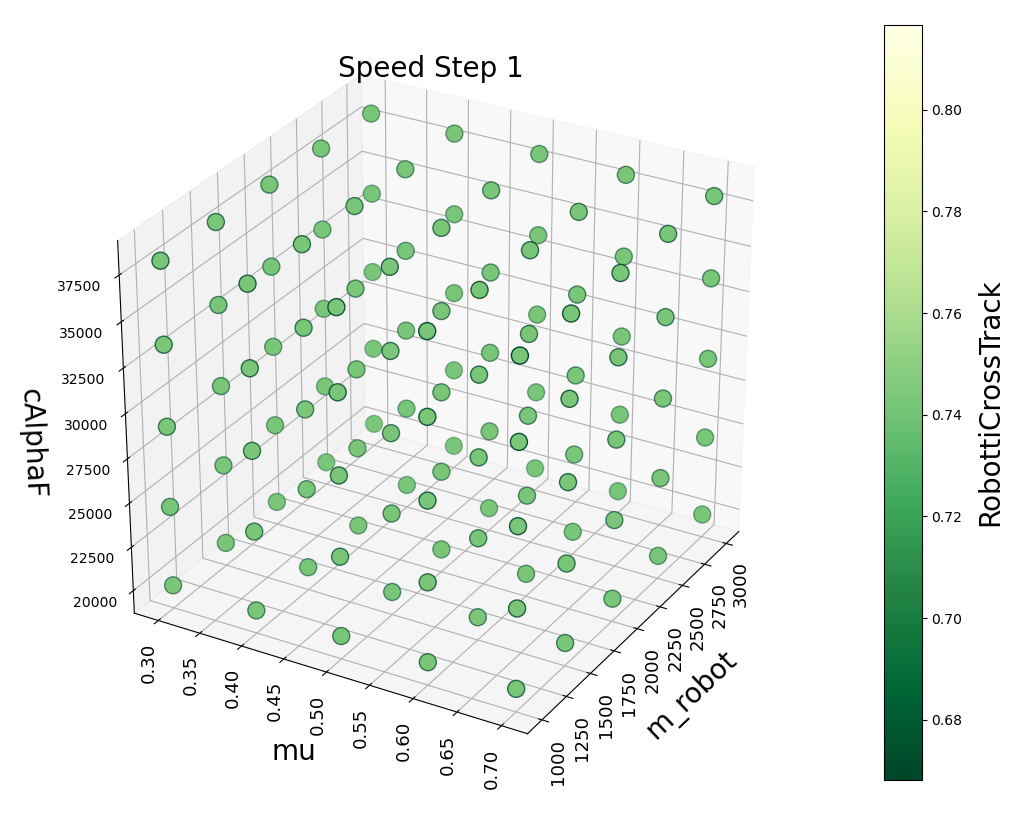}}
	\caption{Mean cross track error variation of scenarios against different parameter values}
\label{fig:res_scenarios}
\end{figure}

\begin{figure}[h!]
	\includegraphics[width=.9\paperwidth]{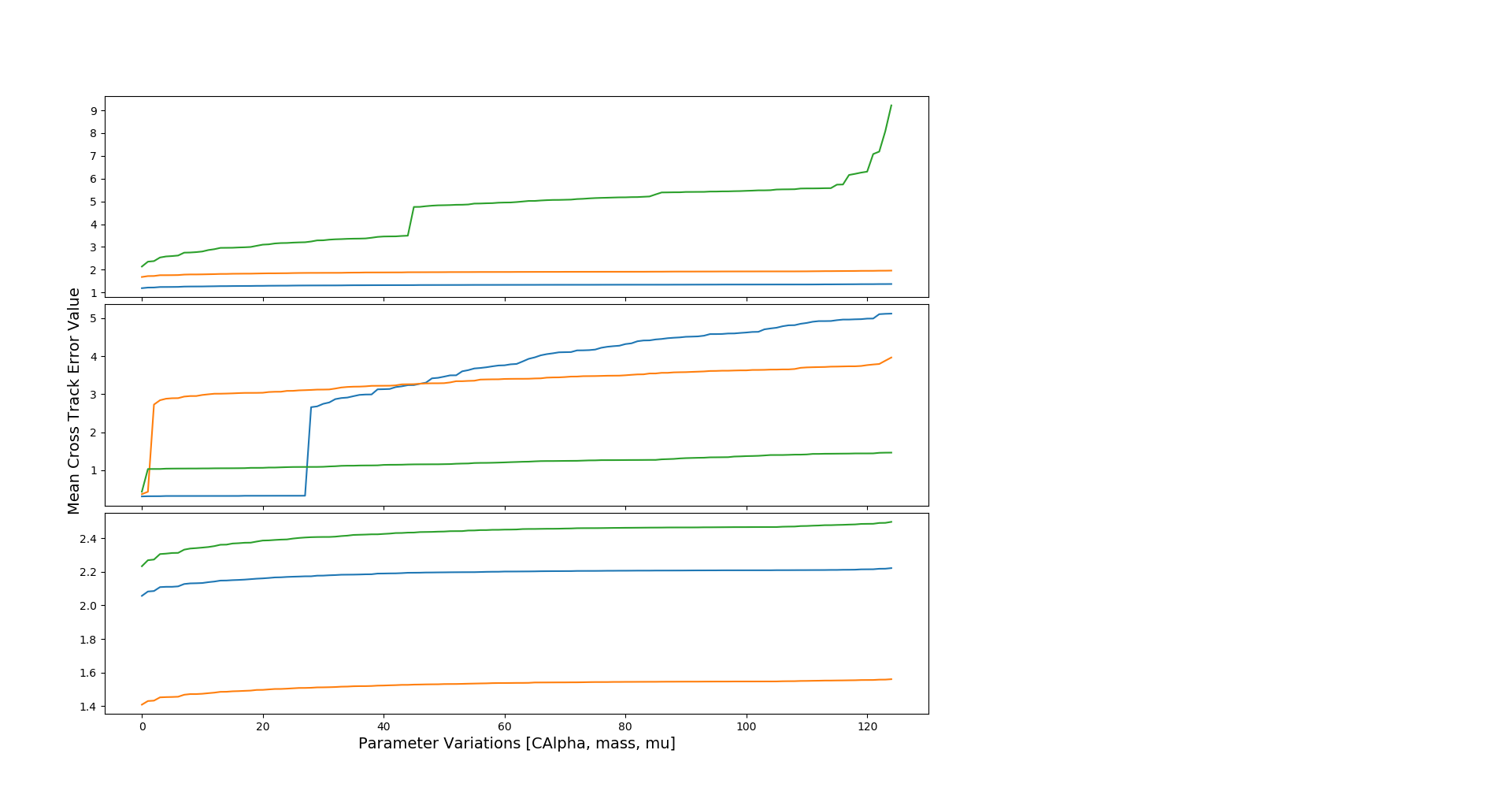}
	\caption{Mean cross track error for different scenario groups (from top $\mathsf{sinN, speed\_rampN, turn\_rampN}$) and scenarios (blue ($\mathsf{N}$=1), orange ($\mathsf{N}$=2), green ($\mathsf{N}$=3)) }
\label{fig:res_scenario_groups}
\end{figure} 
\section{Discussion}
\label{sec:discussion}

We have so far undertaken the data engineering needed to perform co-simulation on the Robotti CT model, and have demonstrated DSE-based tuning of this model over three design parameters against an objective function based on mean deviation between the model and the field data. The study, though so far incomplete, has proved technically viable for this scale of multi-model and volume of data. We here consider two main discussion areas based on the experience gained with DSE, INTO-CPS and the Robotti.

Firstly, as with all tools, there is a need for constant updating to stay relevant and useful in a changing technological landscape. While using INTO-CPS, we found difficulties in two main areas. First, INTO-CPS requires Python 2.7 which for most computers required backdating from the more recent Python 3, which is not backwards compatible. A real work-around was necessary to run INTO-CPS tools. Second, in ranking the data, the Pareto Method is built into the toolchain, it ranks simulations against two parameters to provide a Pareto front of optimal values. For any simulations that want to rank against a single parameter the functionality was not simple to find. We also wanted to compare different scenarios in the same set of ranking which is also not a feature readily available in INTO-CPS. In the end we resorted to using Pareto with a second parameter, max cross track error, but as we were not interested in this parameter we ended up writing our own scripts for ranking and finding the optimal parameters across the scenarios.
	

Secondly, although parameter selection is important when considering the insights that DSE can offer, seeking to optimise performance through DSE, it is equally important to consider the scenarios. 
For example, in the $\mathsf{sin}$ scenarios, all three parameters we looked at affect the cross track error. However, none of them had an impact in the $\mathsf{speed\_step}$ scenarios and whichever parameter combination was selected the same cross track error was measured. It was theorised that the parameters we discussed were only relevant when the robot would be turning, hence the largest impact appears for the sin wave which has many turns and no impact was felt on the $\mathsf{speed\_step}$ which has no turning. To improve our results, it is necessary to consider the parameters which constantly affect the robot, not just when turning, so that the most accurate behaviour can be predicted.

In terms of further work, the next step in the pilot study is to complete the co-simulation of the ADRC controller with the plant model that has been tuned so far, and confirm the extent to which this is faithful to data gathered from field trials. An ADRC controller in itself has about 6 parameters, so exhaustive coverage may not be available within available resources, suggesting the use of genetic algorithms to select co-simulation runs. Aside from the tooling issues above, a priority for future work might be strengthening support for managing scenarios, and more user-friendly methods for generating custom rankings.

\paragraph*{Acknowledgements.}
We are grateful to AgroIntelli for supplying Robotti data, and to Sadegh Soudjani at Newcastle. We acknowledge the European Union's support for the INTO-CPS and HUBCAP projects (Grant Agreements 644047 and 872698). This research was supported in part by the Air Force Office of Scientific Research under award number FA2386-17-1-4065. Any opinions, findings, and conclusions or recommendations expressed in this
material are those of the authors and do not necessarily reflect the views of the United States Air Force.

\bibliographystyle{splncs03}

 \newcommand{\noop}[1]{}

\clearpage
\restorevalues
\endgroup

\begingroup

    \renewcommand{\textfraction}{0.07}	

\title{A Co-Simulation Based Approach for Developing Safety-Critical Systems}
%
%
\author{
Daniella Tola\inst{1} \and
Peter Gorm Larsen\inst{1}
}
\authorrunning{D. Tola and P. Larsen}
%
\institute{
DIGIT, Department of Engineering, Aarhus University, Aarhus, Denmark\\
\email{\{dt,pgl\}@eng.au.dk}}
\maketitle              
\setcounter{footnote}{0} 

\setlength{\textwidth}{12.2cm}
\setlength{\textheight}{19.3cm}

\fontsize{11}{12}\selectfont
\makeatletter
\def\input@path{{5-Tola/}}
\makeatother

\graphicspath{{5-Tola}}

\begin{abstract}
When developing safety-critical systems it is important to provide evidence that such systems do not pose a significant hazard to their surroundings. One way of providing such evidence is known as safety cases and here the Goal Structuring Notation is one of the most popular notations. Evidence can take many different forms and in this paper we explore whether co-simulation of critical scenarios can be used with advantage as a beneficial approach. This is illustrated with a combine harvester using different tools in a Functional Mockup Interface setting. 

\keywords{Goal Structuring Notation  \and Co-simulation \and Functional Mockup Interface.}
\end{abstract}

\section{Process Overview} \label{sec:process_overview}

This section gives a brief overview of the proposed process; more details can be found in~\cite{Tola20}. This process provides clear guidelines on how to incorporate co-simulation in the development of safety-critical systems. The process does not yet include how to ensure the validity of the evidence obtained using co-simulations of the system.

The six main stages of the process are illustrated in Figure~\ref{fig:complete_process_enum}. These stages describe how to start from a hazard analysis and potentially end up with parts of evidence for a safety case. The first two stages are part of the traditional process of creating a safety case, where a safety and hazard analysis is performed, followed by defining the safety goals of the system. Before proceeding to the third stage, the hazards must be studied to determine which method should be used for creating evidence. How to determine this, is described in the following section. If co-simulation is chosen as an appropriate method for creating evidence, then the next step is to specify test scenarios and create models of the different parts of the system. This is followed by co-simulations of the specified test scenarios, leading to results that if prove the system is safe under the specified conditions, can be used as evidence in the safety case. If the results imply the system is not safe under the specified conditions, then safety improvements must be made and the system must be co-simulated in the failed test scenarios again.

\begin{figure}[htbp]
    \centering
    \includegraphics[width=\textwidth]{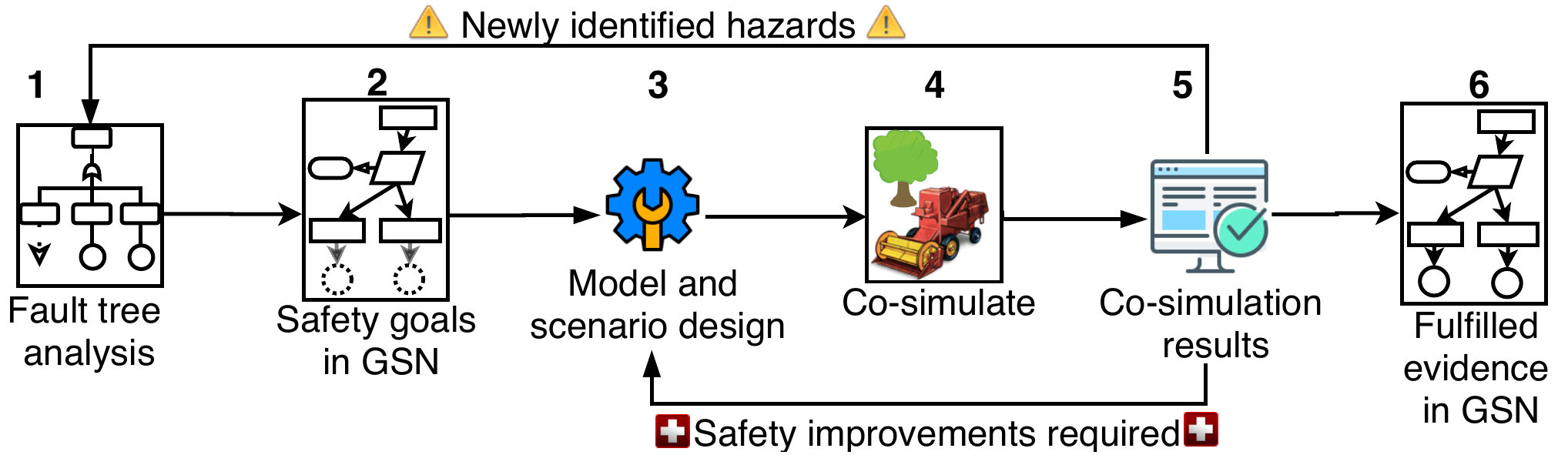}
    \caption{An overview of the method of developing a safety case using co-simulation.}
    \label{fig:complete_process_enum}
\end{figure}

The contributed stages of the process of creating safety evidence are stages three, four and five, meaning that other safety processes construct evidence using other techniques in these exact stages. It is important to note that determining which elements of the safety evidence that can be constructed using co-simulation results, must be defined in stage two, before moving on to stage three. If it appears that no evidence can be created using co-simulation, then another evidence creation technique must be used, for example mathematical analysis.

The following sections describe how the process was used on the case study, where Section~\ref{sec:safety_analysis} describes the safety analysis, and Section~\ref{sec:evidence_cosim} describes how the evidence was created using co-simulation.

\section{Safety Case Analysis} \label{sec:safety_analysis}

This section describes how the safety case analysis of the autonomous combine harvester was performed. The steps illustrated above in Figure~\ref{fig:main_stage_sc} are followed, where the first step of describing the system can be found above in Section~\ref{sec:background:case_study}. The details of the safety case and system are at a high abstraction level, since a Preliminary safety case is the objective.

Hazards were identified by consulting with an agricultural engineer from AGCO and reading articles. Some of the identified hazards are: \textit{H1: collision with object}, \textit{H3: fire ignition}~\cite{Val&19}, and \textit{H5: contamination of harvested crops}. A fault-tree analysis was performed on the hazards to determine potential root causes, see Figure~\ref{fig:ft} for a simplified fault-tree view. This was followed by a risk analysis to determine which hazards posed the highest risks. The risk analysis showed that hazards \textit{H1} and \textit{H3} posed the highest risks, which are the two hazards considered in this paper.

By analysing these two hazards with regards to how they can be mitigated, it was possible to determine that \textit{H1} can be mitigated using software and redundant sensors, while \textit{H3} can be mitigated using mechanical structures for isolating surfaces that may catch on fire. Comparing the mitigation methods of these two hazards, it is clear that \textit{H1} is a relevant safety hazard to provide evidence for using co-simulation since it is possible to model the sensors, vehicle dynamics, and simulate the software for mitigating the hazard. \textit{H3} is somewhat more difficult to co-simulate since the model would be more focused on the mechanics of the system.

A compact fault-tree analysis of the hazard \textit{H1} is demonstrated in Figure~\ref{fig:ft}, where the events of the camera view being obscured or a decreased LiDAR performance could potentially lead to \textit{H1} occurring. Possible root causes of these two events are illustrated in the fault-tree as heavy rain and dense fog. This information is used when describing the safety goals that are formalized using GSN, illustrated in Figure~\ref{fig:gsn}. The evidence for fulfilling the safety goals \textit{G2}, \textit{G3} (from Figure~\ref{fig:gsn}) and \textit{G4: The system can tolerate inaccurate sensor measurements} is defined as follows:

\begin{description}
    \item[Evidence for G2:] Co-simulation results of heavy rain scenarios.
    \item[Evidence for G3:] Co-simulation results of dense fog scenarios.
    \item[Evidence for G4:] Co-simulation results of scenarios with inaccurate sensor measurements.  
\end{description}


\begin{figure}[htbp]
    \centering
    \begin{subfigure}[b]{0.44\textwidth}
        \centering
        \includegraphics[width=\textwidth]{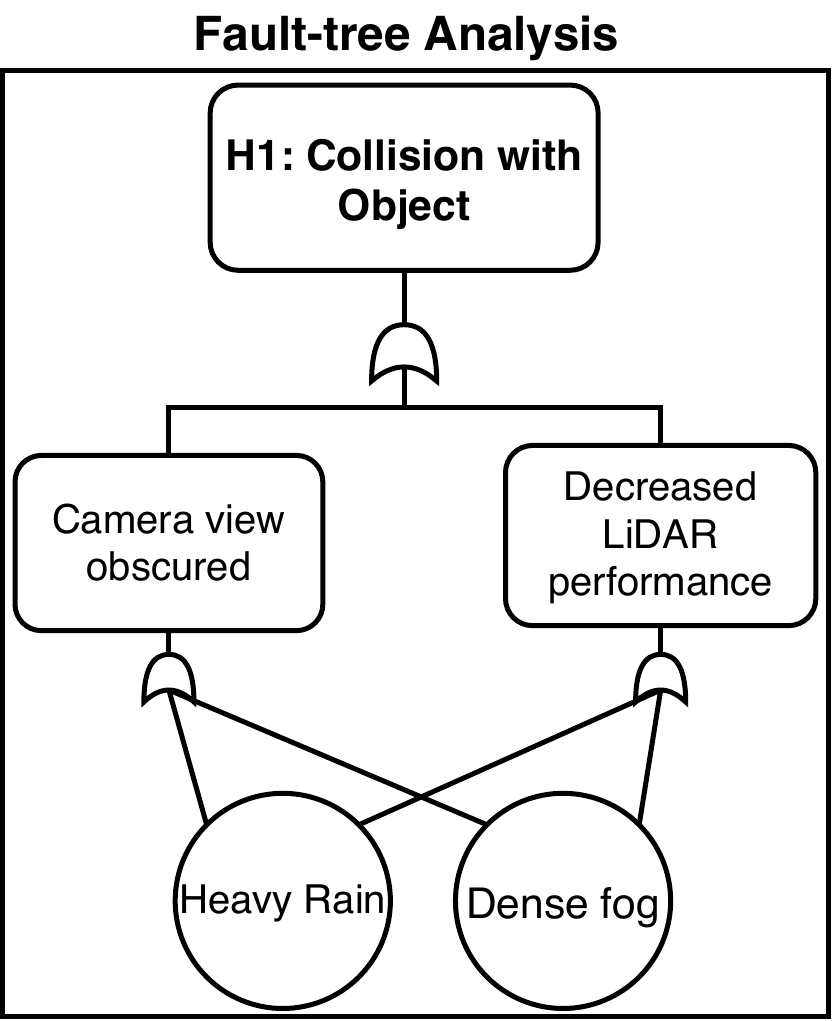}
        \caption{Fault-tree analysis of \textit{H1} with only few components included.}
        \label{fig:ft}
    \end{subfigure}
    \hfill
    \begin{subfigure}[b]{0.454\textwidth}
        \centering
        \includegraphics[width=\textwidth]{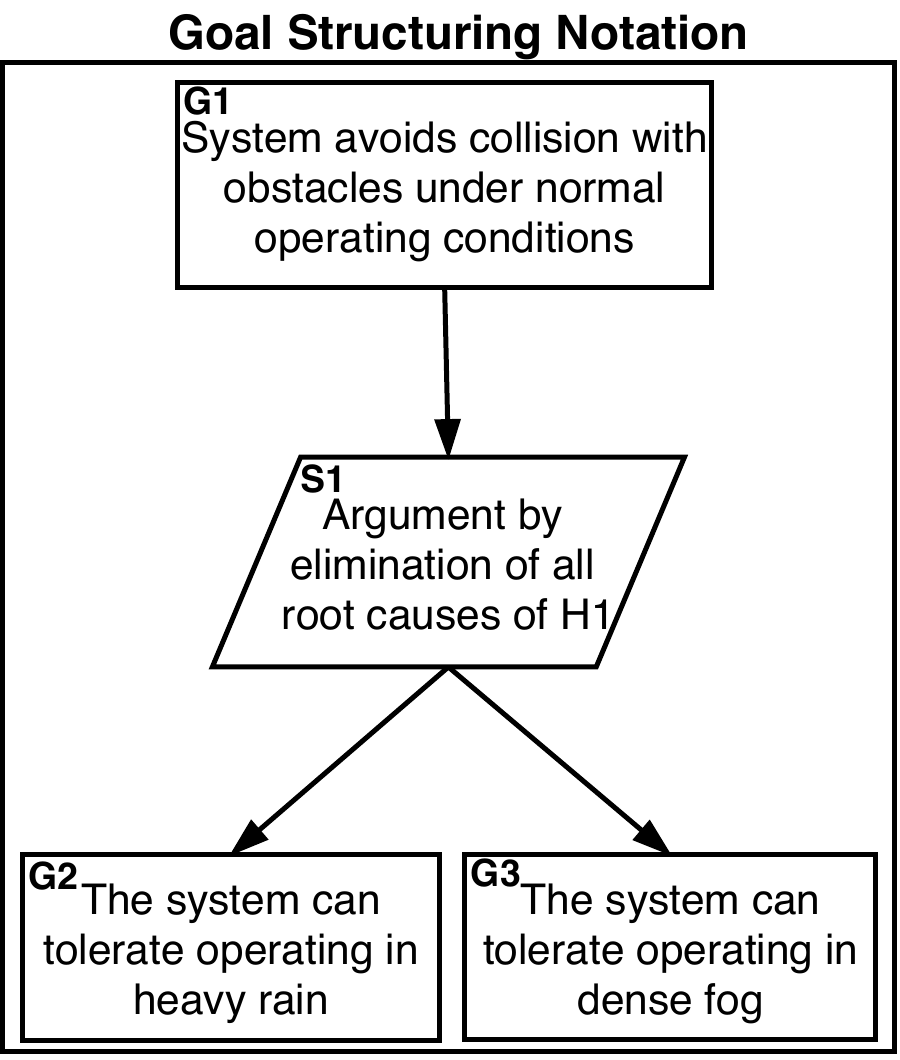}
        \caption{Initial GSN of \textit{H1} with only few components included.}
        \label{fig:gsn}
    \end{subfigure}
   \caption{Example of how the fault-tree analysis can directly be translated to GSN. Further details of the diagrams can be found in~\cite{Tola20}.}
   \label{fig:ft_to_gsn}
\end{figure}

\section{Producing Evidence using Co-simulation} \label{sec:evidence_cosim}






This section describes the process of creating evidence using co-simulation, by demonstrating how to follow the steps 3, 4, 5, and 6 presented above in Figure~\ref{fig:complete_process_enum}.

The first step consists of designing test scenarios and creating the relevant models. The test scenarios must be relevant to the physical system and possible to be executed in a co-simulation. The scenarios should be defined clearly in a way that makes it possible to reproduce the results of the co-simulation. They must also be directly related to the safety goals in a GSN, which in some cases are related to the hazard root causes. Table~\ref{tab:scenarios_overview} describes the defined test scenarios, related to the GSN in Figure~\ref{fig:gsn}, where the specific hazardous events are defined together with the vehicle speed and distance to the obstacle. These scenarios are typically unethical or expensive to test on the physical system, and therefore can beneficially be co-simulated instead. The goal is to co-simulate the system in these scenarios, and demonstrate its safety under the hazardous conditions.

\begin{table}[bt]
    \centering
    \begin{tabular}{|p{3cm}|p{2.7cm}|p{4.3cm}|}
    \hline
    \textbf{Hazardous event} & \textbf{Vehicle speed} & \textbf{Initial distance to obstacle} \\ \hline
    Dense fog             & \{1,2,3\} {[}m/s{]} & \textgreater 1 meter \\ \hline
    Heavy rain            & \{1,2,3\} {[}m/s{]} & \textgreater 1 meter \\ \hline
    Inaccurate sensor     & \{1,2,3\} {[}m/s{]} & \textgreater 1 meter \\ \hline
\end{tabular}
\caption{Defined test scenarios.}
\label{tab:scenarios_overview}
\end{table}

In order to co-simulate these test scenarios, models of the vehicle dynamics, vehicle controller, sensors, environment and a safety controller must be created. Each of the models created in this project are described in Table~\ref{tab:models}. The FMUs containing the models of the vehicle dynamics and controller were taken from an example project from INTO-CPS\footnote{\url{https://github.com/INTO-CPS-Association/example-autonomous_vehicle}} \cite{5-Tola:Fitzgerald&15}. The remaining FMUs were modelled using the python library, PyFMU~\cite{Legaard&20}. The models of the system parts were created with only the necessary details that are relevant in this stage of the project, since a Preliminary safety case is the objective.

\begin{table}[]
    \begin{tabular}{|p{1.9cm}|p{1.5cm}|p{8.6cm}|}
    \hline
    \textbf{Model}         & \textbf{Program} & \textbf{Description} \\ \hline
    Vehicle  & 20-sim\tablefootnote{\url{https://www.20sim.com/}}    & Models the vehicle dynamics using a bicycle model.\\ \hline

    Controller  & Overture\tablefootnote{\url{http://overturetool.org/}} & Controls the vehicle using the pure-pursuit path following algorithm. \\ \hline

    Sensor  & PyFMU\tablefootnote{\url{https://pypi.org/project/pyfmu/}}  & Models the three types of sensors combined, i.e. LiDAR, radar and camera. A more detailed explanation of this model is presented below. \\ \hline

    Environment  & PyFMU  & Models the obstacles in the environment surrounding the vehicle. The environment was modelled using the python library OpenCV\tablefootnote{\url{https://opencv.org/}}. The environment can be loaded using a white background image, with obstacles defined as coloured objects on the image. This allows easily creating different environments. \\ \hline

    Supervisory Controller & PyFMU  & Safety controller that performs a safety stop before crashing into an obstacle. The controller performs a safety stop if a detected obstacle is within the range of the calculated braking distance of the vehicle. The obstacles are detected using the sensor model. Therefore, if the sensor model does not detect an obstacle, then the Supervisory Controller will not receive information about an obstacle, and will therefore not perform a safety stop.\\ \hline 
\end{tabular}
\caption{Models used in the project, together with the program that they were modelled in, and a description of each model.}
\label{tab:models}
\end{table}


The main parts of the process of creating the sensor model are described below. This information is used to illustrate how to combine knowledge from the hazardous events and the sensors of the system. As described above in Section~\ref{sec:background:case_study}, a camera, LiDAR and radar are mounted on the system. The effect of dense fog and heavy rain on cameras and state-of-the-art LiDARs is described as a reduction in the maximum viewing range of the sensors~\cite{Bijelic&18,Heinzler&19,Liu&20,Hennicks&19}. Figure~\ref{fig:fog_rain_model} illustrates how dense fog and heavy rain can be modelled using a single sensor model at a high abstraction level.

\begin{figure}[htbp]
    \centering
    \includegraphics[width=\textwidth]{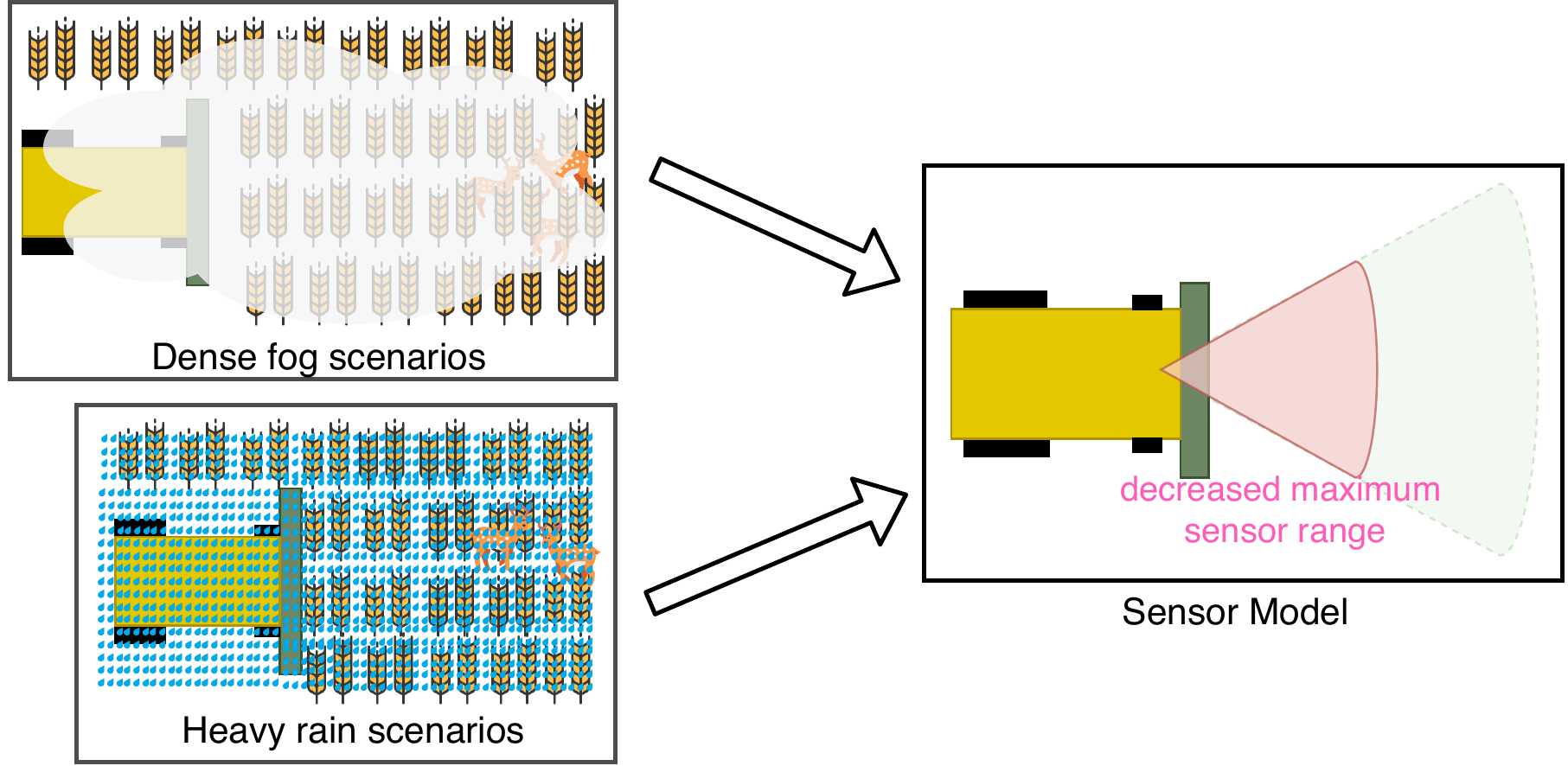}
    \caption{An illustration of how the scenarios related to dense fog and heavy rain may be modelled simultaneously, since they affect the system in similar ways. The figure demonstrates that the scenarios where dense fog or heavy rain occur affect the maximum viewing range of the sensor.}
    \label{fig:fog_rain_model}
\end{figure}

Figure~\ref{fig:sensor_generalization} illustrates how the sensor model combines all the sensors into one, instead of creating a model of each sensor. This allowed modelling the complete effect of the environment, using adjustable parameters defining the minimum and maximum sensor range. The minimum range of the sensor represents the hazard of inaccurate sensor measurements~\cite{Sukhan&12,Clark20}. In the future, when more knowledge about the sensors and the system is obtained, details can be added to the constructed models iteratively~\cite{Maria97}.


\begin{figure}[htbp]
     \centering
     \begin{subfigure}[b]{0.48\textwidth}
         \centering
         \includegraphics[width=\textwidth]{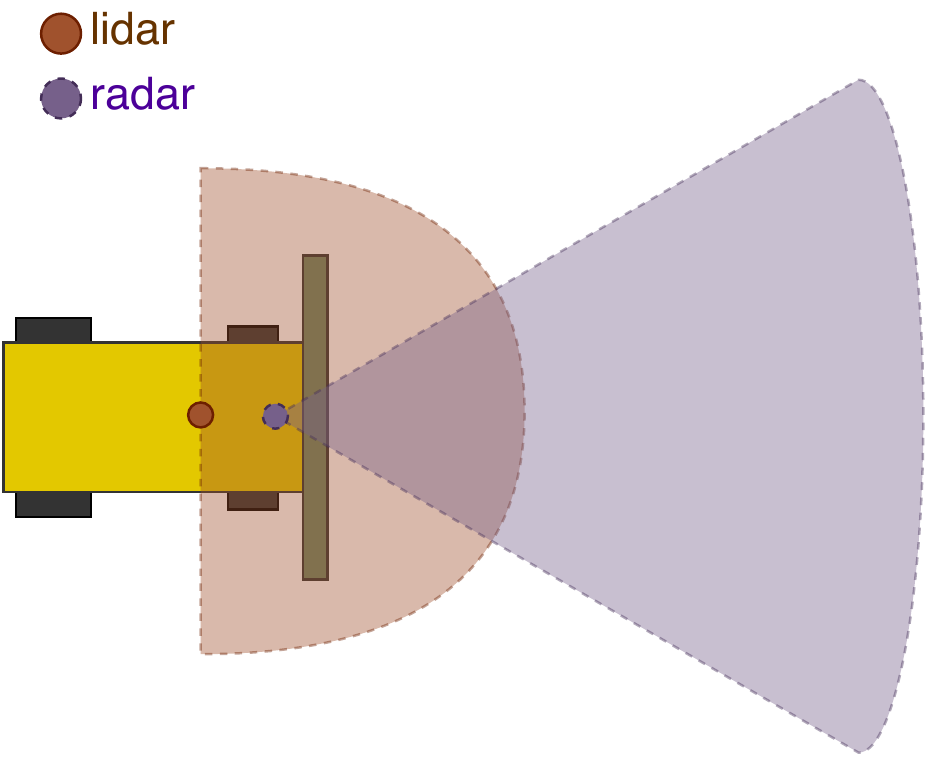}
         \caption{A mounted lidar and radar on a combine harvester, showing the range of each of the sensors. This is just an example, showing two mounted sensors, but in reality there will be several types of sensors mounted in different places on such a system.}
         \label{fig:sensor_true}
     \end{subfigure}
     \hfill
     \begin{subfigure}[b]{0.42\textwidth}
         \centering
         \includegraphics[width=\textwidth]{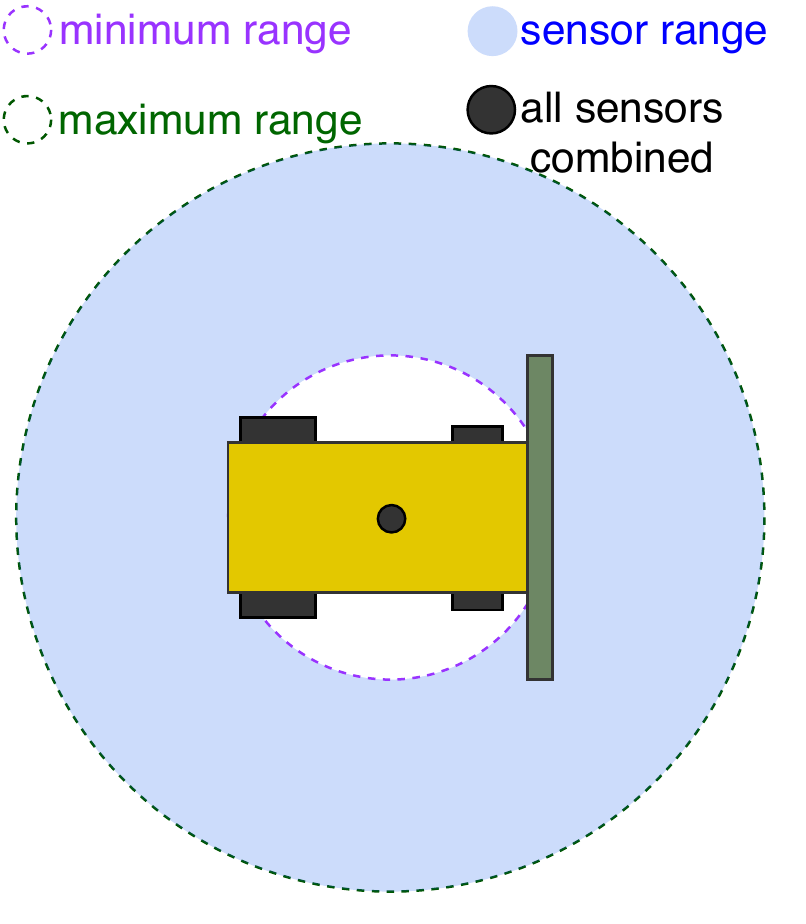}
         \caption{An example illustrating how the sensors are modelled, by generalizing a combination of all of the sensors on a combine harvester. The range of the generalized sensors together is modelled as a circular area around the vehicle, with a minimum and maximum range.}
         \label{fig:sensor_generalized}
     \end{subfigure}
    \caption{Generalization of sensors on the autonomous combine harvester.}
    \label{fig:sensor_generalization}
\end{figure}

In total, 2 scenarios were co-simulated, with the parameters described below, leading to in total 12 co-simulation runs:
\begin{description}
    \item[Environment:] Two different obstacle maps, with varying obstacle locations, densities and sizes.
    \item[Vehicle speed:] 1, 2, and 3 [m/s]. These values are taken from~\cite{Isaac&06}, which describes the optimim combine harvester speed is around 2.2~{[}m/s{]}.
    \item[Minimum sensor range:] When operating correctly, the range was modelled as 0.5m. When operating with inaccurate measurements, the range was modelled as 1m~\cite{Sukhan&12,Clark20}.
    \item[Maximum sensor range:] When operating in dense fog or heavy rain, the range was modelled as 2m, which is the worst case values found in the articles~\cite{Bijelic&18,Heinzler&19,Liu&20,Hennicks&19}.
\end{description}

The different values of the parameters and number of co-simulation runs must be sufficient in order to be able to use the results as evidence. The worst-case values were chosen, but potentially more confidence in the evidence of these hazardous conditions could be obtained using best-case values and at least one value in between.

Figure~\ref{fig:cosim_range} shows one of the live plots of a co-simulation run, where the vehicle successfully performs a safety stop before colliding with the detected obstacles. In the cases where the vehicle failed to perform a safety stop, the system had run over the obstacles.

\begin{figure}[htbp]
    \centering
    \includegraphics[width=0.7\textwidth]{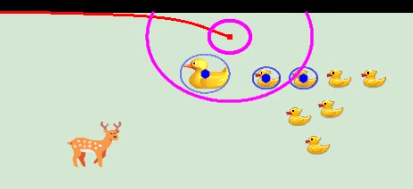}
    \caption{An example of a co-simulation run of the system with visualization of the minimum and maximum range of the sensor, and the currently detected objects. The red lines displays the route the vehicle has driven so far. The two pink circles represent the sensor minimum and maximum range. The blue circles represent the detected objects. The minimum and maximum ranges are set to 0.5m and 2m respectively in this co-simulation, and the vehicle has performed a safety stop.}
    \label{fig:cosim_range}
\end{figure}

The co-simulation results were successful for the scenarios simulating the dense fog and heavy rain, but unfortunately failed in a number of the scenarios simulating the inaccurate sensor measurements. The successful results were used as evidence in the GSN shown in Figure~\ref{fig:gsn_evidence}; notice the validated model is used as part of the argument for justifying the use of co-simulation as evidence. The model is not yet validated, and therefore the away goal, \textit{G0\_ModelValidated}, describing it is greyed out. The unsuccessful co-simulation results mean that improvements on the safety mechanisms must be made, in order to obtain the evidence \textit{Sn3}. It is important to note that the safety properties derived by co-simulation results will only be valid to be used as evidence, if the model of the system is validated and sufficiently represents the true system.

\begin{figure}[htbp]
    \centering
    \includegraphics[width=\textwidth]{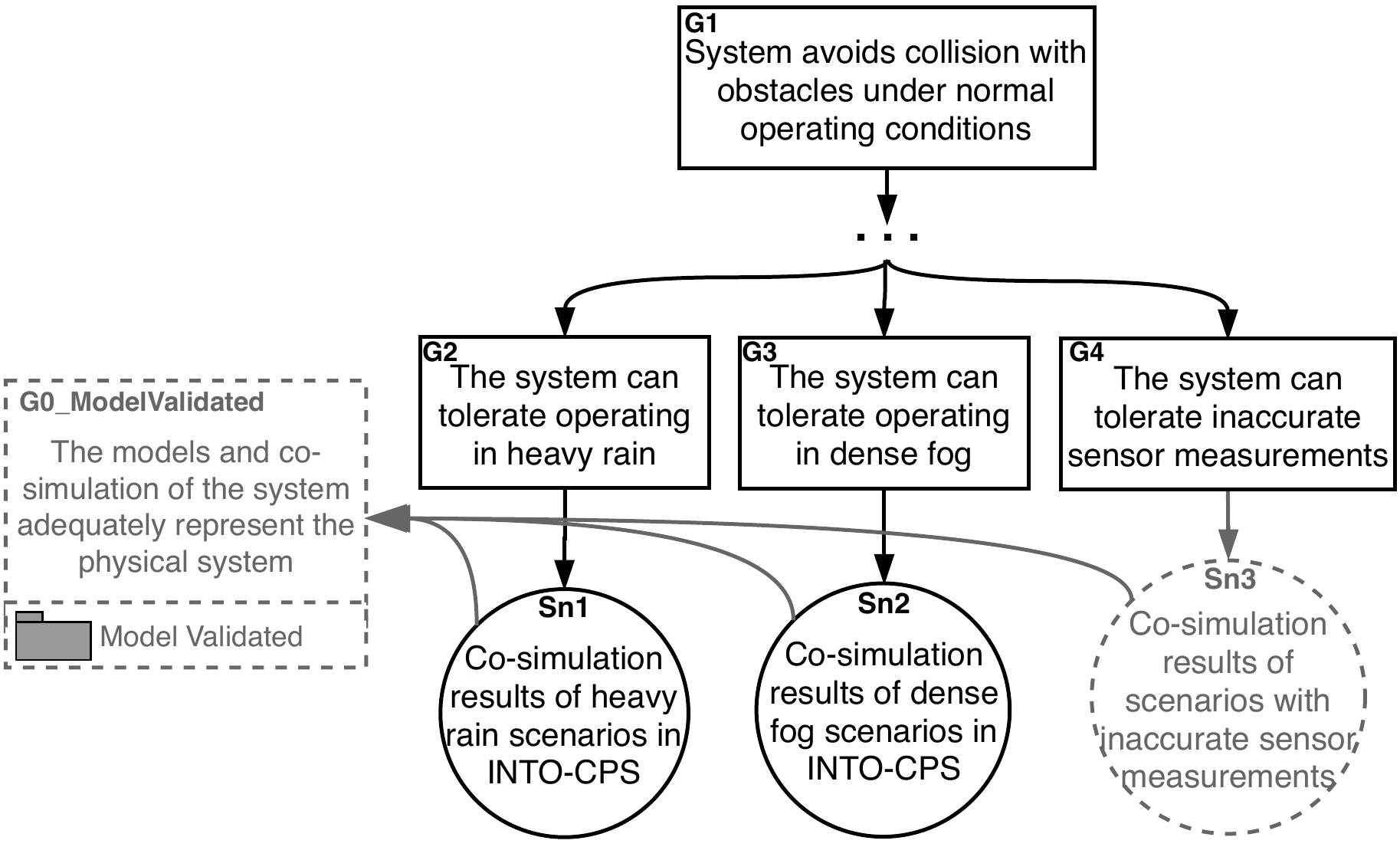}
    \caption{The relevant parts of the GSN diagram, illustrating which evidence was obtained using co-simulation. The dashed border of \textit{Sn3} and \textit{G0\_ModelValidated} illustrates that these components are yet to be constructed, i.e. the evidence for \textit{Sn3} must be obtained, and the model used in the evidence \textit{Sn1} and \textit{Sn2} must be validated against the physical system.}
    \label{fig:gsn_evidence}
\end{figure}


%
%
%

\subsubsection*{Acknowledgements.}
We would like to thank Martin Peter Christensen from AGCO for the discussions about the challenges for an autonomous combine harvester. We would also like to express our thanks to the anonymous reviewers.

\bibliographystyle{splncs04}

\clearpage
\endgroup

\end{document}